\documentclass[preprint,english,american,superscriptaddress,prd,a4paper,nofootinbib,11pt]{revtex4-2}
\pdfoutput=1

\usepackage[usenames,dvipsnames]{xcolor}

\usepackage{amssymb,amsmath}
\usepackage{url,booktabs}
\usepackage{multirow}
\usepackage[utf8]{inputenc}
\usepackage[T1,T2A]{fontenc}
\usepackage{slashed}

\usepackage{graphicx}        
\usepackage{subfig}
\usepackage[export]{adjustbox}

\def\be{\begin{equation}}
\def\ee{\end{equation}}
\def\bea{\begin{eqnarray}}
\def\eea{\end{eqnarray}}

\newcommand{\met}{E^{\rm miss}_T}

\usepackage{amsmath,amssymb,amsfonts,mathrsfs}
\usepackage{tikz}
\usetikzlibrary{arrows,shapes}
\usetikzlibrary{trees}
\usetikzlibrary{matrix,arrows} 				
\usepackage{pgffor}							

\tikzstyle{decision} = [diamond, draw, fill=blue!20, 
text width=4.5em, text badly centered, node distance=3cm, inner sep=0pt]
\tikzstyle{block} = [rectangle, draw, fill=blue!20, 
text width=5em, text centered, rounded corners, minimum height=2em]
\tikzstyle{line} = [draw, -latex']
\tikzstyle{cloud} = [draw, ellipse,fill=red!20, node distance=3cm,
minimum height=2em]
\usepackage{ulem}

\definecolor{deepcarrotorange}{rgb}{0.91, 0.41, 0.17}
\definecolor{dartmouthgreen}{rgb}{0.05, 0.5, 0.06}

\begin{document}

\title{Multilepton Signatures from Dark Matter at the LHC}
\preprint{ADP-22-15/T1186}	
	
\author{Alexander Belyaev}
\email{a.belyaev@phys.soton.ac.uk}
\affiliation{School of Physics and Astronomy, University of Southampton, Highfield. Southampton SO17 1BJ, United Kingdom}
\affiliation{Particle Physics Department, Rutherford Appleton Laboratory, Chilton, Didcot, Oxon OX11 0QX, United Kingdom}

\author{Ulla Blumenschein}
\email{u.blumenschein@qmul.ac.uk}
\affiliation{Particle Physics Research Centre, School of Physical and Chemical Sciences,
Queen Mary University of London, Mile End Road, London E1 4NS, United Kingdom}

\author{Arran Freegard}
\email{acf1g14@soton.ac.uk}
\affiliation{School of Physics and Astronomy, University of Southampton, Highfield. Southampton SO17 1BJ, United Kingdom}
\affiliation{Particle Physics Research Centre, School of Physical and Chemical Sciences,
Queen Mary University of London, Mile End Road, London E1 4NS, United Kingdom}

\author{Stefano Moretti}
\email{s.moretti@soton.ac.uk}
\affiliation{School of Physics and Astronomy, University of Southampton, Highfield. Southampton SO17 1BJ, United Kingdom}
\affiliation{Particle Physics Department, Rutherford Appleton Laboratory, Chilton, Didcot, Oxon OX11 0QX, United Kingdom}
\affiliation{Department of Physics and Astronomy, Uppsala University, Box 516, SE-751 20 Uppsala, Sweden}

\author{Dipan Sengupta}
\email{disengupta@physics.ucsd.edu}
\affiliation{Department of Physics and Astronomy, University of California, San Diego\\
9500 Gilman Drive, La Jolla, California-92037, United States of America}
\affiliation{ ARC Centre of Excellence for Dark Matter Particle Physics, \\
Department of Physics, University of Adelaide, South Australia 5005, Australia }

\date{\today}

\begin{abstract}
Leptonic signatures of Dark Matter (DM) are one of the cleanest ways to discover such a secluded form of matter at high energy colliders. 
We  explore the full parameter space relevant to multi-lepton  (2- and 3-lepton)
	signatures at the Large Hadron Collider (LHC) from  representative minimal consistent models with scalar and fermion DM.
	In our analysis,  we suggest a new parametrisation
 of the model parameter spaces in terms of the DM mass and  mass differences between DM and its multiplet partners. This parametrisation allows us to
explore properties of DM models  in their whole parameter space. This approach  is generic and
	quite model-independent since the  mass differences are related to the couplings
	of the DM to the Standard Model (SM) sector.
	We establish the  most up-to-date LHC limits on the inert 2-Higgs Doublet Model (i2HDM) and Minimal Fermion DM (MFDM) model  parameter
	spaces, by using the complementary information stemming from 2- and 3-lepton signatures.
	We provide a map of  LHC efficiencies and  cross-section limits for such 2- and 3-lepton  signatures allowing one to easily make  model-independent reinterpretations of  LHC
	results for  analogous classes of models.
 We also  present  combined constraints from the LHC, DM relic density and
	direct search experiments indicating the current status of the  i2HDM and MFDM model.
\end{abstract}
	\maketitle
	\newpage
\tableofcontents
\newpage

\section{Introduction}
The nature of Dark Matter (DM) is one of the greatest puzzles of modern particle physics and cosmology.
While overwhelming observational evidences from galactic to cosmological scales point to the existence of DM~\cite{Ade:2015xua,Blumenthal:1984bp,Bullock:1999he}, after decades of experimental effort, only its gravitational interaction has been confirmed. At the moment, we do not have  information about  DM properties, such as its mass, spin, interactions (other than gravitational), symmetry responsible for its stability, number of states associated to it and possible particles that would mediate the interactions between DM and the Standard Model (SM) particles.

If DM is light enough and interacts with SM particles directly, or via some mediators with a strength beyond the gravitational one, its elusive nature can be detected or constrained in  direct production at colliders, complementing direct and indirect DM searches in non-collider experiments. Therefore, the search for DM is one of the top priorities of the Large Hadron Collider (LHC)~\cite{LHC_ref} programme and that of future collider experiments.

The most general  DM signature  at colliders is the  mono-$X$ one, where $X$ stands for a SM object,  such as jet, Higgs, $Z$, $W$, photon, top-quark, etc. that recoils against the missing energy from the DM pair. This signature has  limitations, though, 
especially if  there are no light mediators decaying into a DM pair which could enhance the corresponding  signal to compete with the large SM background from $Z\to \nu\nu$. If such mediators are absent or very heavy even the High-Luminosity LHC
(HL-LHC) \cite{Gianotti:2002xx} could probe the DM mass only up to about 250 GeV, as shown, e.g., for the case of higgsino DM in the Minimal Supersymmetric Standard Model
(MSSM)~\cite{Schwaller:2013baa,Barducci:2015ffa,ATLAS:2018diz,CMS:2018qsc}. This limitation motivates us to look for signatures beyond the mono-$X$ one.

This study is devoted to  a  generic class of DM models,
where the DM is a part of an Electro-Weak (EW) multiplet
and the mass splitting   between DM ($D_1$) and its charged partner(s)
($D^+$, ...) or next-to-lightest neutral partners ($D_2$, ...)
is large  enough so as to give rise to multi-lepton signatures at the LHC
from processes such as
$pp\to D_1 D^\pm$,
$pp\to D^+ D^- $,
$pp\to D_1 D_2$,
$pp\to D^\pm D_2$ and 
$pp\to D_2 D_2$,
which are then followed by the 
$D^\pm$ and $D_2$ decays to gauge or Higgs bosons which in turn decay into leptons.

These signatures with visible 1-4 lepton(s) and Missing Transverse Energy (MET)  can be observed for $\Delta m^+=
m_{D^\pm}-m_{D_1}$ and/or $\Delta m^0=m_{D_2}-m_{D^\pm}$
above a few GeV.
The case  when  $\Delta m^+$ is of the order of the pion mass --  when $D^+$ and $D_1$ are degenerate at tree-level and their mass split is generated radiatively, due to the quantum corrections involving a loop of the photon and the charged DM multiplet partner(s) 
 -- leads to a compelling complementary signature with disappearing tracks which was recently studied in detail in~\cite{Belyaev:2020wok}.

The 2- and 3-lepton signatures from DM models  have been previously explored in the context of the  inert 2-Higgs-Doublet Model (i2HDM)~\cite{Belanger:2015kga},  Supersymmetric scenarios \cite{andreev2007using,1998} (including those with sneutrino DM~\cite{Arina:2013zca}) 
or models with vector-like leptonic DM~\cite{Bhattacharya:2015qpa}.
These earlier studies  have mainly explored  the one or two-dimensional parameter space of specific DM models, or just  selected benchmarks points from their multi-dimensional parameter space.
Our study presents the following  new results on multi-lepton signatures from DM at the LHC:
 \begin{itemize}
	\item 
	we  explore the full  three-dimensional parameter space
	relevant to the LHC
	for two representative minimal consistent DM models
	with  DM of spin-0 and -1/2, respectively:
the 	i2HDM and  Minimal Fermion DM (MFDM) model, respectively, 
	and present the  sensitivity of 
	recent LHC data to  these;
	\item 
	we suggest a  new parametrisation for both models which allows one 
	to better understand and interpret their
	properties  and visualise the  no-loose theorem 
	in their full parameter space;
	\item 
	we find that the 3-lepton signature
	 becomes relevant for large
	$D^+-D$ mass gaps and adds  important and complementary   sensitivity to the LHC searches;
	\item 
	we  implement and validate the 8 TeV ATLAS~\cite{Collaboration2008}  
	multi-lepton analysis of Ref.~\cite{B_langer_2015} in CheckMATE~\cite{Dercks:2016npn} relevant to our study and made available those analyses to the community;
	\item
	we create a map of the LHC efficiencies and the cross-section limits
	for 2- and 3-lepton signatures in the
	simple parameter space expressed through the DM  mass and the $D^+-D$ and $D_2-D^+$ mass differences  which can be used for generic reinterpretation of  spin-0 and spin-1/2  DM  models.
	\end{itemize}
	
The rest of the paper is organized as follows:
in section~\ref{sec:models} we present the DM models under study,
in section~\ref{sec:signal} we discuss details of the signal processes leading to 2- and 3-lepton signatures,  in section~\ref{sec:results} we present details of the analysis, including the  implementation into the CheckMATE package, and the new LHC sensitivity to the parameter space of the two models under study while in section~\ref{sec:conclusions}
we draw our conclusions. Furthermore, details of our study 
are given in Appendices A--E, including the respective plots on the exclusion of the parameter space (Appendix~\ref{app:plots}) and tables with the efficiencies and cross-section limits 
expressed in terms of universal parameters such as the DM mass and  the $D^+-D$ and $D_2-D^+$ mass differences  which can be applied for generic reinterpretation of  spin-0 and spin-1/2  DM  models (Appendices~\ref{app:i2HDM-exc}--\ref{app:MFDM-exc}).

\section{Models}
\label{sec:models}

In the following subsections, \ref{sec:MFDMmod} and \ref{sec:i2HDMmod}, we present the models under study in this work. Both of these  are phenomenologically well-motivated and can be mapped onto well-known Supersymmetric scenarios. They are minimal in nature, due to the small number of parameters that impact the phenomenology,  in terms of both collider and DM considerations. 

For the purposes of this paper, where we are interested primarily in the collider phenomenology of the model, we do not  initially impose strict requirements on relic density and (in)direct detection of DM. Rather, we first set out the collider constraints independently of DM considerations.  Only eventually  we show how the latter  impact the parameter space, in order
to illustrate that these are  complementary to collider constraints.

\subsection{MFDM Model}
\label{sec:MFDMmod}

The Lagrangian for the MFDM model, which augments the SM  by an  EW fermion ($\psi$)  doublet of $\chi_1^0$, $\chi_2^0$ and $\chi^+$ as well as a Majorana singlet fermion $\chi_s^0$, is~\cite{Belyaev:2020xxx} 
\begin{equation}
    \mathcal{L}_{\rm MFDM} = \mathcal{L}_{\rm SM}+\overline{\psi}(i \slashed{D}-m_\psi)\psi+\frac{1}{2}\overline{\chi}^0_s(i\slashed{\partial}-m_s)\chi^0_s-(Y_{\rm DM}(\overline{\psi}\Phi\chi^0_s)+h.c.).
\end{equation}

  The fermion $SU(2)$ doublet is 
 \begin{align}
     \psi=\begin{pmatrix}
     \chi^+\\
     \frac{1}{\sqrt{2}}(\chi_1^0+i\chi_2^0)
     \end{pmatrix}
 \end{align}
 while the Higgs doublet $\Phi$, after EW Symmetry Breaking (EWSB), is 
  \begin{align}
     \Phi=\begin{pmatrix}
     0\\
     \frac{1}{\sqrt{2}}(v+h)
     \end{pmatrix},
 \end{align}
where $h$ is the  SM-like Higgs boson with a mass of $\rm 125$ GeV, while $v=246$ GeV is Vacuum Expectation Value (VEV) of the SM Higgs potential. 
We impose a $\mathcal{Z}_{2}$ odd parity on the EW fermion doublet as well as the Majorana singlet fermion, such that the lightest particle after diagonalisation of the fermion mass matrix is stable and a DM candidate.  Note that, in the model as written above, we assume that $\chi_{1}^{0},\chi_{2}^{0}$ are Majorana fermions. The DM relic density requirements are satisfied by the usual freeze-out mechanism by scatterings mediated by the $Z$ or  Higgs boson. The minimality of this model is manifest in the fact there are only 3 new parameters, $m_{\psi}, Y_{\rm DM} $ and $m_{S}$.

Note that $\chi_1^0$ and $\chi_s^0$ mix via Yukawa couplings while $\chi^0_2$ and $\chi^+$ are mass degenerate. The physical masses are then obtained by diagonalisation of the mass matrix from the $(\chi_s^0,\chi_1^0,\chi_2^0)$ interaction basis to the $(D,D_2,D^{'})$  mass basis as
 \begin{equation}
      M =\begin{pmatrix}
      m_{\psi} & Y_{\rm DM}v &0\\
      Y_{\rm DM}v & m_s & 0 \\
      0&0&m_{\psi}.
      \end{pmatrix}.
 \end{equation} 
 The mass matrix is thus diagonalised by one rotation angle $\theta$ of the upper $2\times 2$ block, while we identify $m_{\psi}=m_{D^{+}}$.

 The rotation angle $\theta$  that diagonalises the mass matrix can be expressed in terms of two useful relations, 
     \begin{eqnarray}
  \tan{2\theta} &=& \frac{2Y_{\rm DM}v}{m_\psi -m_s},  \\
  \sin 2\theta &=& - \frac{2 Y_{\rm DM} v }{ m_{D_{2}} - m_{D}} 
  \label{eq:tan2theta}
  \end{eqnarray}
 and the connection between gauge eigenstates $\chi_s^0$
 and $\chi^0_1$ and mass eigenstates $D$ and $D_{2}$
 is given by
 \begin{eqnarray}
  \chi_s^0 & =   &D \sin{\theta}+D_2\cos{\theta},\nonumber\\
  \chi_1^0 & =  & D \cos{\theta}-D_2\sin{\theta}. \ \ 
    \label{eq:D_3Mixing}
  \end{eqnarray}  
 The connection between physical masses and mass parameters in the Lagrangian is given by
 \begin{eqnarray}
   m_s		&=&m_{D}+m_{D_{2}}-m_{D+}, \nonumber \\
   m_\psi	&=& m_{D+} = m_{D^{'}} \ .
 \end{eqnarray}  
The mass  splitting  between $m_{D_1}$ and $m_{D_3}$
is related to the Yukawa coupling $Y_{\rm DM}$ via
\begin{equation}
	Y_{\rm DM}=\frac{\sqrt{(m_{D_{2}}-m_{D+})(m_{D+}
			-m_{D})}}{v}=\frac{\sqrt{\Delta m^0\Delta m^+}}{v} 
	\label{eq:YDM} \ ,
\end{equation}
where 
\begin{equation}
	\Delta m^0=m_{D_{2}}-m_{D+},\ \ \Delta m^+=m_{D+}-m_{D} \ .
	\label{eq:i2hdm-dm}
\end{equation} 
The requirement that $Y_{\rm DM}$ must be real leads to the following mass ordering: 
\begin{equation}
m_{D_{2}} \geq m_{D_{+}} =m_{D^{'}} > m_{D}.
\end{equation}

The parameter space of the model can be expressed in terms of three physical masses:
\begin{equation}
\left\{ m_{D},\ m_{D^{'}}=m_{D\pm},\ m_{D_{2}} \right\} .
\end{equation}
However, since the Yukawa coupling controls the {mass splitting} as well as DM mixing and therefore the strength of the interactions between DM and SM, it is also convenient
to trade parameters and parametrise the model space using the following set
\begin{equation}
	\left\{m_{D},\ \Delta m^+,\ \Delta m^0\right\}_{\rm MFDM},
\end{equation}
which consists of the DM mass and two mass splittings 
between $D^+$ and DM  as well as  between $D_{2}$ and $D^+$.
As we will see, the latter parametrisation allows to better visualise and interpret collider search results for the whole parameter space.
The model is also subject to various theoretical constraints, like perturbativity and radiative stability, which are satisfied for the couplings and masses adopted  in this paper. 

As far as DM constraints are concerned, we first assess direct detection constraints. Since there is no tree level $Z$ boson interaction with the DM candidate,  the spin-independent direct detection constraint only arises from DM-nucleon scattering controlled by the Higgs coupling to DM. By controlling the mass splitting, this coupling can be made small, leaving a viable parameter space. In order to obtain the correct (not over-abundant) relic abundance, one needs an efficient annihilation mechanism via the Higgs funnel $m_{D}\simeq m_{h/2}$.  These considerations were previously taken into account in a previous  paper \cite{Belyaev:2020xxx}.


\subsection{\rm i2HDM}
\label{sec:i2HDMmod}
This scenario \cite{Gunion:2002zf,Eriksson:2009ws,Deshpande:1977rw}  introduces a
second scalar doublet $\phi_2$ to the SM Higgs sector, with the same
quantum numbers as the SM Higgs doublet $\phi_1$. However, a discrete $Z_{2}$ odd symmetry 
is imposed on the second doublet such that there  are no direct coupling to fermions, rendering it `inert'.

In the unitary gauge,  the two scalar doublets can be written as
\begin{equation}
	\phi_1=\frac{1}{\sqrt{2}}\begin{pmatrix}0\\v+H\end{pmatrix}, \phi_2=\frac{1}{\sqrt{2}}\begin{pmatrix}\sqrt{2}D^+\\D_1+iD_2\end{pmatrix} \ ,
\end{equation}
where $D_1$, $D_2$ and $D^+$ are the lightest, next-to-lightest and charged inert scalars, respectively.
Within this parametrisation, the $Z_2$ symmetry is preserved  due to the fact that
$v_2=0$ and the absence of Yukawa couplings of the 
$\phi_2$ to fermions.  This `inert minimum', for $<\phi_i^0>=\frac{v_i}{\sqrt{2}}$, corresponds to $v_1=v$ and $v_2=0$. 
  This symmetry provides the stability of  the lightest inert boson against  decaying into SM particles, which makes $D_1$ a DM candidate.

The scalar Lagrangian is then given by
\begin{equation}
	\mathcal{L}_\phi = |D_\mu \phi_1|^2+|D_\mu\phi_2|^2 - V(\phi_1,\phi_2),
\end{equation}
where the most general scalar potential is written as
\begin{align}
	V(\phi_1,\phi_2)= -m_1^2(\phi_1^\dagger\phi_1)  -m_2^2(\phi_2^\dagger\phi_2)+\lambda_1(\phi_1^\dagger\phi_1)^2+\lambda_2(\phi_2^\dagger\phi_2)^2+\lambda_3(\phi_1^\dagger\phi_1)(\phi_2^\dagger\phi_2) \nonumber\\
	+\lambda_4(\phi_2^\dagger\phi_1)(\phi_1^\dagger\phi_2)+\frac{\lambda_5}{2}[(\phi_1^\dagger\phi_2)^2+(\phi_2^\dagger\phi_1)^2]
\end{align}
and contains all of the scalar interactions allowed by the $Z_2$ symmetry. 
Here, all mass and coupling parameters are defined to be positive. 
Under the
\be
m_{D_{1}}^2 > (\lambda_{345}/2-\sqrt{\lambda_1\lambda_2})v^2
\ee
condition~\cite{Belyaev:2016lok}, where $m_{D_1}$ is the mass of $D_1$ and $\lambda_{345}=\lambda_3+\lambda_4+\lambda_5$,

The physical masses are given by ($\lambda_5$ can always  be chosen positive due to the  
$\phi_2 \to i\phi_2, \ \ \lambda_5\to -\lambda_5$ symmetry)
\be
m_{D_1}^2=\frac{1}{2}
(\lambda_3+\lambda_4-\lambda_5)v^2-m_2^2, \ \
m_{D_{2}}^2=\frac{1}{2}
(\lambda_3+\lambda_4+\lambda_5)v^2-m_2^2, \ \
m_{D^+}^2=\frac{1}{2}\lambda_3 v^2-m_2^2,
\ee
which, together with $\lambda_{345}$ and 
$\lambda_2$, parametrise completely the DM sector of the model, described by five parameters:
\begin{equation}
	m_{D_1},\ m_{D_2},\ m_{D^{\pm}},\ \lambda_{345},\ \lambda_2 .
\end{equation}
These can be further reduced for this analysis, by noting that  $\lambda_2$ does not affect the LHC phenomenology under study at all since it controls DM the self-interaction only.

Furthermore, we also exclude $\lambda_{345}$
from our analysis for the following reason.
The relevant coupling for the
$gg\to H^* \to D^+ D^-$ and $gg\to H^* \to D_2 D_2$
processes are $\lambda_3$ and $\lambda_3+\lambda_4-\lambda_5$,  respectively,
which are limited by EW precision tests plus perturbativity~\cite{Goudelis:2013uca,Belyaev:2016lok} and related to the mass splitting   between $D^+$ and $D_2$.
We have checked that, even if we maximise the cross section of either of the two processes by increasing the respective coupling up to the maximal allowed value, the respective cross section\footnote{Both cross sections cannot be simultaneously made large since either $D^+$ or $D_2$ should be heavier than the other, which would make one of the  cross sections much smaller than the other.} will still be below
the signal  production cross section via the weak coupling.
Moreover, the value of the $\lambda_{345}$ coupling is also strongly limited by DM direct detection constraints
as well as by limits on the invisible  Higgs boson decay branching ratio  in the case when  $m_H>2m_{D_1}$ (see, e.g., \cite{Belyaev:2016lok,Belyaev:2018ext}).
 Therefore, in order to establish a conservative and generic limit on the i2HDM parameter space, 
we exclude  $\lambda_{345}$ from the current study
and set its value to zero.
Therefore, the parameter space we focus on is three-parametric:
\begin{equation}
	m_{D_1},\ m_{D_2},\ m_{D^{\pm}},
\end{equation}
which can  easily be visualised in 2D planes, while fixing one parameter.
In analogy with the MFDM model we also use the 
\begin{equation}
	\left\{m_{D_1},\ \Delta m^+,\ \Delta m^0\right\}_{\rm i2HDM} 
\end{equation}
parametrisation, where
\begin{equation}
	\Delta m^0=m_{D_{2}}-m_{D^{\pm}},\ \ \Delta m^{+}=m_{D^{\pm}}-m_{D_{1}} \ ,
	\label{eq:i2hdm_DM}
\end{equation}
which allows an even better visualisation and interpretation of the model parameter space.
We note one important caveat here. Note that we deliberately choose the mass hierarchy
$m_{D_{2}} > m_{D^{\pm}}>m_{D_{1}}$, such that $D_{1}$ is the lightest state. This choice 
is motivated by phenomenological reasons in order to bring out the intricacies of the collider constraints
set by 2- and 3-lepton searches.  The other hierarchy $m_{D^{\pm}} > m_{D_{2}}>m_{D_{1}}$
is entirely feasible and will lead to constraints comparable to the ones described in this paper. 
We leave this possibility for a future work.

\section{Signal Processes}
\label{sec:signal}

{In this section, we study  production and decay processes inclusively, the combination of which  gives an idea about the LHC event rate of the signatures under study.} Specifically,  we study lepton signatures 
from DM multiplet partners generically appearing when the mass splitting   between DM and those partners 
is large  enough to give rise to sufficiently energetic leptons, which can originate from the following processes: 
\begin{itemize}
	\item 
	$pp\to D_1 D^\pm \to  D_1 D_1 W^{\pm(*)}\to D_1 D_1 \ell^\pm \nu $
	\item 
	$pp\to D^+ D^- \to D_1 D_1 W^{+(*)} W^{-(*)}\to D_1 D_1 \ell^+ \ell^-\nu \bar{\nu} $
	\item 
	$pp\to D_1 D_n \to D_1 D_1 Z^{(*)}\to D_1 D_1 \ell^+ \ell^-$

	\item 	
	$pp\to D^\pm D_n \to D_1 D_1Z^{(*)} W^{\pm(*)}\to D_1 D_1 \ell^+ \ell^-\ell^\pm\nu $
	
	\item 	
$pp\to D^\pm D_n \to D_1 D_1 W^\pm H \to
D_1 D_1 W^{\pm} W^{\pm} W^{\mp*}
\to 
D_1 D_1 \ell^\pm \ell^\pm\ell^\mp\nu\nu\nu $
	
	\item 
	$pp\to D_n D_n \to D_1 D_1 Z^{(*)}Z^{(*)}\to D_1 D_1 \ell^+ \ell^- \ell^+ \ell^-/ D_1 D_1 \ell^+ \ell^- q\bar{q}$ 
\end{itemize}
where $D_n=D_2$ or $D_n=D',D_2$
depending on the scenario, i2HDM or MFDM model, respectively.
Since $D^\pm$ and $D_n$  decay via $Z$ or $W^\pm$ either hadronically or
leptonically in association with the DM candidate, 
they can provide  signatures with several charged leptons plus MET, which are the subject of this study.

\begin{figure*}[htbp]
	\centering
	\subfloat[]{
	\includegraphics[width=0.25\textwidth]{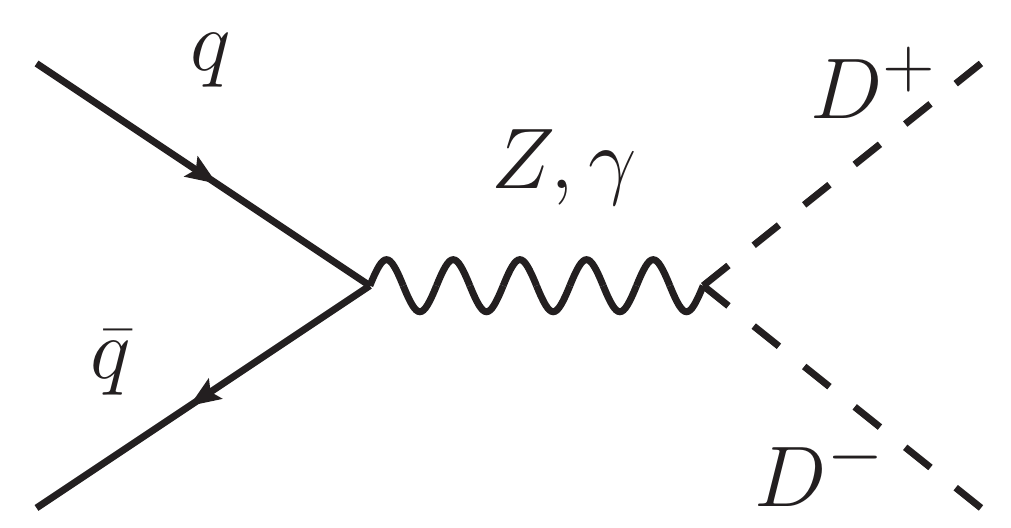}
}
	\subfloat[]{
	\includegraphics[width=0.25\textwidth]{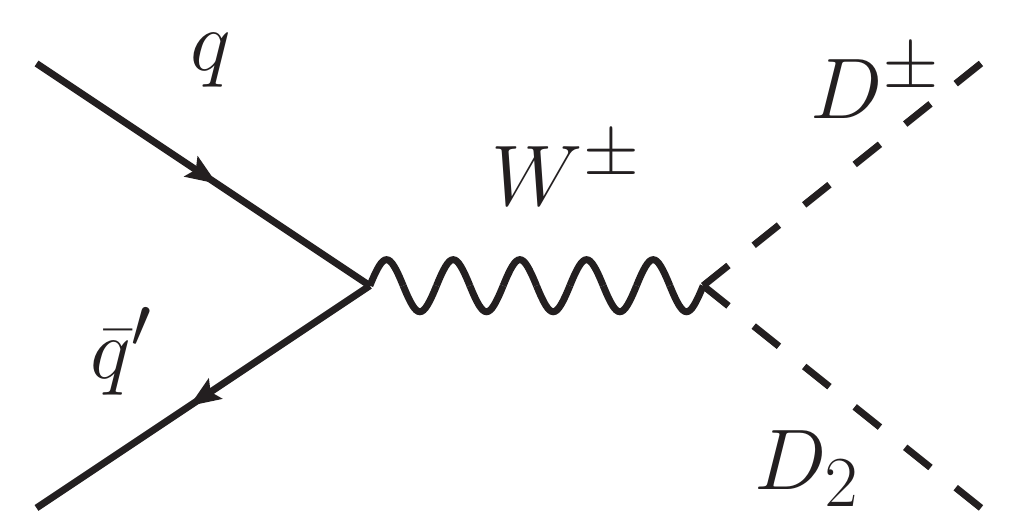}
}
\\
	\subfloat[]{
		\includegraphics[width=0.25\textwidth]{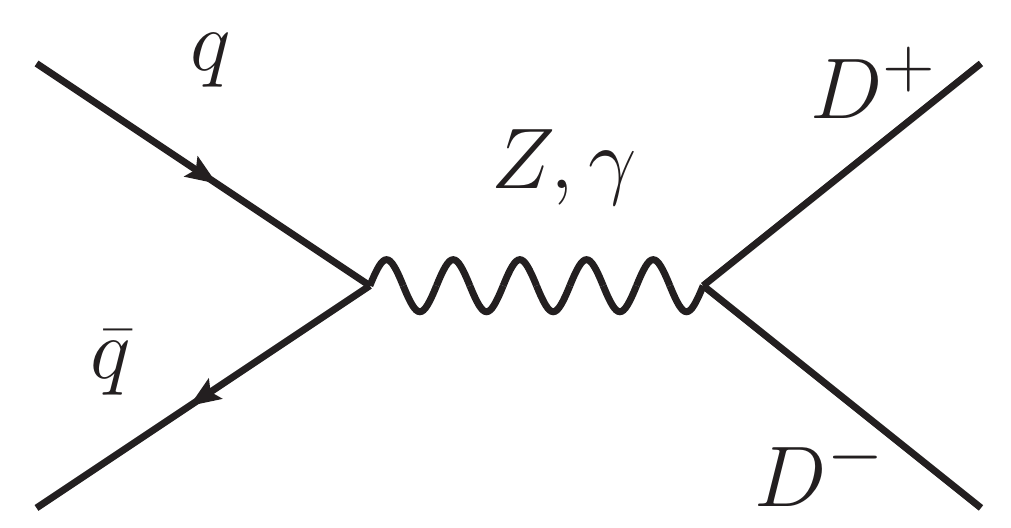}
	}
	\subfloat[]{
		\includegraphics[width=0.25\textwidth]{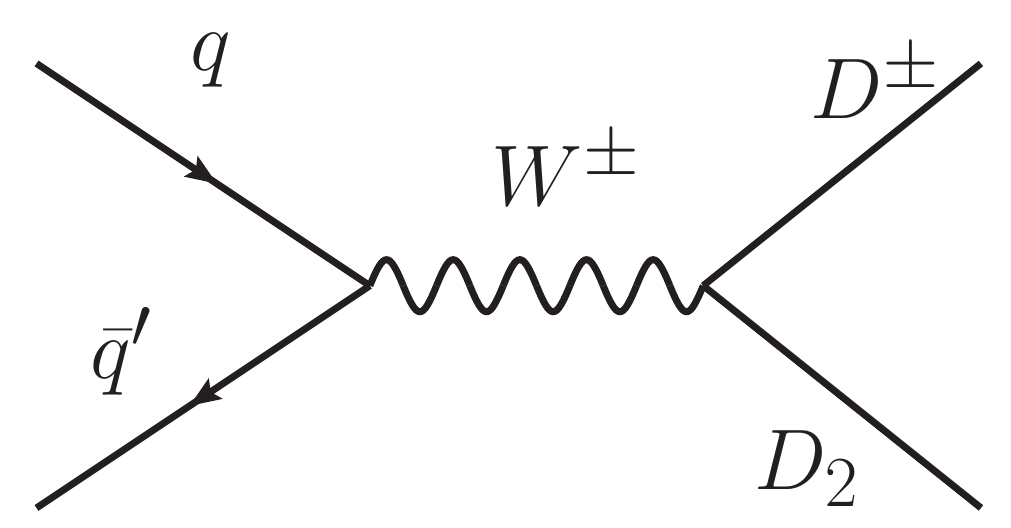}
	}
	\caption{Feynman diagrams for $D^+ D^-$ and $D^\pm D_2$ production
	common to the   i2HDM (top) and MFDM model (bottom).}
	\label{fig:FeynProd1}
\end{figure*}

\begin{figure*}[htbp]
	\centering
		\subfloat[]{
		\includegraphics[width=0.25\textwidth]{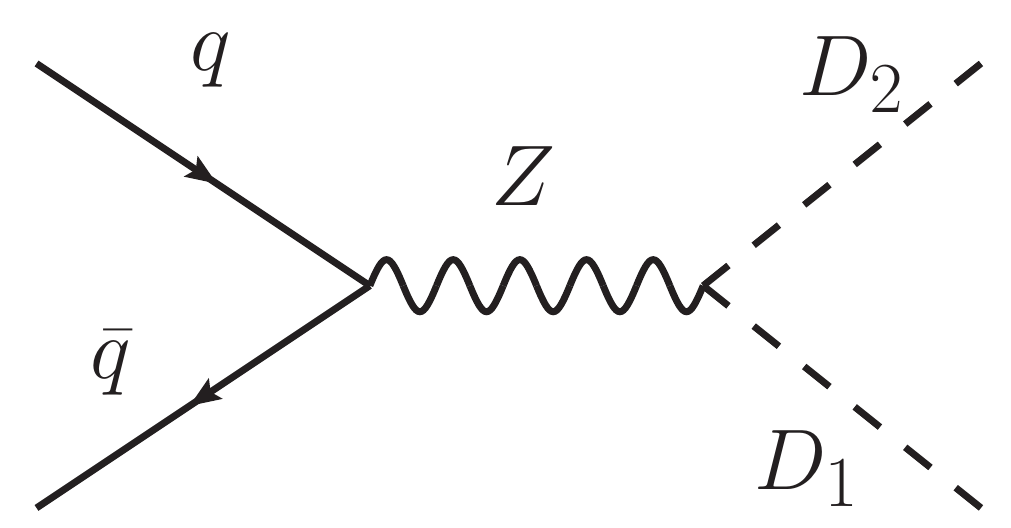}
	}	
		\subfloat[]{\includegraphics[width=0.25\textwidth]{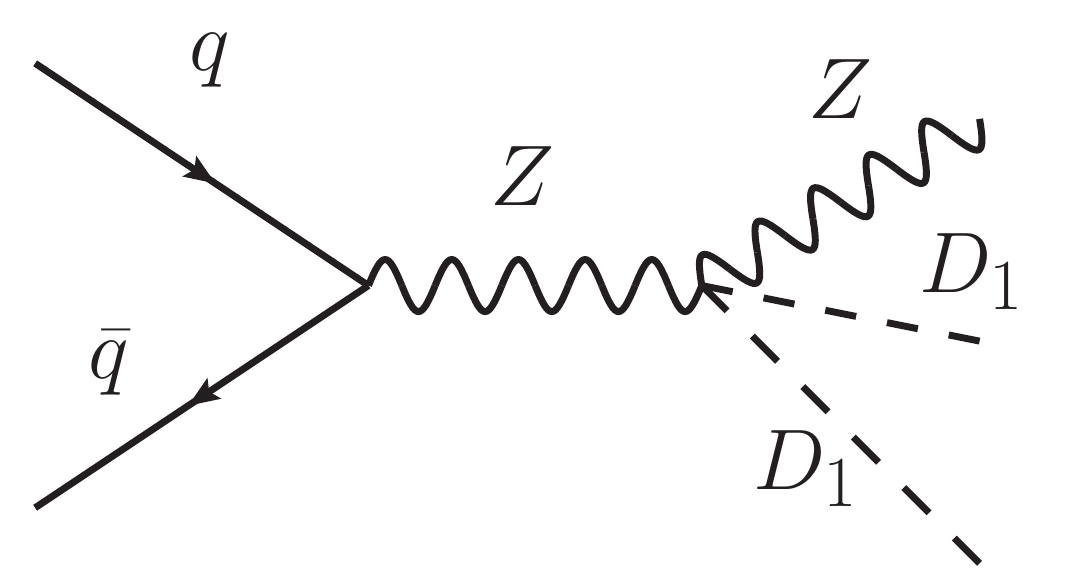}}
		\\
			\subfloat[]{
			\includegraphics[width=0.25\textwidth]{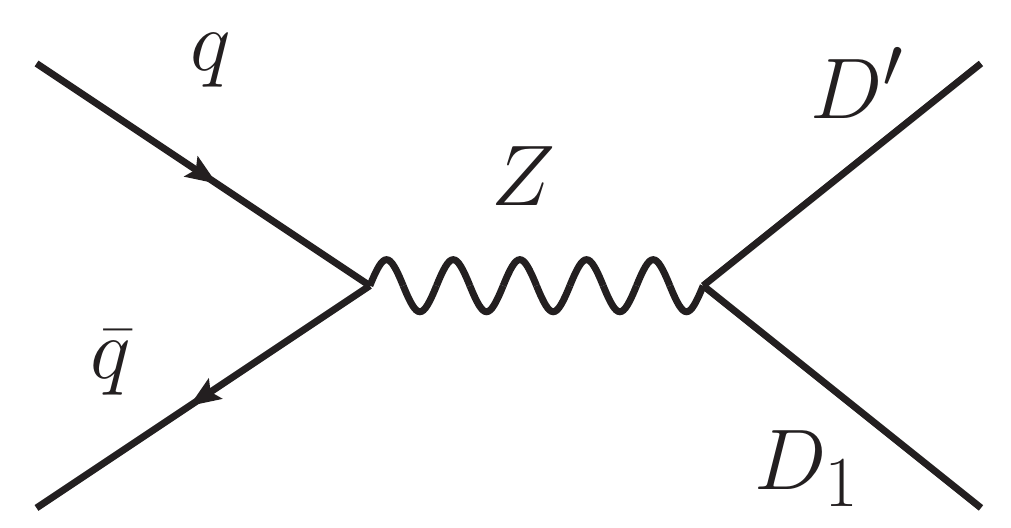}
		}
		\subfloat[]{
			\includegraphics[width=0.25\textwidth]{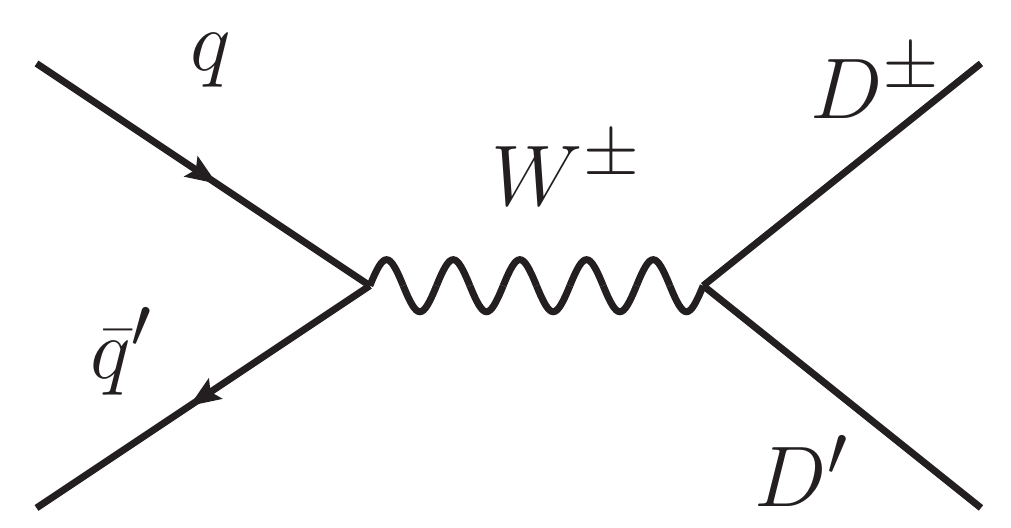}
		}
			\subfloat[]{
		\includegraphics[width=0.25\textwidth]{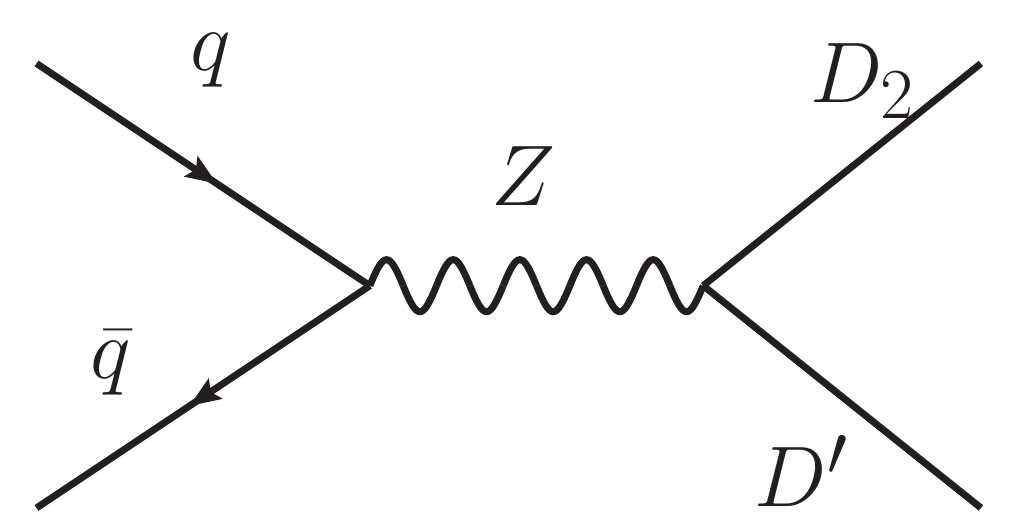}
	}
	\caption{Feynman diagrams for $D_2D_1$,  $ZD_1 D_1$ production exclusive to the  i2HDM, (a) and (b), and for $D_1 D'$, $ D^\pm D'$ and $ D_2 D'$ production  exclusive to the  MFDM model, (c), (d) and (e).}
	\label{fig:FeynProdExclusive}
\end{figure*}

The common  diagrams for both models under study 
for the signal production processes 
\bea 
q\bar{q} \to  D^+ D^-
& \ \  \mbox{and} \ \ 
      q\bar{q}' \to  D^\pm D_2
\eea
that provide the multi-lepton signatures we study here are presented in Fig.~\ref{fig:FeynProd1}.
In addition to these common Feynman diagrams, the $ZD_2D_1$ and $ZZD_1D_1$ vertices,
specific only to the i2HDM, provides additional Feynman diagrams 
and corresponding new kinematic topologies, shown in Fig.~\ref{fig:FeynProdExclusive}(a) and (b).
Conversely, since the MFDM model has an additional neutral DM
 partner, $D'$, in contrast to the i2HDM, there are additional MFDM processes\footnote{{MFDM also allows for $D_2D_1$ production through the $HD_2D_1$ vertex, but we do not study this production mode  as it is highly suppressed compared to the other processes discussed here.}}, 
 \bea
 q\bar{q}  \to  D' D_1 
 & \mbox{,} \ \ 
  q\bar{q}  \to  D_2 D' 
 & \ \  \mbox{and} \ \ 
 q\bar{q}' \to  D^\pm D',
 \eea
 providing the required multi-lepton signatures through the topologies  shown in Fig.~\ref{fig:FeynProdExclusive}(c), (d) and (e), respectively.
 
\begin{figure*}[htb]
	\subfloat[]{
		\includegraphics[width=0.5\textwidth]{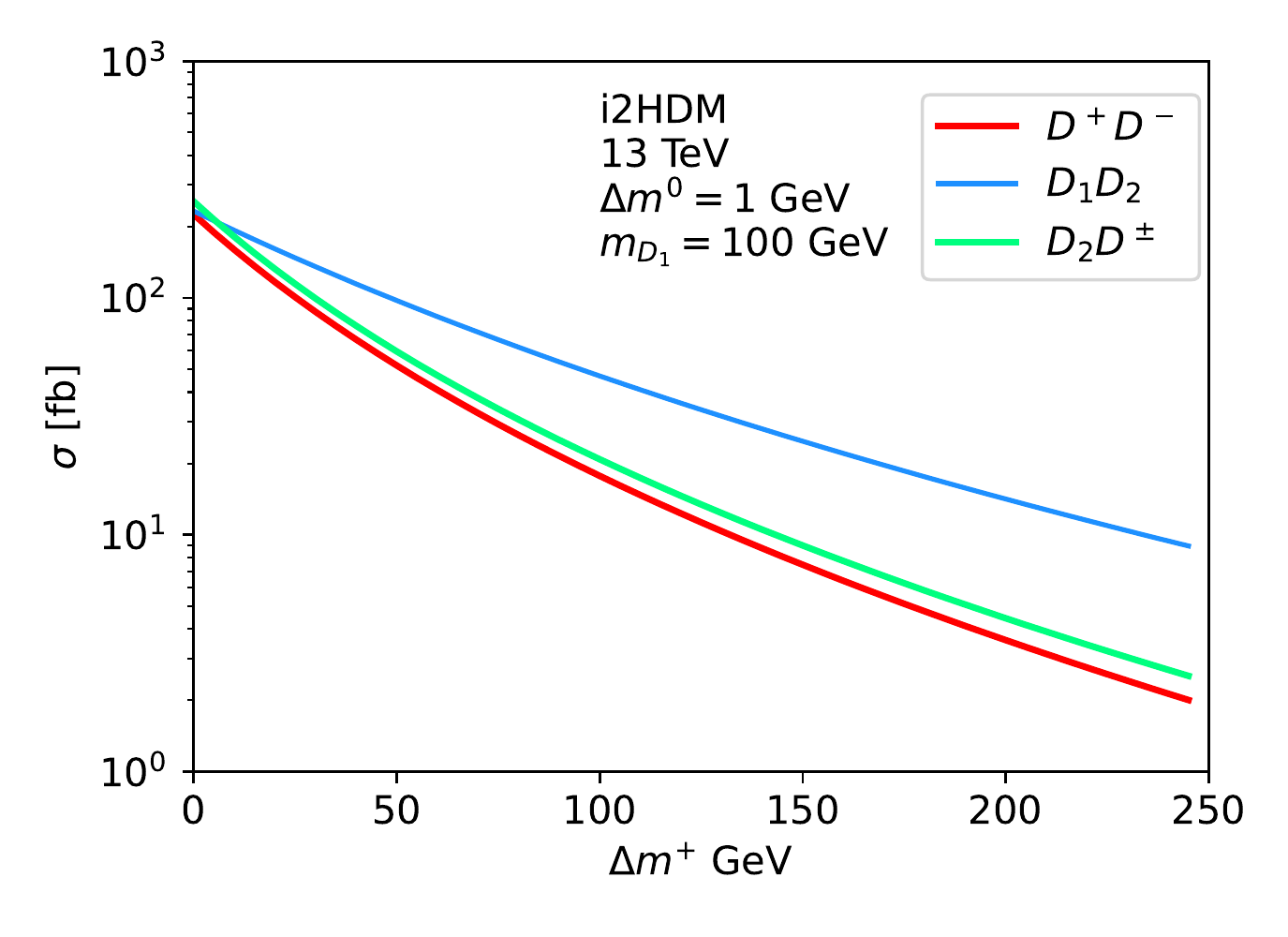}
	}
	\subfloat[]{
		\includegraphics[width=0.5\textwidth]{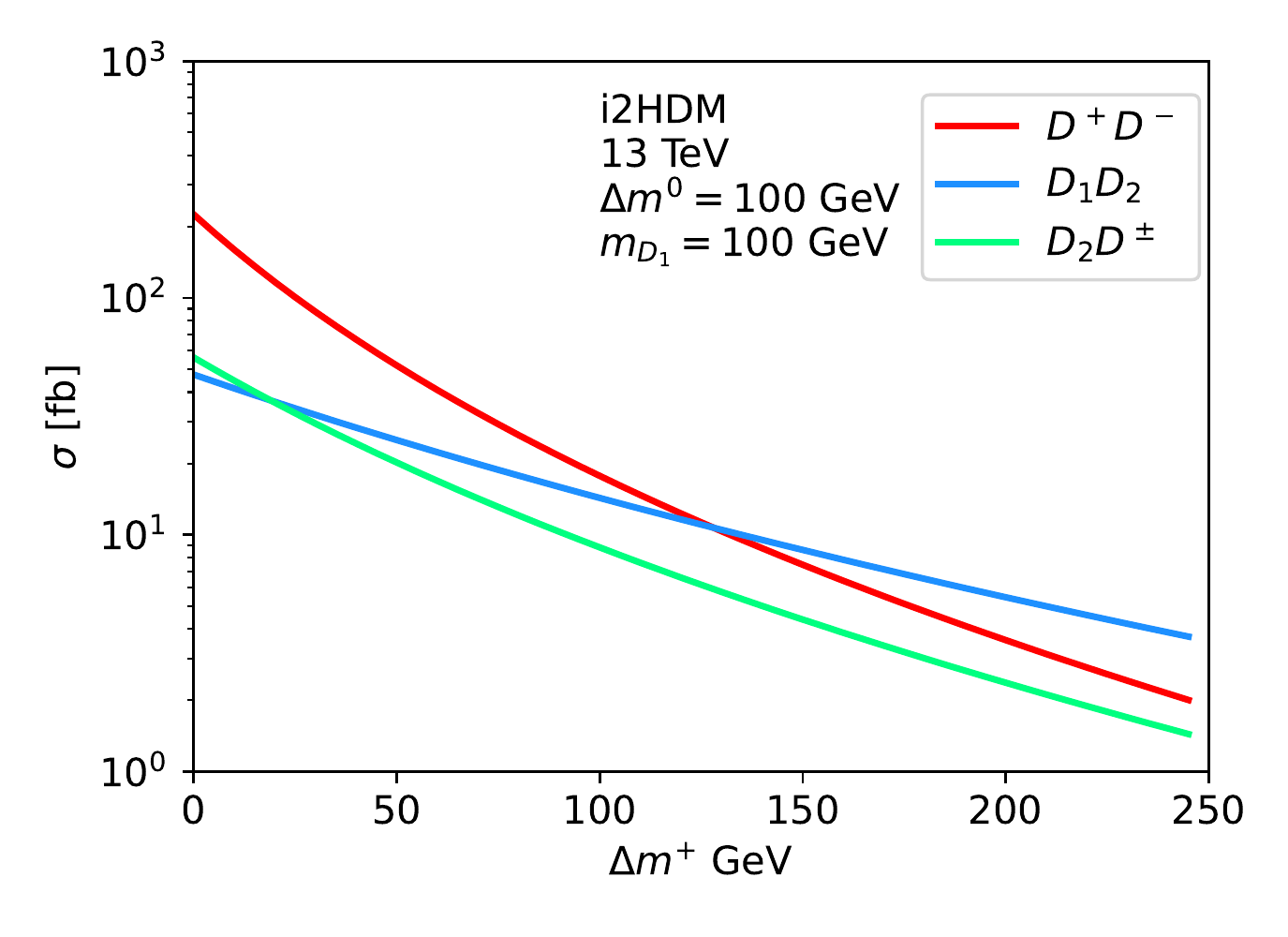}
	}
\vspace*{-0.3cm}
	\\
	\subfloat[]{
		\includegraphics[width=0.5\textwidth]{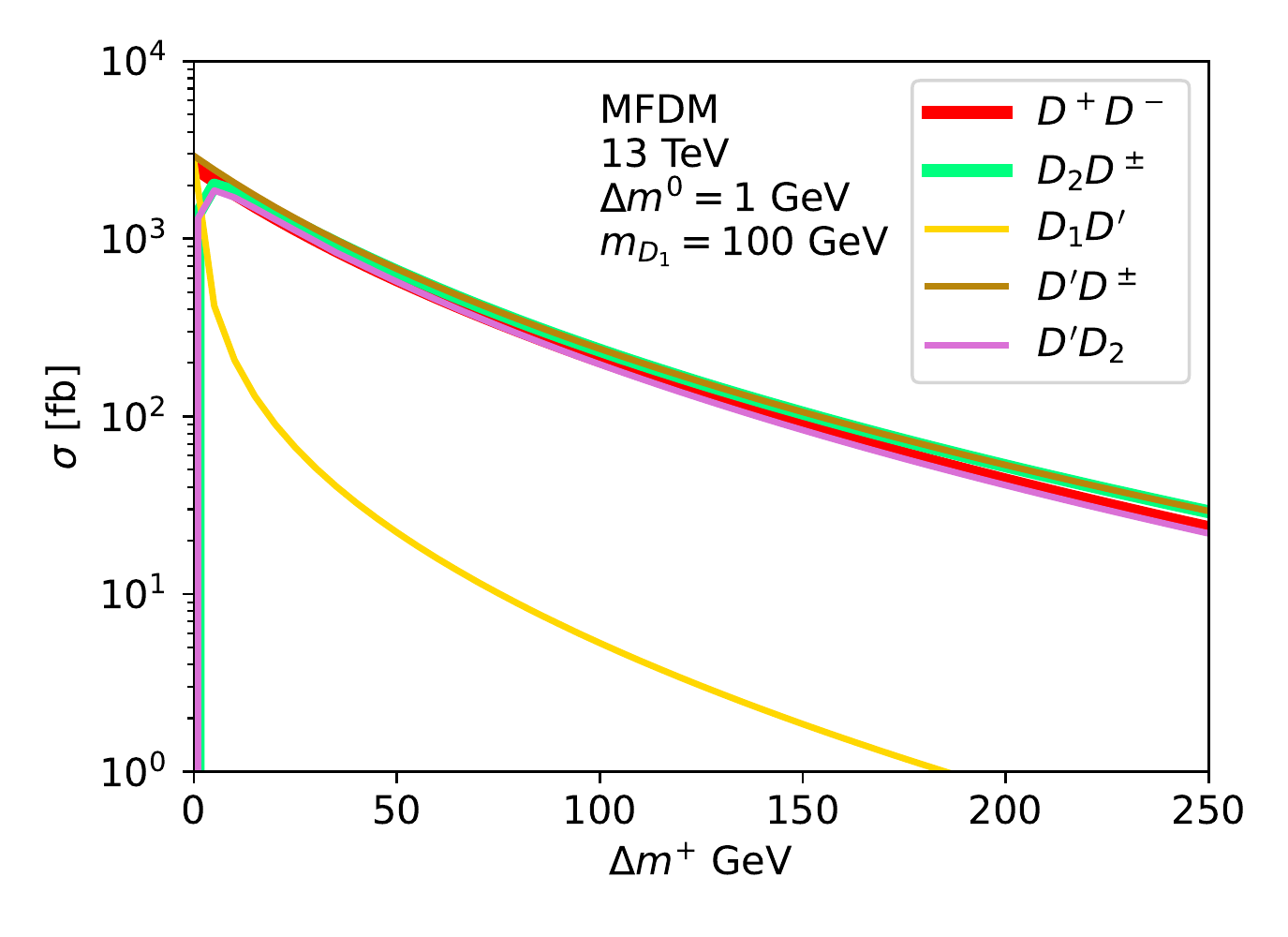}
	}
	\subfloat[]{
		\includegraphics[width=0.5\textwidth]{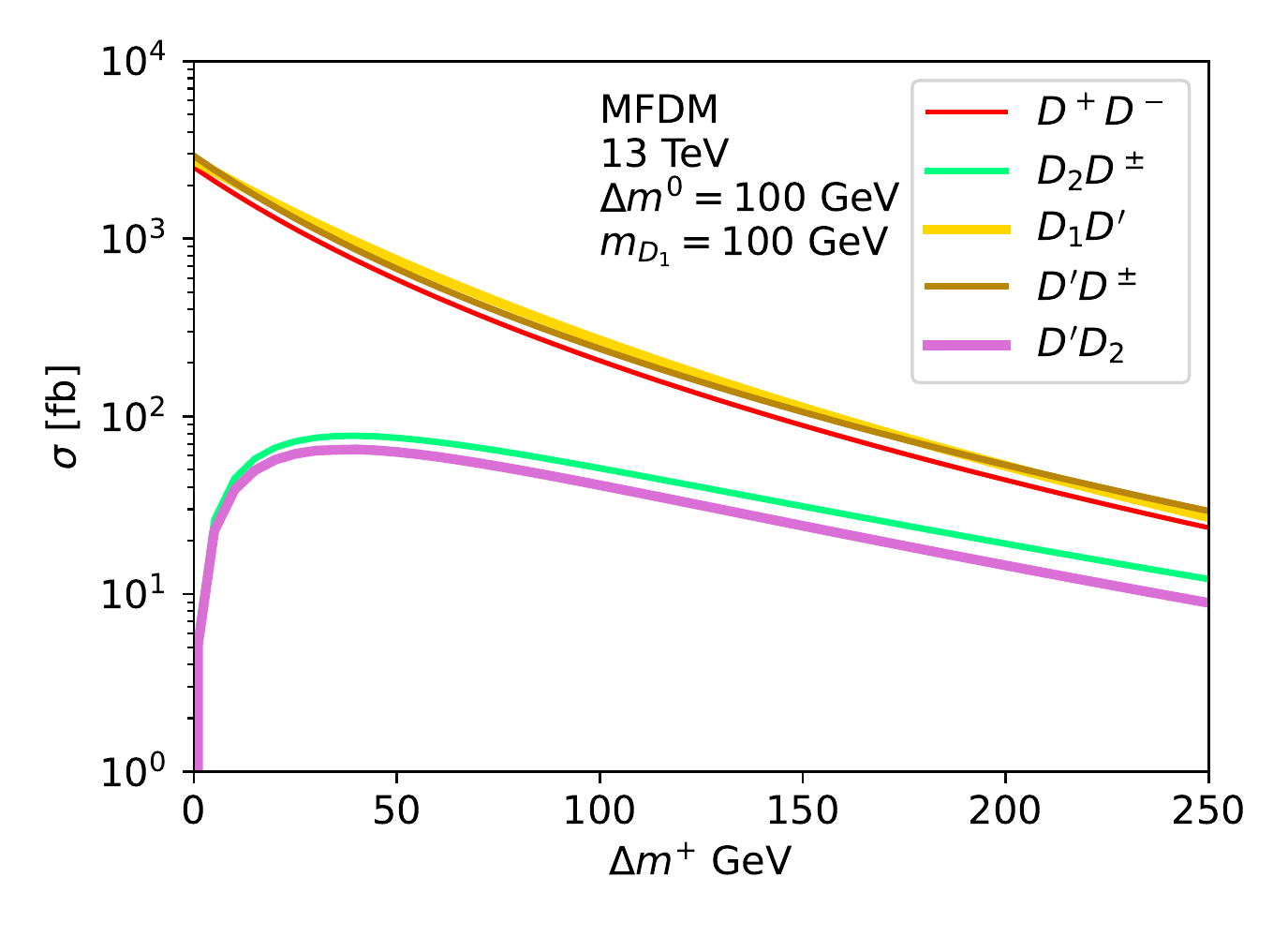}
	}
	\caption{The cross sections for pair production of DM partners for the  i2HDM (top) and MFDM model (bottom)  for $\Delta m^0=1$ and $100$ GeV (left and right panels, respectively).
}
	\label{fig:XSecsTeV}
\end{figure*}

The cross sections for the above production processes presented here are  calculated using CalcHEP~\cite{Belyaev:2012qa}, a package designed for the evaluation of the  tree-level processes and  their respective Monte Carlo (MC) simulation.
CalcHEP exploits the Beyond the SM (BSM) scenarios implemented on the HEPMDB \cite{hepmdb}, allowing to cross check these in different gauges for the purpose of validating the relevant implementation, and further has  an interface to LHAPDF~\cite{Buckley:2014ana} for a wide selection of Parton Distribution Functions (PDFs), allowing to ascertain the dominant systematic error on the theoretical side\footnote{Note that the production and decay processes exploited here are EW in nature, so that the QCD corrections to these are small, typically of order 20\%, and their residual uncertainty negligible.}.
 The events generated by CalcHEP in  Les Houches Event (LHE) format 
 can then be used  by other tools enabling further simulations of the parton shower and detector effects.
For the latter, we use here the combination PYTHIA \cite{Sjostrand:2014zea} and DELPHES \cite{deFavereau:2013fsa}, respectively.

\begin{figure*}[!t]
	\subfloat[]{
		\includegraphics[width=0.49\textwidth]{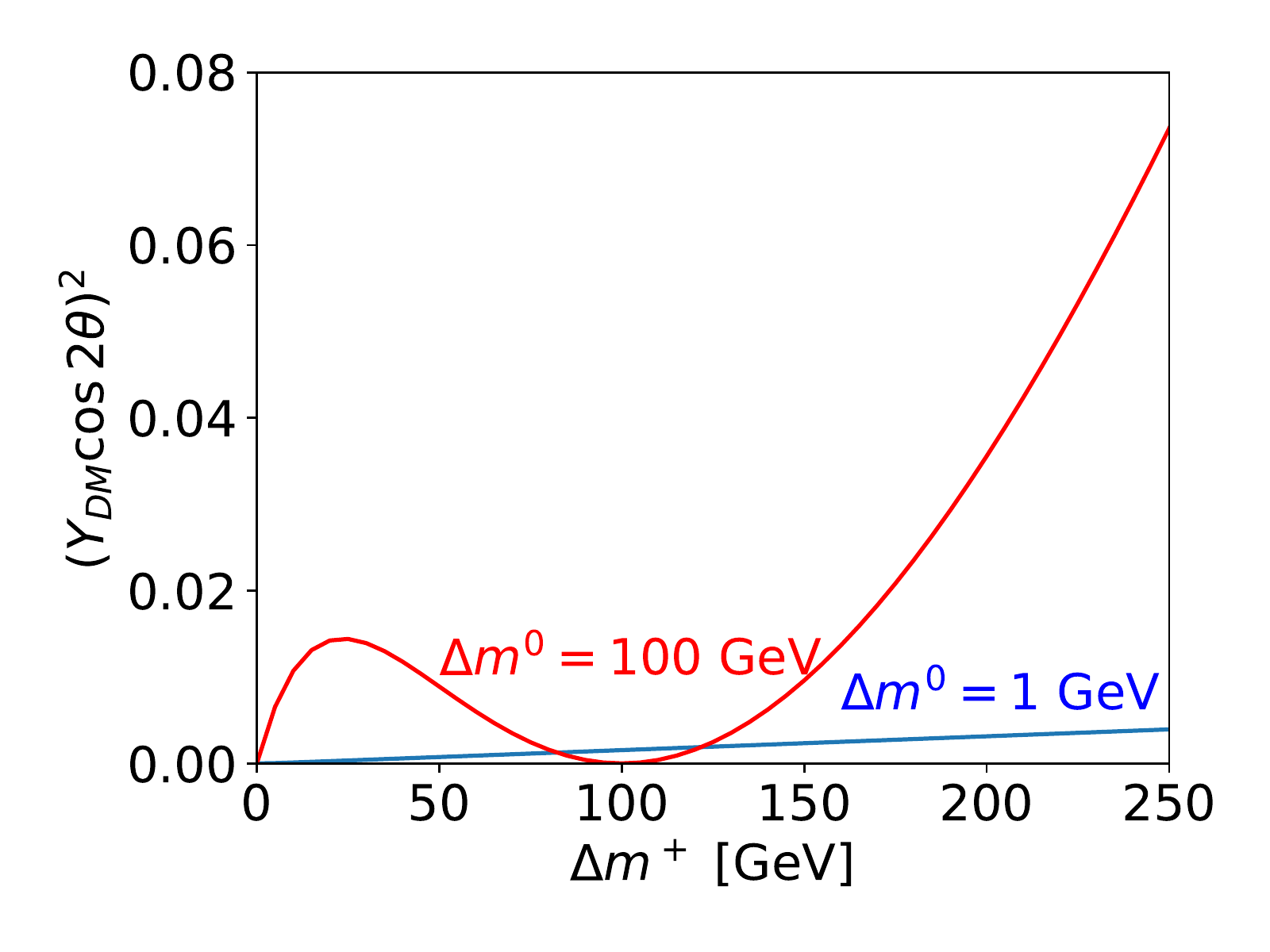}
	}
	\subfloat[]{
		\includegraphics[width=0.49\textwidth]{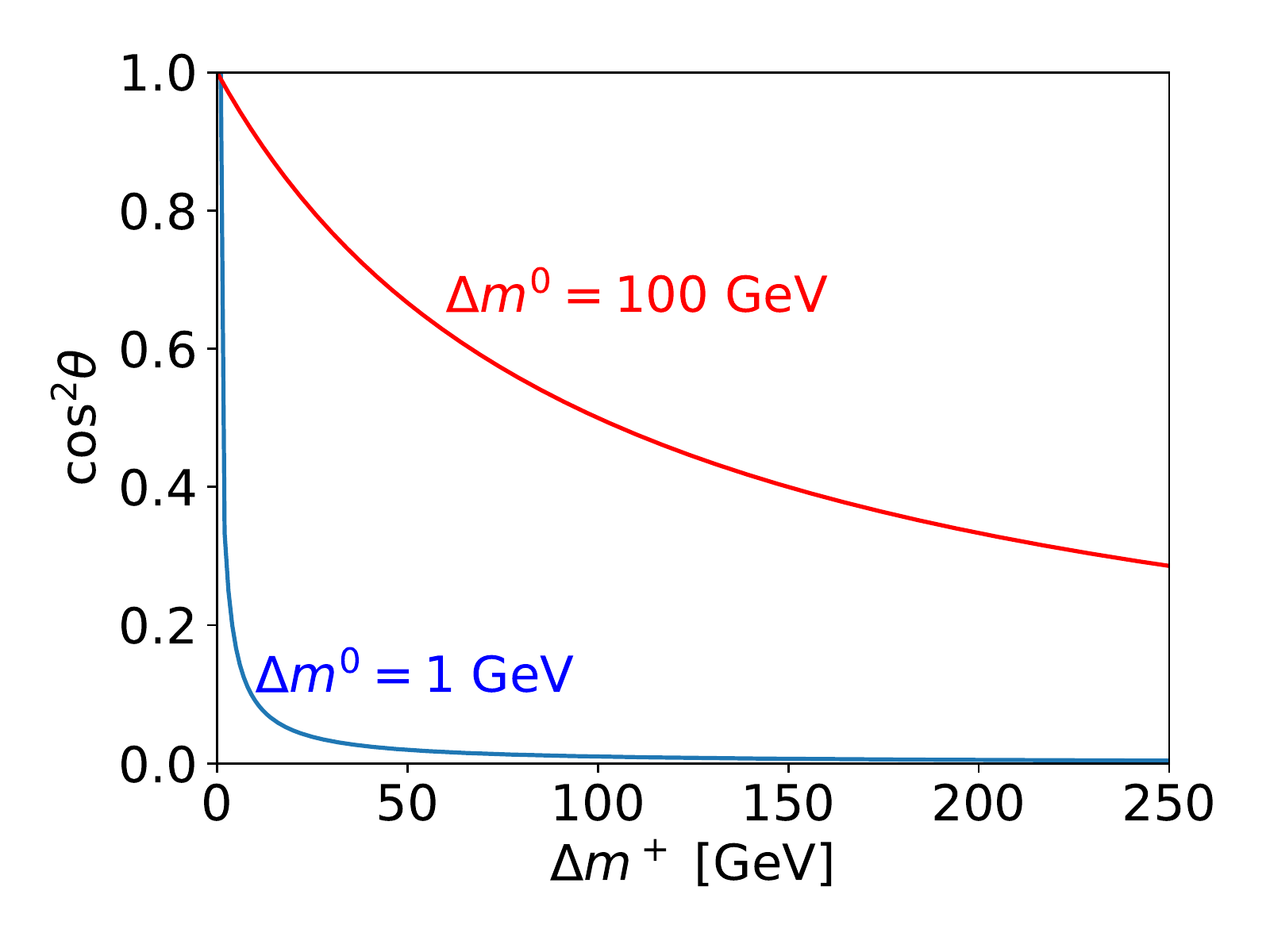}
	}
	\caption{$D_1 D_2 H$ coupling squared (a) and  $\cos^2\theta$ of the $\chi_1^0-\chi_s^0$ mixing (b)  as a function of $\Delta m^+$
		for two values of $\Delta m^0=$ 1 GeV (blue) and 100 GeV (red).}
	\label{fig:couplings}
\end{figure*}

The relevant cross sections are presented in Fig.~\ref{fig:XSecsTeV} as a function of the $\Delta m^+$
parameter for a 100 GeV  DM mass and two values of $\Delta m^0$=1 and 100 GeV (left and right panels, respectively) for the i2HDM (top panels) and MFDM model (bottom panels).
Here, one can see that for $\Delta m^0$=1,  
 $D_1D_2$ production has the highest rate  for the i2HDM while,
 in the case of the MDFM model, {the analogous $D_1D'$} channel is highly suppressed.
 This can be explained by the fact that the 
 $Z D_1 D_2$ coupling controlling this process in the i2HDM model is just  a pure (weak) gauge coupling while 
 in the MFDM model the  $Z D_1 D'$ coupling is the product of a (weak)  gauge coupling and the cosine  of the $\chi_1^0-\chi_s^0$ mixing
 angle, which is suppressed when $\Delta m^+ \gg \Delta m^0 =1$, as shown by the blue line in Fig.~\ref{fig:couplings}(b).
 
 In contrast, for  $\Delta m^0$=1 GeV, the production cross
 sections for  $D^+D^-$, $D_2 D^\pm$ and $D' D^\pm$ (represented by red, green and brown lines, respectively) 
 are close to each other in each model. This can be explained by the fact that the  $D^+$ and $D_2$ masses are about the same 
as well as the couplings controlling these processes (which  are purely (weak) gauge ones). 
{Furthermore, one should note that the $ZZD_1 D_1$ coupling 
	(unique to the  i2HDM)  contributes to  $ZD_1D_1$
	production, the  $2\to 3$ process
	with a subdominant cross section  in comparison with $2\to 2$ production. 
	However, when the $D_2\to Z D_1$ decay is open,
	the  $ZD_1D_1$ process  will include the corresponding 
	resonant $2\to 2$ production and decay.
	Therefore, to avoid  double counting,
	 we do not present the cross section for this
	$2\to 3$ process   in Fig.~\ref{fig:XSecsTeV}(a). 
}

For the MFDM model, the additional production process
 $D' D_2$ (pink line),  characteristic 
 of this scenario,  
 has a similar cross section
 to $D^+D^-$ and $D'D^\pm$ for the very same reason.
 One should also note that the cross sections 
 for  scalar DM production are smaller than those  for 
 fermion DM production by the spin factor 
 $\beta^2/4$ (where $\beta=\sqrt{1-(m_{D_i}+m_{D_j})^2/\hat{s}}$ with $m_{D_i}$ and  $m_{D_j}$ being the masses 
of DM particles in the  final state), which ranges from about $1/4$   to about $1/10$,
 as one can see from  Fig.~\ref{fig:XSecsTeV} (left panels).
 
 Let us now consider  the $\Delta m^0$=100~GeV case
 presented in  the right panels of 
 Fig.~\ref{fig:XSecsTeV}.
 In the i2HDM, the   $\Delta m^0=1 \to 100$~GeV change  (which increases the $D_2$ mass)
 equally suppresses   $D_1 D_2$ and $D_2 D^\pm$ production by about a factor of  4.
 For the MFDM  model,  the  $\Delta m^0=1 \to 100$ GeV change affects both the  $D_2$ mass (see Eq.~(\ref{eq:i2hdm-dm})) 
and  the $\chi_1^0-\chi_s^0$ mixing angle $\theta$ (the $Z D_2 D'$ and $W^+ D_2 D^- $ couplings are both proportional to $\sin \theta$), 
leading to a suppression of $D' D_2$ and $D_2 D^\pm$ 
 production from both causes (see pink and green lines, respectively, in Fig.~\ref{fig:XSecsTeV}(d)).
 In contrast, $D_1 D'$ production is enhanced,
 since it is proportional to $\cos^2\theta$,
 which  increases with $\Delta m^0$ (see Fig.~\ref{fig:couplings}(b)).
 Finally, the  $D^+D^-$ processes are not affected by a  $\Delta m^0$ variation  in both models. 
One should also note that in the MFDM model 
 $D'D^\pm$ is also not affected by  a $\Delta m^0$ variation since such a variation  does  change neither the $D'$ mass nor the $W^+ D'D^-$ 
coupling (which is purely weak and is not affected by the  $\chi_1^0-\chi_s^0$ mixing).

\begin{figure*}[!t]
\centering
\subfloat[]{
		\includegraphics[width=0.2\textwidth]{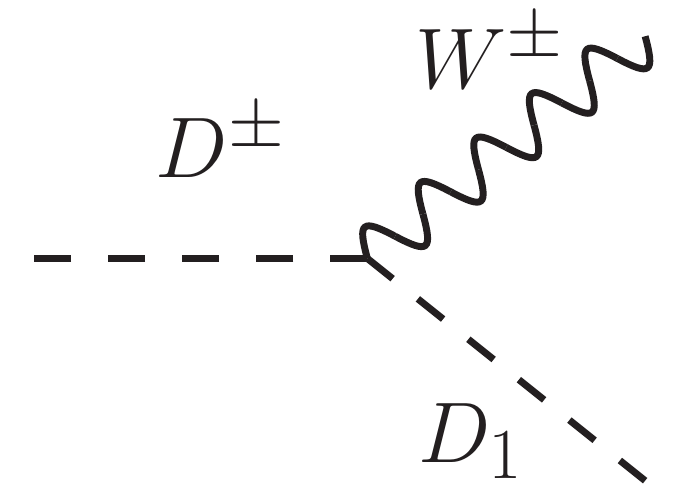}}
	    \subfloat[]{
		\includegraphics[width=0.2\textwidth]{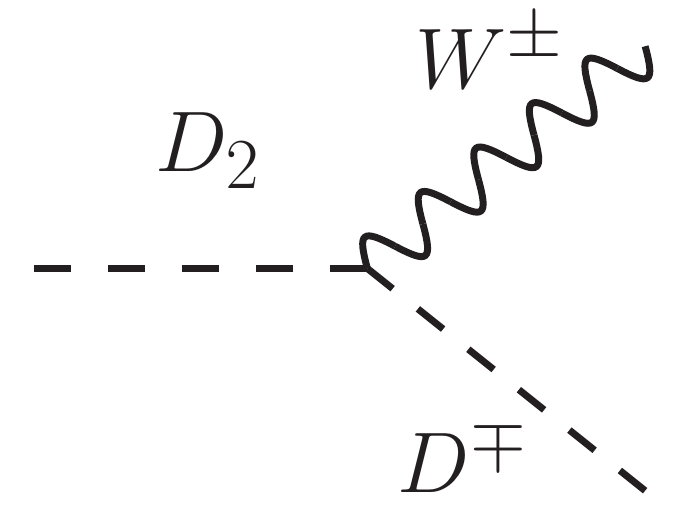}
	}
\\
\subfloat[]{\includegraphics[width=0.2\textwidth]{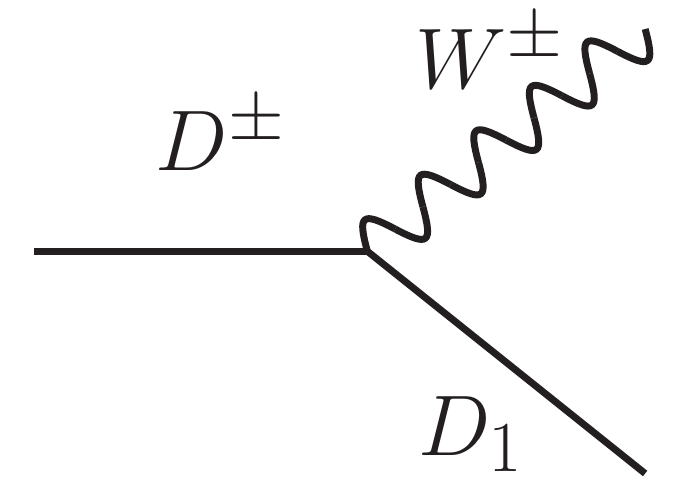}}
	\subfloat[]{
	\includegraphics[width=0.2\textwidth]{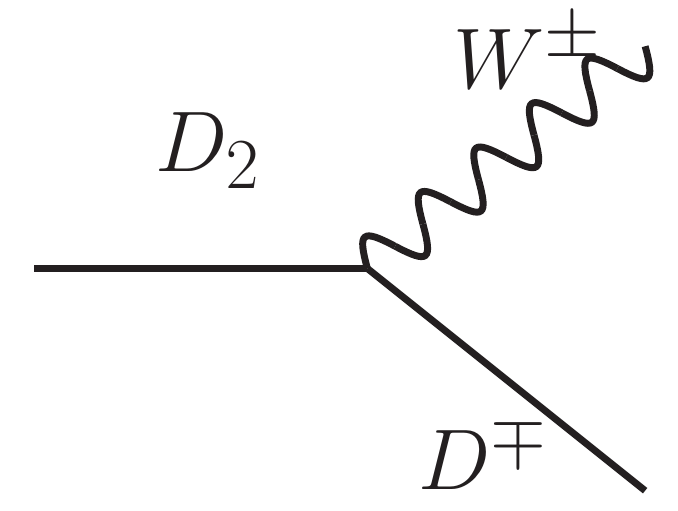}
}
\caption{Decays analogous between the i2HDM (left) and the MFDM model (right).}
\label{fig:FeynDecay1}
\end{figure*}
\begin{figure*}[!h]
	\centering
		\subfloat[]{
		\includegraphics[width=0.2\textwidth]{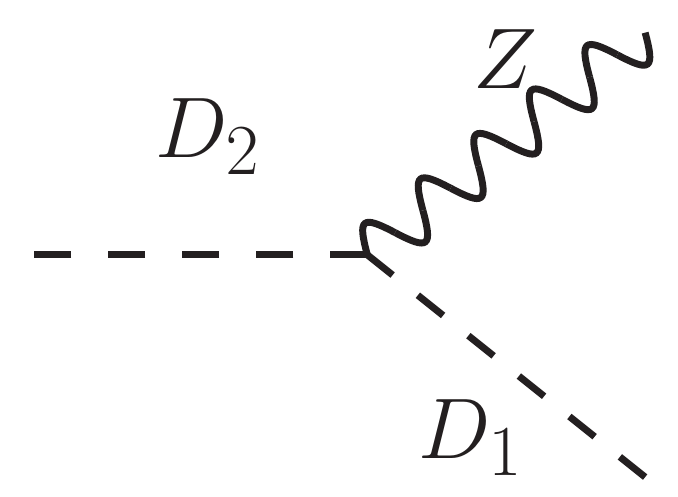}
	}
\\
	\subfloat[]{
	\includegraphics[width=0.2\textwidth]{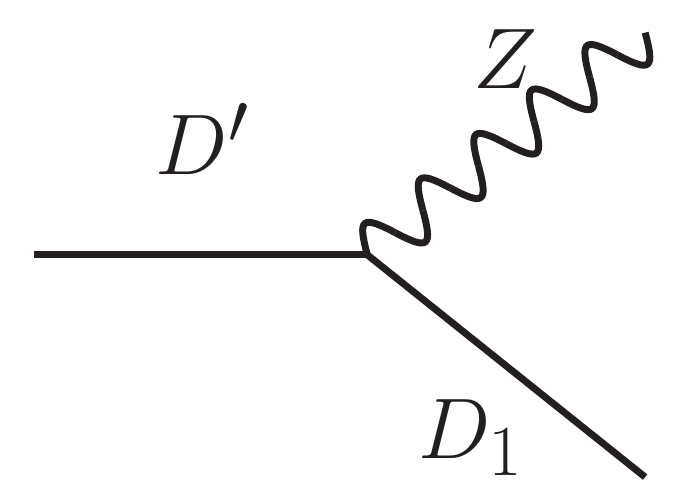}
}
	\subfloat[]{
		\includegraphics[width=0.2\textwidth]{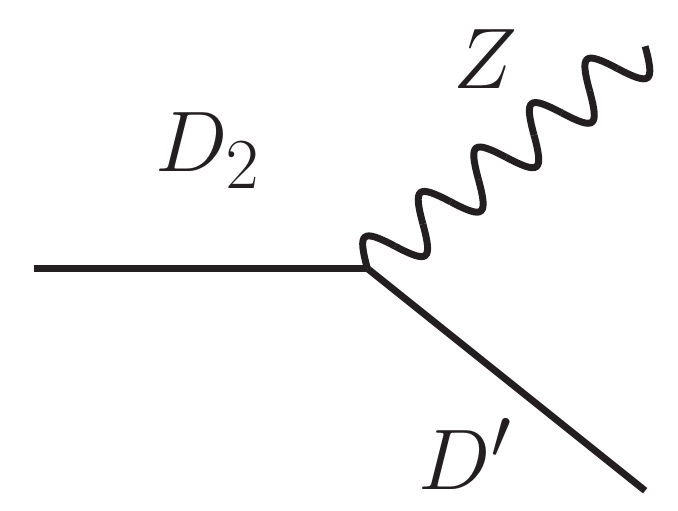}
	}
	\subfloat[]{
		\includegraphics[width=0.2\textwidth]{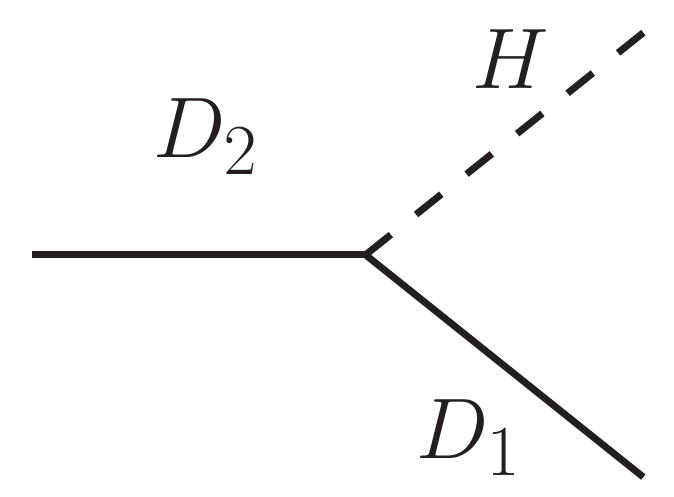}
	}
		\caption{Decays exclusive to the i2HDM (top) and MFDM model (bottom).}
\label{fig:FeynDecayMFDM}
\end{figure*}
\begin{figure*}[!h]
	\centering
	\subfloat[]{
		\includegraphics[width=0.49\textwidth]{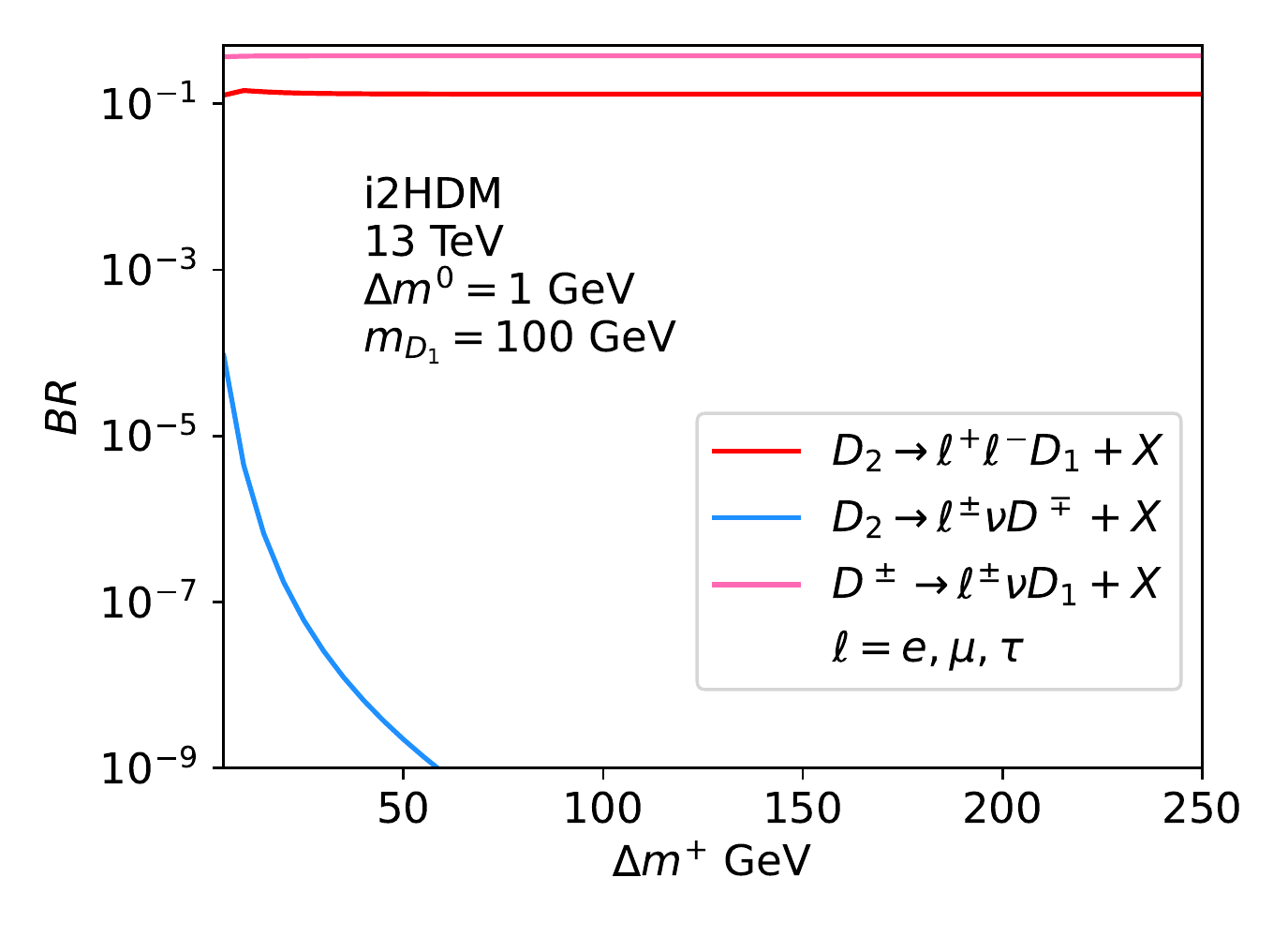}
	}
	\subfloat[]{
		\includegraphics[width=0.49\textwidth]{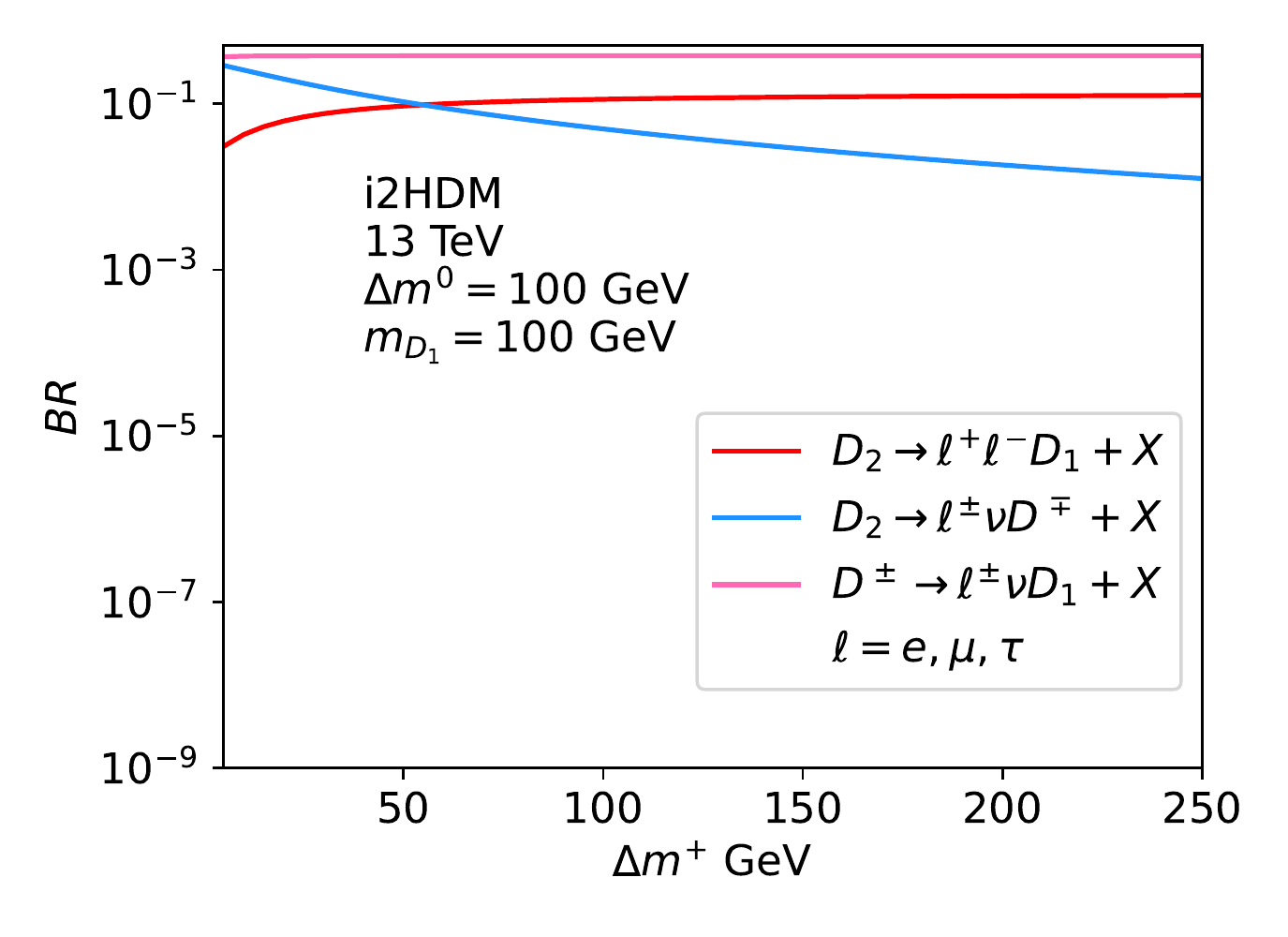}
	}
	\\
	\subfloat[]{
		\includegraphics[width=0.49\textwidth]{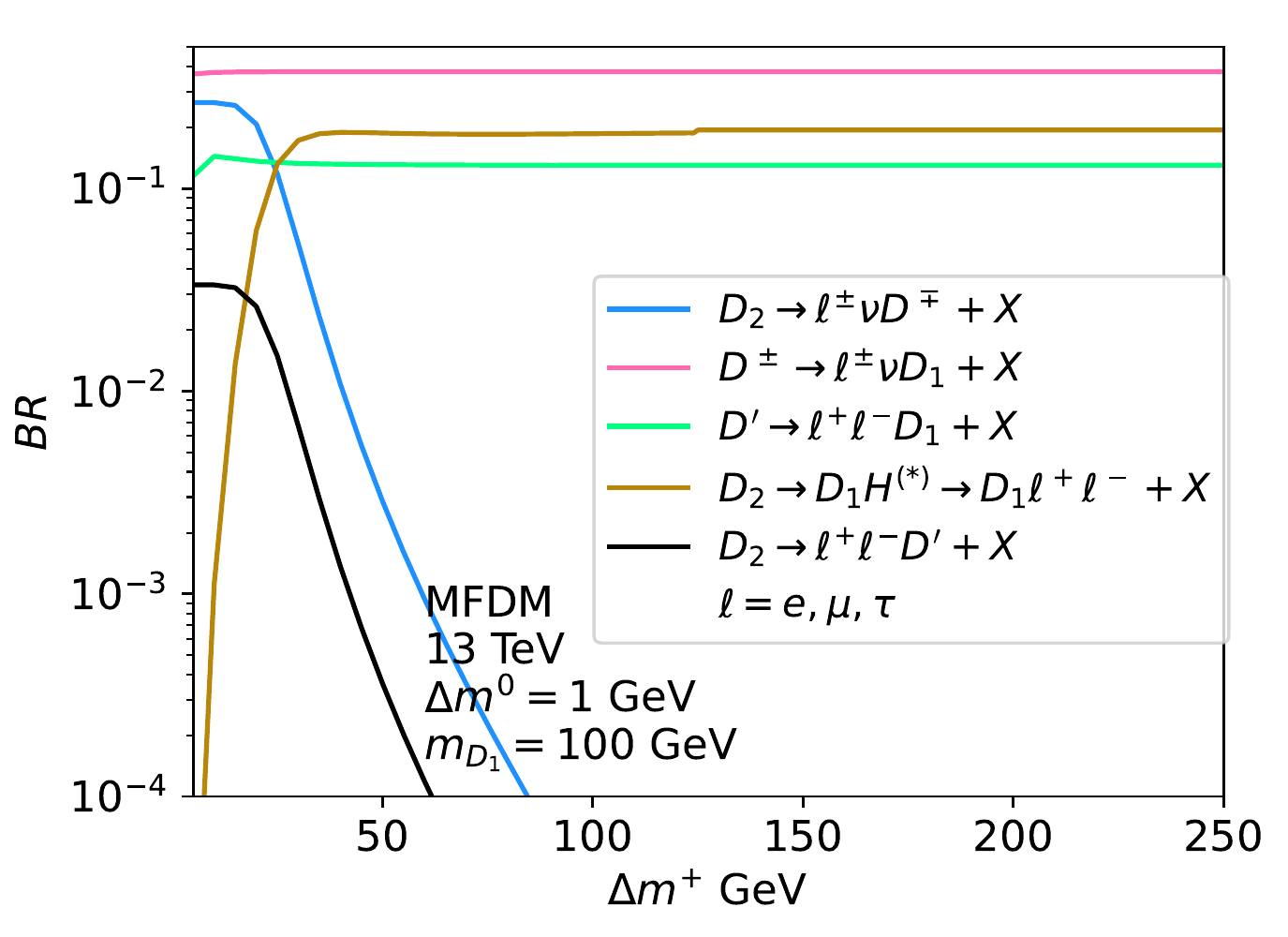}
	}
	\subfloat[]{
		\includegraphics[width=0.49\textwidth]{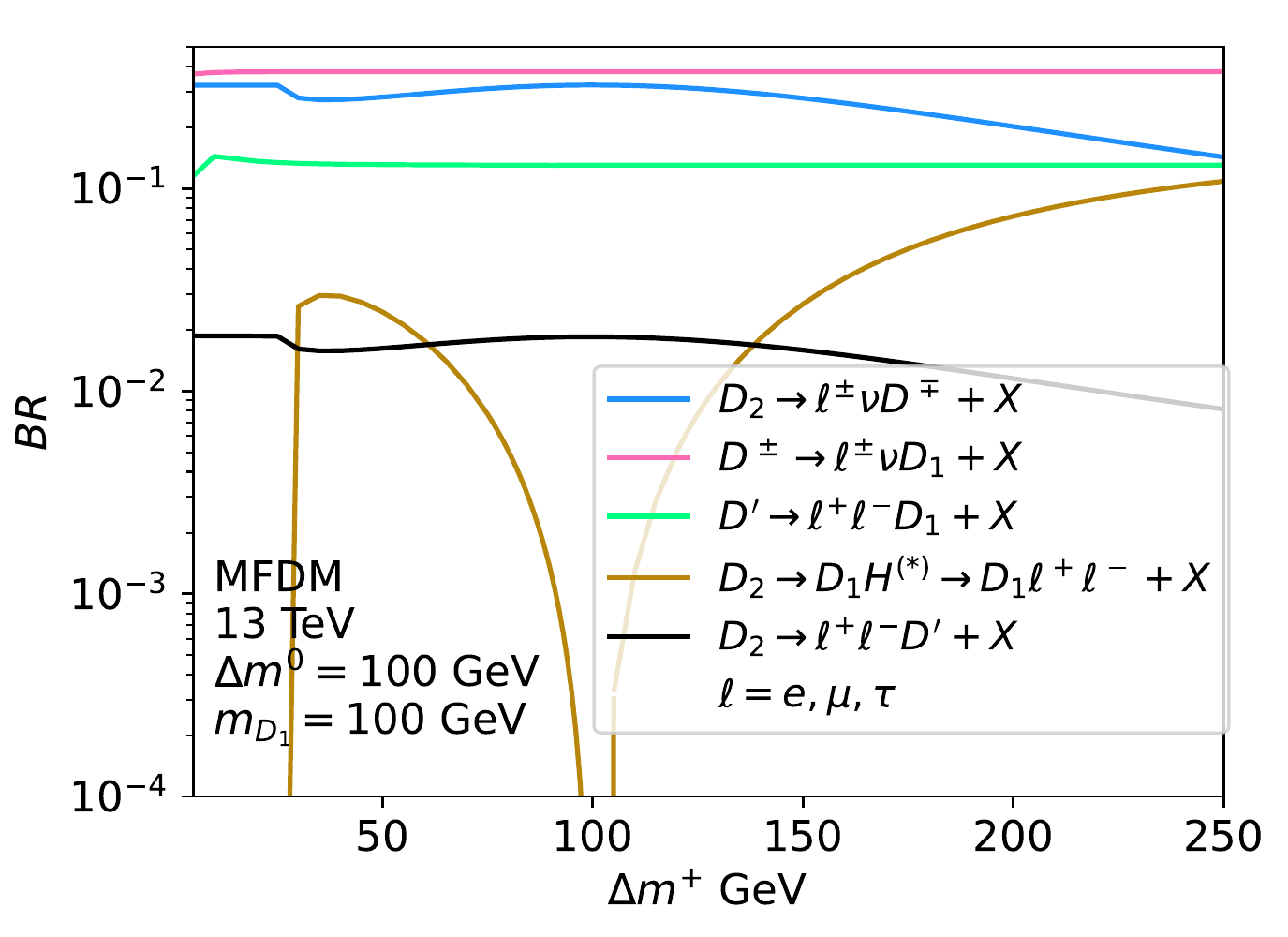}
	}
	\caption{The BRs of leptonic decays and DM particles for the i2HDM (top) and MFDM model (bottom)  for $\Delta m^0=1$ and $100$ GeV (left and right panels, respectively).}
	\label{fig:Decays13TeV}
\end{figure*}

In order to get an idea of the rate at which leptons are produced in the final state
we also need to discuss Branching Ratios (BRs) of decay chains leading to DM particles and connect these with their pair production cross sections. In Fig.~\ref{fig:FeynDecay1}, we present the decay patterns common to both the i2HDM and MFDM model. In Fig.~\ref{fig:FeynDecayMFDM}, we then present the unique decays of each model: the $D_2 \to Z D_1$ decay permitted in the i2HDM while the $D_2 \to H D_1$ and $D'$ decays are so in the MFDM model.

The BRs for the decays leading to our DM signatures (see Figs.~\ref{fig:FeynDecay1} and~\ref{fig:FeynDecayMFDM}) are presented in Fig.~\ref{fig:Decays13TeV} 
as a function of the $\Delta m^+$ parameter, again, for a 100 GeV DM mass and the two values of $\Delta m^0$=1 and 100 GeV.
The $D^\pm$ decay to $W^\pm$ dominates in all cases. This is because, for the i2HDM, only $m_{D_2}>m_{D^\pm}$ is considered and so $D^{\pm}$ can only decay to $W^\pm D_1$ while, for the MFDM model, $D'$ and $D^\pm$ are mass degenerate and their decay to one another is not permitted. 
For $\Delta m^0$=1, in the i2HDM (Fig.~\ref{fig:Decays13TeV}(a)), the $D_2$ decay to leptonically decaying $Z$  and $D_1$ is favoured over the $D_2$ decay to leptonically decaying $W^+$ and $D^-$ which is suppressed due to a small $D_2-D^\pm$ mass splitting. This BR falls rapidly with increasing $\Delta m^+$, due to $D^+$ in the final state increasing in mass, while the mass of the $D_1$ in the $D_2\rightarrow ZD_1$ final state is fixed. 

The behaviour of the $D_2$ BRs changes when considering $\Delta m^0=1 \to 100$~GeV (Fig.~\ref{fig:Decays13TeV}(b)) since the $D_2-D^\pm$ mass gap is increased, initially suppressing the $D_2$ decay to $ZD_1$. However, the trend over increasing $\Delta m^+$ remains the same and the two $D_2$ BRs cross for $\Delta m^+$  around 60 GeV. 
Meanwhile, for the MFDM model, the $D'$ $\to$ $W^+D^-$ decay is not permitted and so the BR  of the $D'$ decay to a leptonically decaying $Z$ and $D_1$ remains unchanged between Fig.~\ref{fig:Decays13TeV}(c) and (d).

The MFDM model decays from $D_2$ include a unique Higgs interaction, $D_2\rightarrow D_1H^{(*)}$, where the $D_2D_1H$ coupling is a product of the Yukawa $Y_{\rm DM}$ (which is proportional to $\sqrt{\Delta m^0\Delta m^+}$) and $\cos(2\theta)$ of the $\chi_1^0-\chi_s^0$ mixing angle $\theta$. This interplay is seen in Fig.~\ref{fig:couplings}(a) (blue line), where $Y_{\rm DM}$ suppresses this coupling at small $\Delta m^0$ and $\Delta m^+$, but the latter rises with increasing $\Delta m^+$. For $\Delta m^0$=1 the shape of the $D_2\to D_1H^*$  (brown) line in Fig.~\ref{fig:Decays13TeV}(c) follows the trend just described, then it  levels out as $\sqrt{\Delta m^+}$. As the Higgs boson becomes on-shell with increasing $\Delta m^+$, a jump in the BR of  leptonic decays via $D_2\rightarrow D_1H$ is observed in Fig.~\ref{fig:Decays13TeV}(c), due to the additional contribution $H\rightarrow W^+W^-$. Meanwhile, the $D_2$ decays to $W^+D^-$ and $ZD'$ are both controlled by the sine of the mixing angle, i.e., the amount by which the $D_2$ is $\chi_1^0$. Therefore, their shapes are similar to one another but the blue and black lines fall as the $D_2\rightarrow D_1H$ brown line increases with $\Delta m^+$ (note that the $D_2\rightarrow W^+D^-$ BR is a factor $\sim 8$ larger than the $D_2\rightarrow D'Z$ one  due to combinatorics and the ratio $m_Z/m_W$ ).

Now, considering $\Delta m^0=100$ for the MFDM model, $D_2$ decays are strongly affected by any $\Delta m^0$ variation. In Fig.~\ref{fig:Decays13TeV}(d), $D_2\rightarrow ZD'$ and $D_2\rightarrow W^+D^-$ are now on-shell and these decays are then  enhanced. When $\Delta m^+=30$ GeV and  $m_{D_2}-m_{D_1}=130$ GeV,  the $D_2\rightarrow HD_1$ decay (brown line)  becomes on-shell and boosted, reducing the other $D_2$ decays proportionally. However, $D_2$ changes from mostly $\chi_s^0$ to mostly $\chi_1^0$ as $\Delta m^+$ reaches $\Delta m^0$ from below, where the coupling reaches zero at $\Delta m^+=\Delta m^0=100$ GeV (see Fig.~\ref{fig:couplings}(a) red line). Beyond this value of $\Delta m^+$, the coupling then increases with $\sqrt{\Delta m^+}$  as $\Delta m^+>\Delta m^0$ which in turn reduces the $D_2\rightarrow ZD'$ and  $D_2\rightarrow W^+D^-$ leptonic BRs above $\Delta m^+>100$ GeV as seen in Fig.~\ref{fig:Decays13TeV}(d).

The discussed combinations of  production and decay rates provide the 
	expected rates for the 2- and 3-lepton  signatures which we study below at the (fast) detector  level and compare to  published LHC data.

\section{Results\label{sec:results}}

In order to understand the constraints on the models we study in this paper, we reinterpret existing multi-lepton searches at the LHC. We first provide a brief summary of the models, then describe briefly the reinterpretation tools for this work, finally followed by an assessment of the impact of these searches on our two benchmark models. 

\subsection{LHC Searches and Tools}

We begin by identifying the LHC searches that can potentially be useful to constrain the latter.  
For the i2HDM, constraints were derived previously in  \cite{Belanger:2015kga} by reinterpreting  8 TeV 2- and 3-lepton searches  using MadAnalysis5  \cite{MadAnal2013}. In preparation for this publication, the previous work is verified using the public recast  tool CheckMATE  \cite{Dercks:2016npn}   (see Appendix~\ref{app:i2HDM8}). 
The corresponding recast software is publicly available at  \cite{ArranCheckMATESUSYCode,ArranCheckMATEHIGGSCode}. In this  paper, we extend this result to 13~TeV for both the i2HDM and MFDM model based on searches available in CheckMATE.

\begin{table}[!htb]
	\centering
	\resizebox{\columnwidth}{!}{%
	\begin{tabular}{l| c|c|c}
		Analysis&Description&Final States&Lumi. [fb$^{-1}$]\\
		\hline
		\hline
{\tt atlas\_1609\_01599}  \cite{analysis_ttz}  &       $t\bar t V$ cross-section measurement at 13 TeV&  two or three leptons(one OSSF pair)+bjets&  3.2      \\
{\tt atlas\_conf\_2016\_076}  \cite{ATLAS-CONF-2016-076}   & direct top squark pair + DM production & two leptons + jets + $\met$   &    13.3 \\    
{\tt atlas\_conf\_2016\_096}  \cite{ATLAS-CONF-2016-096}  & EW production of charginos and neutralinos& two or three leptons $+\met$  &  13.3      \\   
{\tt atlas\_1712\_08119}   \cite{ATLAS:2017vat}   &   EWinos search with soft leptons &two soft OSSF leptons $+\met$&   36.1     \\
{\tt cms\_sus\_16\_039}  \cite{cms_sus_16_039}      &  EWinos in multilepton final state &$\geq$two leptons + $\tau$ $+\met$ &	35.9\\
{\tt cms\_sus\_16\_025}   \cite{CMS-PAS-SUS-16-025}    &   EWino and stop compressed spectra&two soft OS leptons $+\met$&	12.9	\\
{\tt cms\_sus\_16\_048}   \cite{CMS-PAS-SUS-16-048}    &  Search for new physics in events with soft leptons&two soft OS leptons $+\met$&	35.9	
\end{tabular}
}
\caption{The relevant 13 TeV ATLAS and CMS analyses which are sensitive to the DM signatures under study in this paper.}
\label{tab:13TeV_Analyses}
\end{table}

In Tab.~\ref{tab:13TeV_Analyses}, we identify the 13~TeV searches that are relevant for the reinterpretations in this paper.  The most stringent constraints for these models are expected to emerge from 
 LHC Supersymmetry searches targeting EW gauginos. Since we are investigating models with a variety of mass gaps in this paper, we look for searches that constrain both small and large mass gaps.  We therefore note the following.

\begin{itemize}

\item For small mass gaps the CMS  \cite{CMS-loi} soft lepton searches {\tt cms\_sus\_16\_025}  \cite{CMS-PAS-SUS-16-025} 
and  {\tt cms\_sus\_16\_048}  \cite{CMS-PAS-SUS-16-048} can potentially constrain the parameter space under study here.  The search {\tt cms\_sus\_16\_025} constrains $\chi_{1}^{+}\chi_{2}^{0}$ pair production followed by decay to leptons and missing energy via off-shell $W^\pm$ and $Z$ bosons  for mass gaps  $\simeq 5-50$ GeV. It also constrains direct stop pair production followed by a decay to leptons via off-shell $W^\pm$'s for mass gaps up to 70 GeV.   The search {\tt cms\_sus\_16\_048} targets the same EWino pair production as above and constrains mass gaps up to about 50~GeV at an integrated luminosity of $\rm 35.9 ~ fb^{-1}$. Both of the above searches require Opposite Sign (OS) $ee/\mu\mu/e\mu$ with leading leptons $p_T < 20$ GeV, $\met<200$ GeV and at least one jet.

\item For large mass gaps 
the ATLAS  \cite{Collaboration2008}  search  {\tt atlas\_conf\_2016\_096}  \cite{ATLAS-CONF-2016-096} as well as the CMS search {\tt cms\_sus\_16\_039} \cite{cms_sus_16_039} are the most constraining ones. These publications target EWino pair production with the same decay pattern as the soft lepton searches. The searches look for OS leptons and constrain mass gaps above 50 GeV.

\end{itemize}

The  i2HDM and MFDM model input LHE files were produced with different mass parameters,  as described in section~\ref{sec:signal}, with CalcHEP  \cite{Belyaev:2012qa}. This is  followed by showering and hadronisation using PYTHIA8  \cite{Sjostrand:2014zea}. Jets, with final-state hadrons are constructed using FASTJET  \cite{Cacciari_2012}, while detector simulation is performed using DELPHES  \cite{deFavereau:2013fsa}. The entire process (barring parton level event generation) is performed within CheckMATE. The built-in AnalysisHandler processes the detector-level events with  the user selected analyses. The signal size is determined based on  the efficiency,  acceptance, signal cross section and integrated luminosity  \cite{Lumi2013} of the analysis.

\subsection{Constraints on the i2HDM and MFDM Model}

In this section, we present the results of our study for the i2HDM and MFDM model. For each of these scenarios, we constrain the parameter space in the ($m_{D_{1}}, \Delta m^+$) plane using  
the searches quoted above to calculate the  {\it r-value}:
\begin{equation}\label{eqn:rvalue}
r=\frac{\rm S_{DM}}{\rm S_{95}},
\end{equation}
where $\rm S_{DM}$ is the number of DM events expected to pass the signal selection and $\rm S_{95}$ is the 95$\%$ Confidence Level (CL)  upper limit on the number of selected events.
Any point with $r\geq1$ is excluded by current LHC limits.

In general we observe the following.
\begin{itemize}
\item {\bf{Low Mass Gap: 2${l}$ + $p_{T}^{\rm miss}$ channel - }}  For low mass gaps the soft lepton analyses {\tt cms\_sus\_16\_025}  and {\tt cms\_sus\_16\_048} constrain a large part of the parameter space in the 2${l}$ + $p_{T}^{\rm miss}$ channel. For the MFDM model,  they exclude phase space within $30<m_{D_1}<160$~GeV and  $10<\Delta m^+<50$~GeV for   $\Delta m^0=1,10, 100$~GeV. For the i2HDM, the two soft lepton analyses exclude phase space within $1<m_{D_1}<60$~GeV and  $1<\Delta m^+<60$~GeV for  $\Delta m^0=1$~GeV as well as  phase space within  $1<m_{D_1}<40$~GeV and $1<\Delta m^+<40$~GeV for  $\Delta m^0=10,100$~GeV. 

\item {\bf{Large Mass Gap:  2${l}$ + $p_{T}^{\rm miss}$ channel - }} The harder 2-lepton signatures for the MFDM model are excluded by the {\tt atlas\_conf\_2016\_096}  \cite{ATLAS-CONF-2016-096} signal regions {\tt 2LASF} ({\tt 2LADF}) which require two same (different) flavours $e/\mu$ with $m_{{l}{l}} - m_Z> 10$~GeV and $m_{T2}>0$~GeV. These signal regions exclude a phase space within $0<m_{D_1}<30$~GeV and $50<\Delta m^+<80$~GeV for  $\Delta m^0=1,10, 100$~GeV.  In the i2HDM case, the {\tt atlas\_conf\_2016\_096} analysis excludes a phase space within $0<m_{D_1}<45$~GeV and  $60<\Delta m^+<90$~GeV for  $\Delta m^0=1,10$~GeV.

\item {\bf{ 3${l}$ + $p_{T}^{\rm miss}$ channel : }} For the MFDM model, the majority of the 3-lepton final state signal is excluded by the analysis  {\tt cms\_sus\_16\_039}  \cite{cms_sus_16_039}, by various signal regions, including {\tt SR\_C03} ($150<p^{\rm miss}_T<200$~GeV), {\tt SR\_C04} ($200<p^{\rm miss}_T<250$~GeV) and  {\tt SR\_C05} ($p^{\rm miss}_T\geq250$~GeV). They all require three leptons with at least two $e$ or $\mu$ forming an opposite-sign same-flavour (OSSF) pair with $m_{{l}{l}} < 75$~GeV and including a $\tau$-lepton. In the MFDM model, these signal regions exclude a phase space within $30<m_{D_1}<110$~GeV and  $20<\Delta m^+<300$~GeV for $\Delta m^0=1,10,100$~GeV. 
For the i2HDM, substantial $r$-value contributions  are  provided by the {\tt cms\_sus\_16\_039} analysis with signal regions for three leptons, {\tt SR\_A30} (three leptons with one OSSF pair $75<m_{{l}{l}} < 105$~GeV, $250<p^{\rm miss}_T<400$~GeV), {\tt SR\_K02} (four leptons, including two $\tau$-leptons, $50< p_T^{\rm miss}<100$~GeV), {\tt SR\_F10} (three leptons including two $\tau$-leptons) and   {\tt SR\_I03} (four or more leptons, including one $\tau$),
within the phase space $1<m_{D_1}<20$~GeV and   $5<\Delta m^+<70$~GeV for $\Delta m^0=1,10$~GeV.

\end{itemize}

\subsubsection{Constraints on the i2HDM channels}

\begin{figure*}[!ht]
	        	\subfloat[]{
\includegraphics[width=0.49\textwidth]{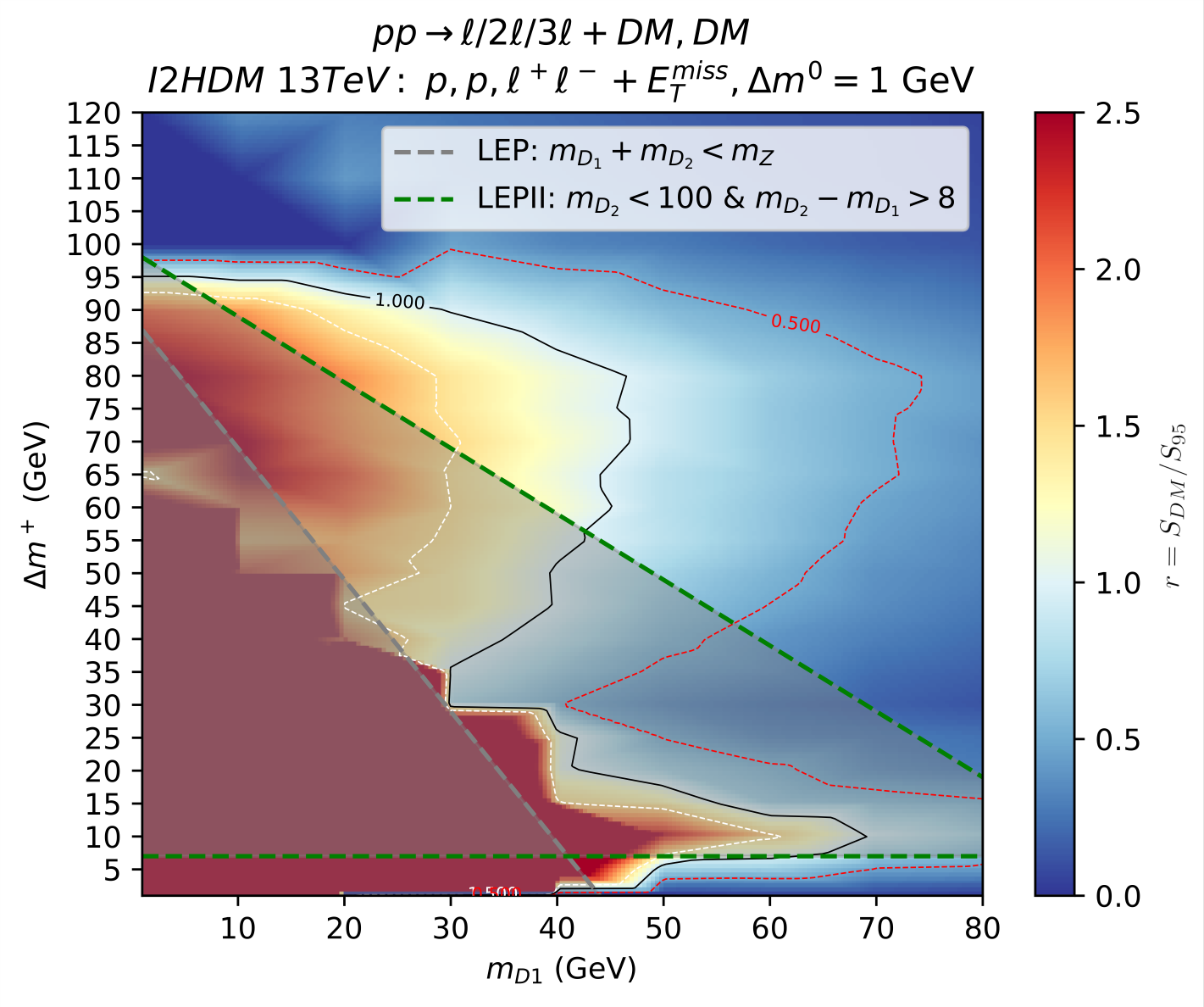}
}
        	\subfloat[]{
\includegraphics[width=0.49\textwidth]{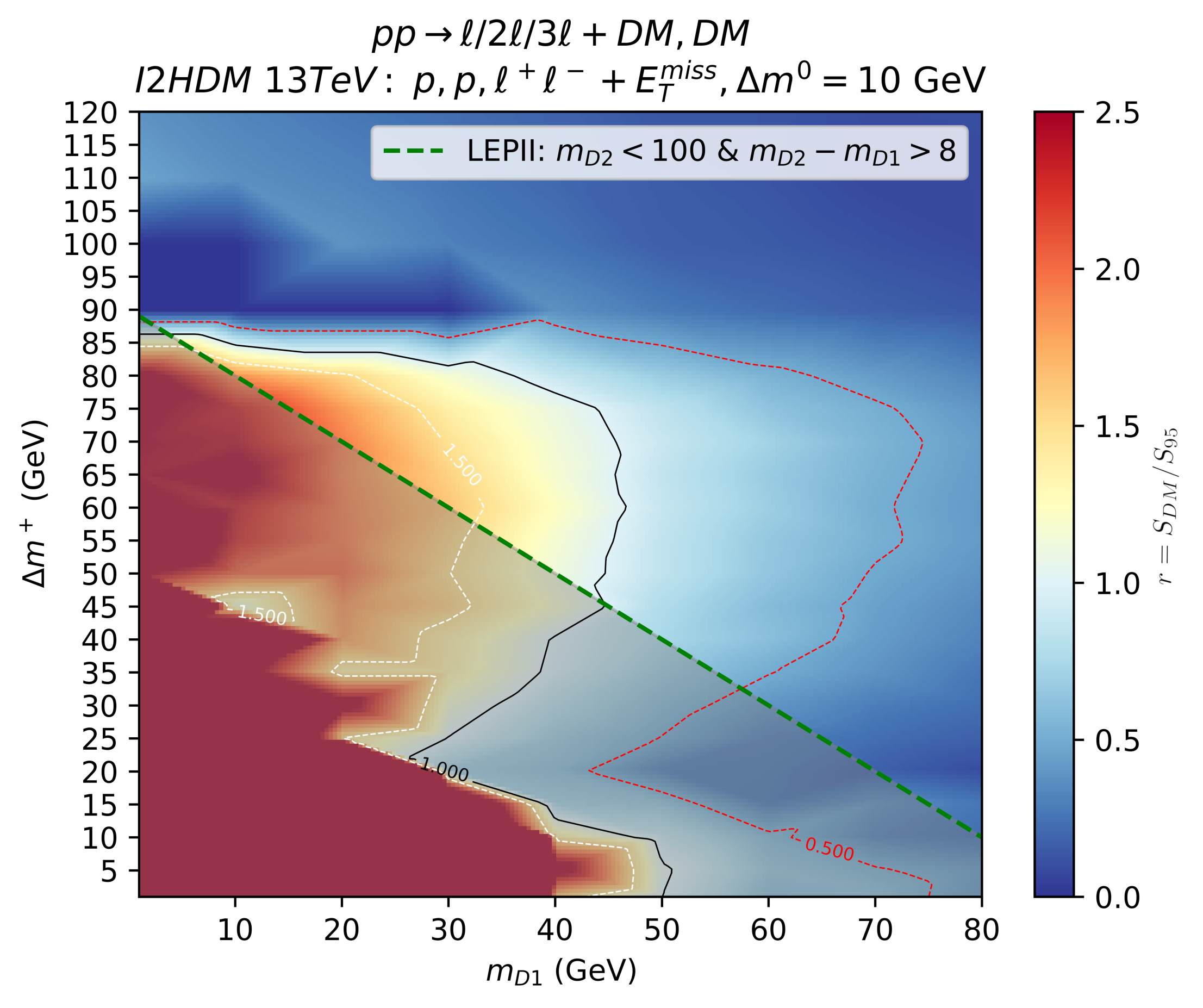}
}\\
        	\subfloat[]{
\includegraphics[width=0.49\textwidth]{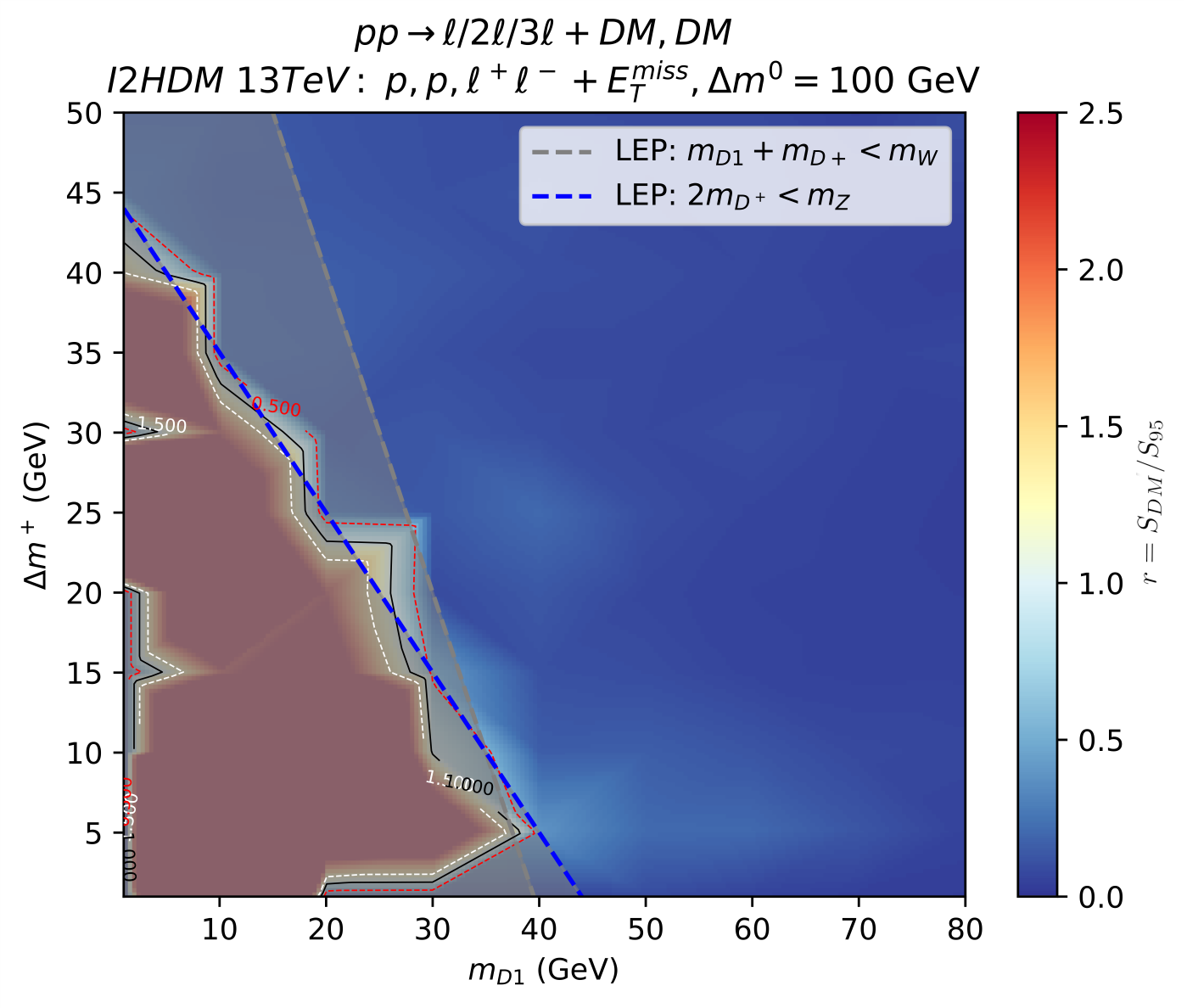}}
	\caption{i2HDM 13 TeV 
		$r$-value contour plots for 2- and 3-lepton final states as a function of $\Delta m^+$  and $m_{D1}$ for $\Delta m^0=1$ (a), $\Delta m^0=10$ (b) and $\Delta m^0=100$ (c) GeV. These are overlaid with limits from the LEP-I and LEP-II experiments  \cite{Lundstr_m_2009}, \cite{Pierce_2007}.  This includes $m_{D_1}+m_{D_2}<m_Z$ (grey line) and $2m_{D^+}<m_Z$(blue), from LEP-I $Z$ boson width measurements excluding on-shell $Z\rightarrow D_1D_2$ and $Z\rightarrow D^+D^-$ decays. The same applies to the $W$ width measurement excluding on-shell $W^\pm\rightarrow D^\pm D_1$ decays. The  region $m_{D_1} < 80$ GeV {\it and} $m_{D_2} < 100$ GeV {\it and} $m_{D_2}-m_{D_1} > 8$ GeV (green lines) is excluded by LEP-II observations. Where these lines are absent, they are overlapped completely by the other LEP limits.
}
	\label{fig:i2HDM-DeltaM}
\end{figure*}

Using the re-parametrisation of Eq.~(\ref{eq:i2hdm_DM}),  Fig.~\ref{fig:i2HDM-DeltaM} (overlaid with total cross-section yields in Fig.~\ref{fig:i2HDM-DeltaM_Numb} of Appendix~\ref{app:plots}), show 
the $r$-value (see Eq.~(\ref{eqn:rvalue})) of the i2HDM  2- and 3-lepton signatures discussed in section~\ref{sec:signal} as a function of $\Delta m^+$  and $m_{D1}$ for various choices of $\Delta m^0$. The 3-lepton contributions to the $r$-value  are important, reaching up to $\sim70\%$ of the 2-lepton-only  $r$-values. However, 2-lepton-only  $r$-values still dominate throughout the i2HDM contour plots.

For $\Delta m^0=1$~GeV, the dominant DM pair production process is  $D_2D_1$ (see Fig.~\ref{fig:XSecsTeV}(a)) and the dominant $D_2$ decay contribution is $D_2 \to ZD_1$ (see Fig.~\ref{fig:Decays13TeV}(a)), providing two leptons. The 3-lepton final states can be provided by the second largest DM production process, $D_2D^\pm$, with decays as $D_2 \to ZD_1$ and $D^\pm \to W^\pm D_1$. This process and its decays can also fulfil the 2-lepton criteria if the $W$ decays hadronically, or a lepton is misidentified.

The horizontal wedge of large $r$-value in Fig.~\ref{fig:i2HDM-DeltaM}(a) within $\Delta m^+<60$~GeV, $m_{D1}<70$~GeV  is excluded by analyses {\tt cms\_sus\_16\_025}  \cite{CMS-PAS-SUS-16-025} and {\tt cms\_sus\_16\_048}  \cite{CMS-PAS-SUS-16-048}, both with signal region {\tt SR1\_weakino\_1low\_mll\_2}, requiring two leptons with $m_{{l}{l}} < 20$~GeV and at least one jet. 
In the i2HDM, this signature would be mostly provided by the leptonic $Z$ decay and the hadronic $W$ decay in the  $D_2D^\pm$ pair production.
Most of this phase space is already excluded by LEP-II observations  \cite{Lundstr_m_2009} (green line) and LEP-I limits from $Z$ width measurements (grey line), excluding on-shell $Z\to D_2D_1$ decays.
However, at $\Delta m^+<8$~GeV, there is a small wedge of allowed phase space from soft leptons, when $m_{D_1}>50$~GeV and $\Delta m^+<8$~GeV which is not covered by the LEP-I or LEP-II limits.
The second, broader 
region of large $r$-value around $40<\Delta m^+<100$~GeV is excluded by analysis {\tt atlas\_conf\_2016\_096}  \cite{ATLAS-CONF-2016-096}, signal region {\tt 2LASF}. 
The dominant $D_2D_1$ process with leptonic $Z$ decay provides a signal that strongly contributes to the $r$-value in this region. 
Note that a significant portion in Fig.~\ref{fig:i2HDM-DeltaM}(a) within the region $60<\Delta m^+<95$~GeV and $m_{D_1}<50$~GeV not excluded by LEP, is excluded by these LHC limits. 

Similar excluded regions are found for $\Delta m^0=10$~GeV in Fig.~\ref{fig:i2HDM-DeltaM}(b), but without the region of allowed phase space at small $\Delta m^+$, for which the LEP-II limit now overlaps with this region (the grey line corresponding to the LEP limit $m_{D_1}+m_{D_2}<m_Z$  in Fig.~\ref{fig:i2HDM-DeltaM}(a) would be overlapped completely in Fig.~\ref{fig:i2HDM-DeltaM}(b) by the green line, so is not plotted here). Again, we see a significant contribution from LHC limits in $60<\Delta m^+<95$~GeV and $m_{D_1}<50$~GeV where LEP does not already exclude.

As $\Delta m^0$ is increased further to $100$~GeV in Fig.~\ref{fig:i2HDM-DeltaM}(c), the $D_2D_1$ and $D_2D^\pm$ cross sections are now suppressed at small $\Delta m^+$ (see Fig.~\ref{fig:XSecsTeV}(b)) compared to the dominant production $D^+D^-$ for $\Delta m^+<130$~GeV, with decays $D^\pm \to W^\pm D_1$ (see Fig.~\ref{fig:Decays13TeV}(b)). The $M_{{l}{l}}>100$ GeV cut applied in the  {\tt atlas\_conf\_2016\_096}  \cite{ATLAS-CONF-2016-096} analysis now removes all events for larger $\Delta m^+$ in Fig.~\ref{fig:i2HDM-DeltaM}(c), as heavier $D_2$ production cross section with harder lepton decays has decreased.
In addition, the combination of LEP limits covers the totality of LHC limits for $\Delta m^0=100$ GeV in Fig.~\ref{fig:i2HDM-DeltaM}(c).

\subsubsection{Constraints on the MFDM Model Channels}
 
Based on the re-parametrisation of Eq.~(\ref{eq:i2hdm-dm}), Fig.~\ref{fig:DMD3plotsMFDM} (overlaid in Fig.~\ref{fig:DMD3plotsMFDM_Numb}  of Appendix~\ref{app:plots} with cross-section yields) shows the $r$-value exclusion contours of the MFDM model  2- and 3-lepton signatures discussed in section~\ref{sec:signal} as a function of $\Delta m^+$  and $m_{D1}$ for various choices of $\Delta m^0$. 
The LEP-II limit in the MFDM model case corresponds to bounds on fermionic DM  \cite{Pierce_2007}, which covers $m_{D^+}<100$~GeV, a smaller $\Delta m^+$ range than the LEP limits for the i2HDM. Since we allow DM-Higgs coupling, the Higgs-to-invisible limit  \cite{Sirunyan_2019} of  $\sim 0.15$ BR (magenta region) is also plotted.

Contributions to the 2-lepton $r$-value are provided by the dominant $D^+D^-$ production (see Fig.~\ref{fig:XSecsTeV}(c)-(d)) with its leptonic decays $D^\pm \to W^\pm D_1$ (see Fig.~\ref{fig:Decays13TeV}(c)-(d)), by $D'D^\pm$ and, specifically for $\Delta m^0=1$~GeV, $D_2D^\pm$ productions, where the latter two require $D'$ and $D_2$ to decay leptonically. Cascades from  $D'D_2$ production can also contribute to 2-lepton final states for $\Delta m^0=1$~GeV.
However,  $D_2D^\pm$ and $D'D_2$ production become suppressed with $\Delta m^0=100$~GeV (see Fig.~\ref{fig:XSecsTeV}(d)). Meanwhile, $D'D_1$ production, which is suppressed for $\Delta m^0=1$~GeV, becomes enhanced
with $\Delta m^0=100$~GeV as detailed in section~\ref{sec:signal}, contributing to 2-lepton final states.
Contributions to the 3-lepton $r$-value are provided by the $D'D^\pm$ production with fully leptonic decays, although these are less likely to satisfy the signal regions that require $\tau$-leptons than $D'D_2$ and $D_2D^\pm$ production with $D_2\to D_1H^*\to D_1\tau\tau$ decay. However, as stated, $D_2$ productions become suppressed with $\Delta m^0=100$~GeV (see Fig.~\ref{fig:XSecsTeV}(d)).

For $\Delta m^0=1$~GeV, the sharp excluded region of large $r$-value in Fig.~\ref{fig:DMD3plotsMFDM}(a) within $\Delta m^+<30$~GeV, $m_{D_1}<150$~GeV was seen similarly for the i2HDM case in Fig.~\ref{fig:i2HDM-DeltaM}. As with the i2HDM case, this signal of soft leptons for the MFDM model in Fig.~\ref{fig:DMD3plotsMFDM}(a) is also excluded by the 2-lepton analyses {\tt cms\_sus\_16\_048} and {\tt cms\_sus\_16\_025} both with signal regions {\tt stop\_1low\_pt}, {\tt weakino\_1low\_mll}. These require two leptons and at least one jet, a signal also provided by the $D'D^\pm$ pair production, dominant here in the MFDM model (see Fig.~\ref{fig:XSecsTeV}(c)).

As $\Delta m^+$ increases, the 3-lepton $r$-value becomes larger than the 2-lepton-only $r$-value in the region $30<\Delta m^+<100$~GeV and $m_{D_1}<80$~GeV, excluded by the {\tt cms\_sus\_16\_039} \cite{cms_sus_16_039} analysis signal regions requiring three leptons including $\tau$. This is where the most dominant decay of $D_3$ changes from 1-lepton final states to 2-lepton final states including $\tau$-leptons through virtual Higgs decays, as detailed in section~\ref{sec:signal} and Fig.~\ref{fig:Decays13TeV}(c). This gives more 3-lepton final states, with $\tau$-leptons, from $D^+D_2$ production. 
The additional third excluded area of smaller $r$-value than the previous two areas in Fig.~\ref{fig:DMD3plotsMFDM}(a), but still within an excluded region $120<\Delta m^+<260$~GeV, $m_{D_1}<100$~GeV 
has contributions to the signal from on-shell Higgs decay, where $D_2$ changes from decaying through $H^*$, into decaying to real Higgs which further decays to two $W$ bosons or $\tau$-leptons. This contributes more to the $r$-value as $\Delta m^+>M_H$ for real Higgs production as detailed in section~\ref{sec:signal}. This region is also excluded by analysis {\tt cms\_sus\_16\_039} \cite{cms_sus_16_039}, with its 3-lepton (including one $\tau$) signal regions.

\begin{figure*}
	\centering
	\includegraphics[width=0.49\textwidth]{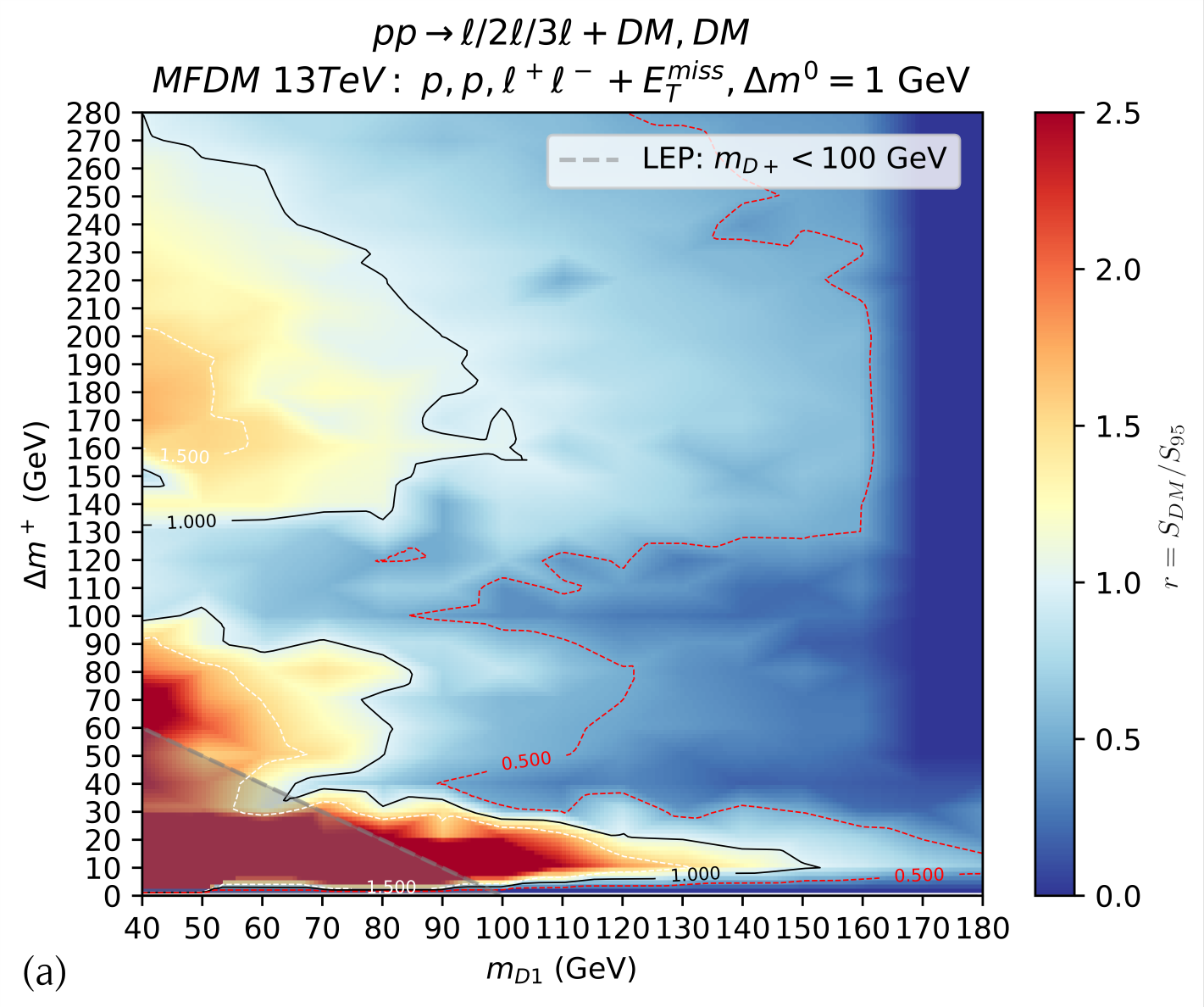}
	\includegraphics[width=0.49\textwidth]{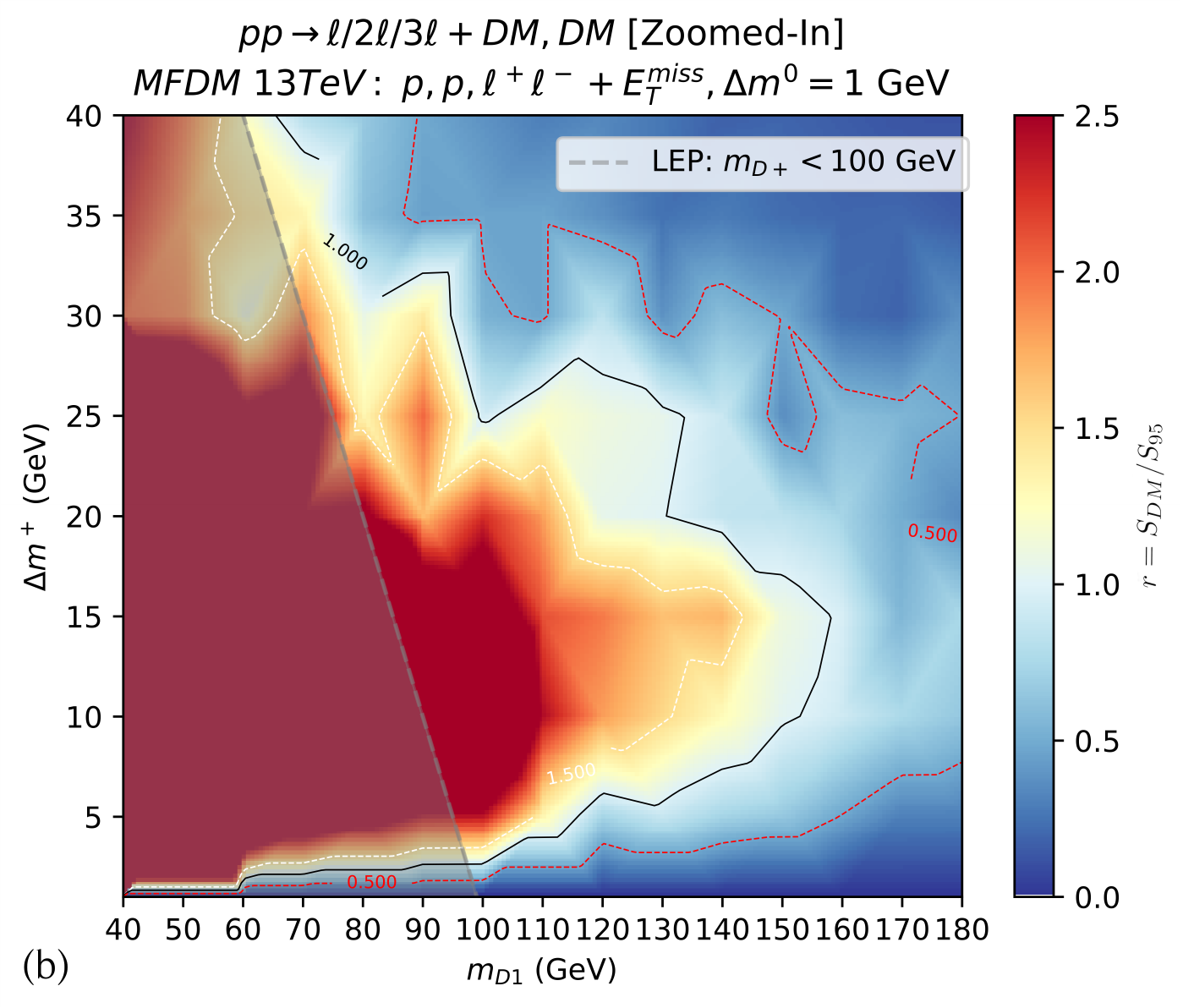}
\\
	\includegraphics[width=0.49\textwidth]{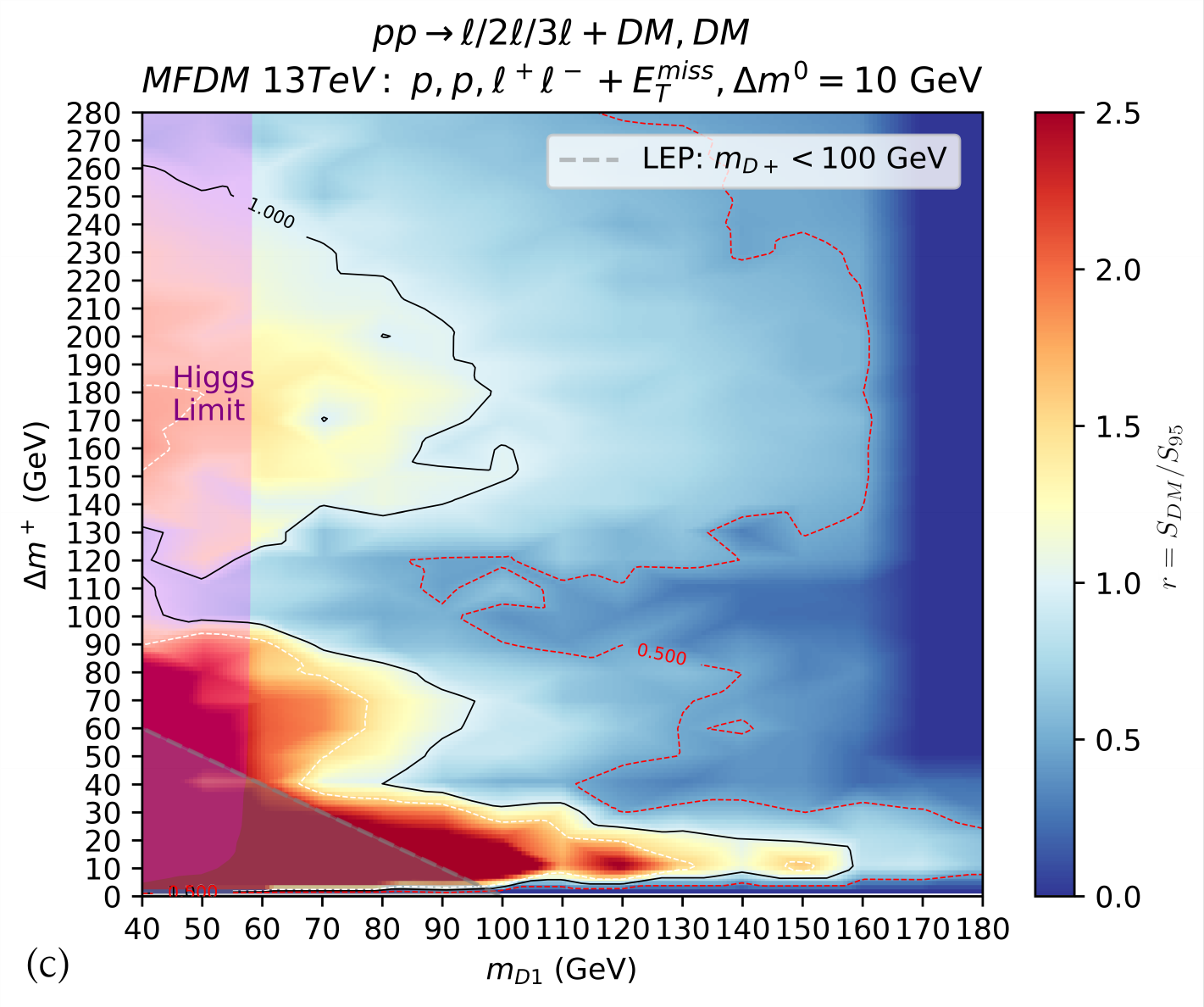}
\includegraphics[width=0.49\textwidth]{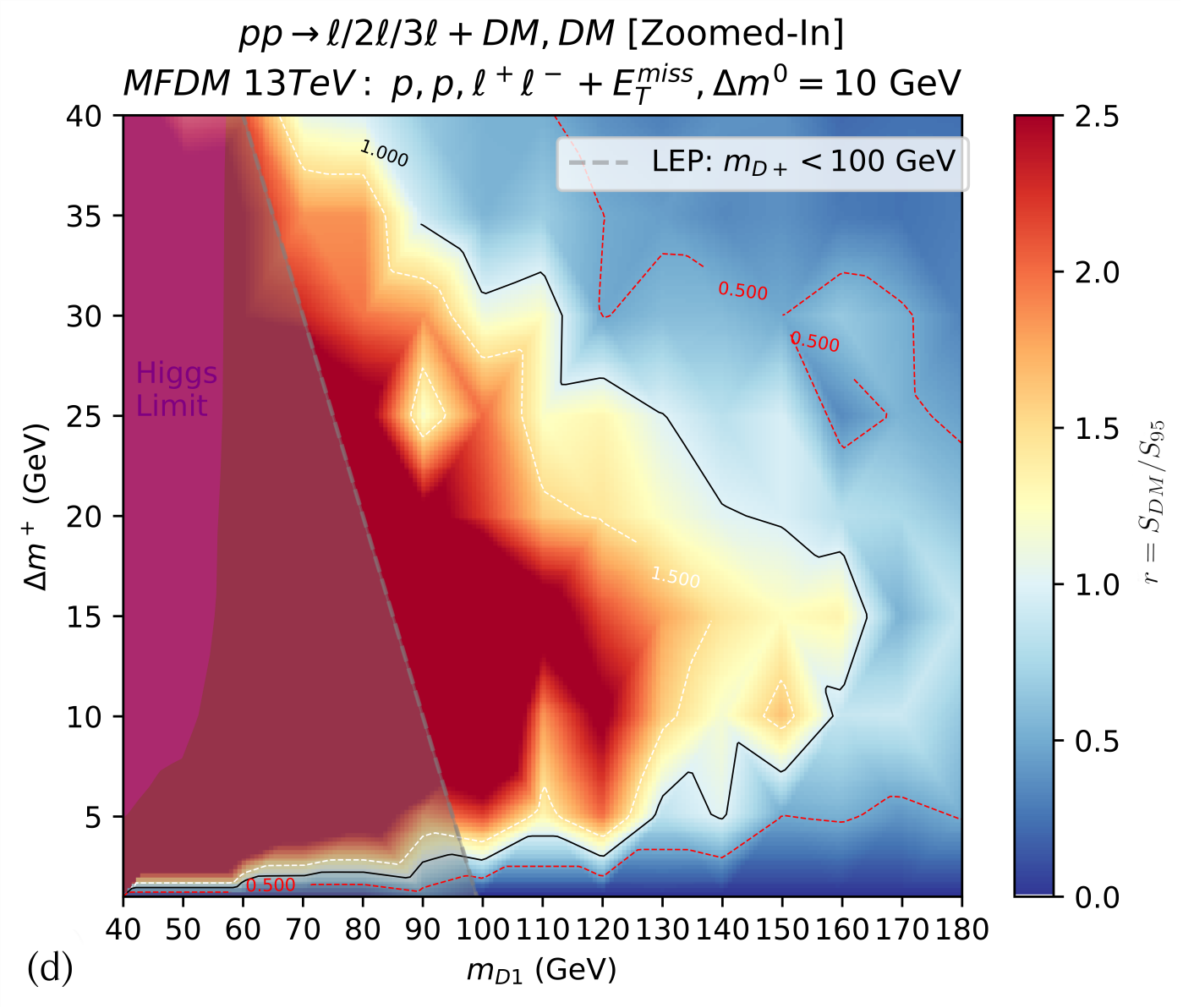}
\\
		\includegraphics[width=0.49\textwidth]{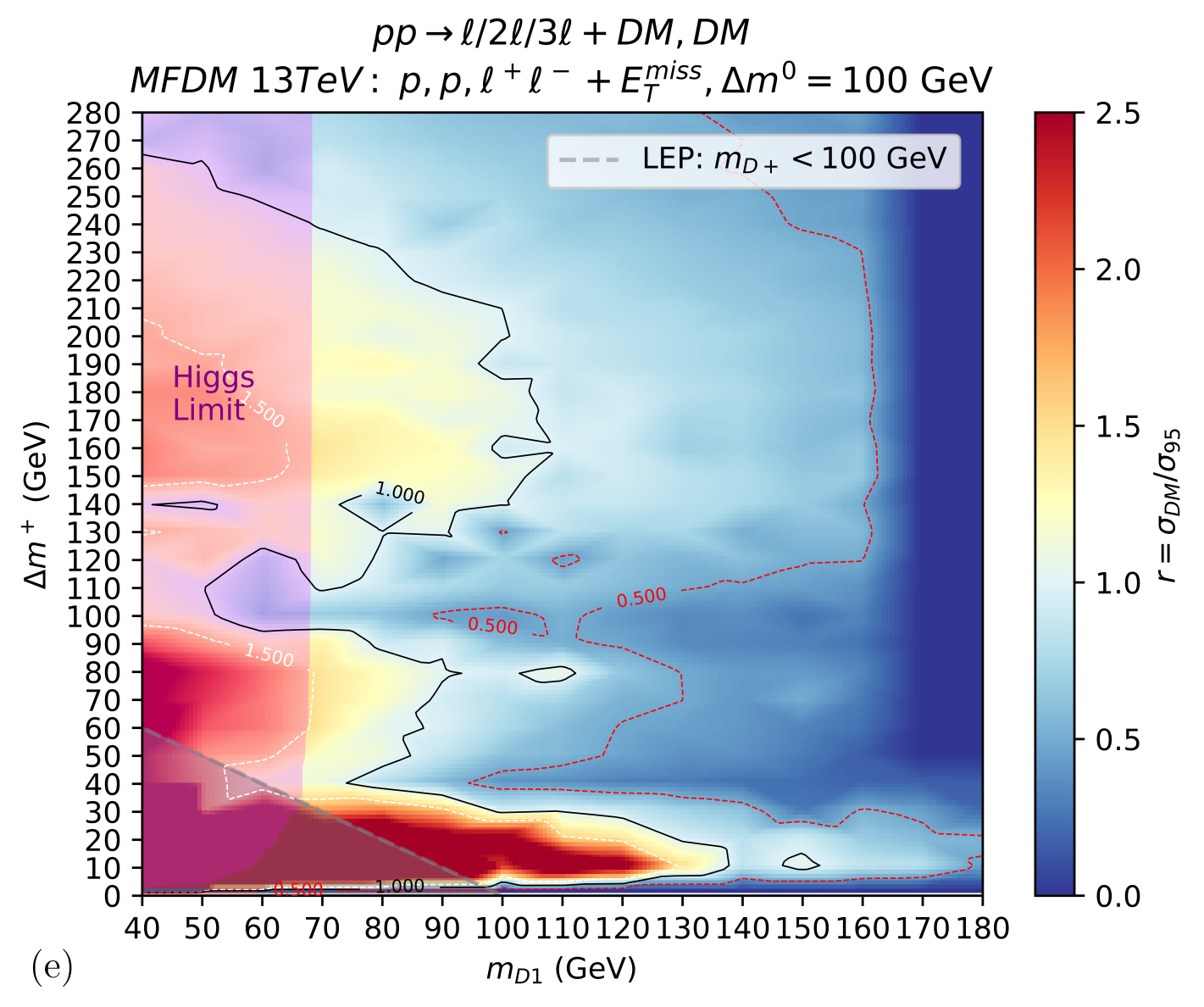}
		\includegraphics[width=0.49\textwidth]{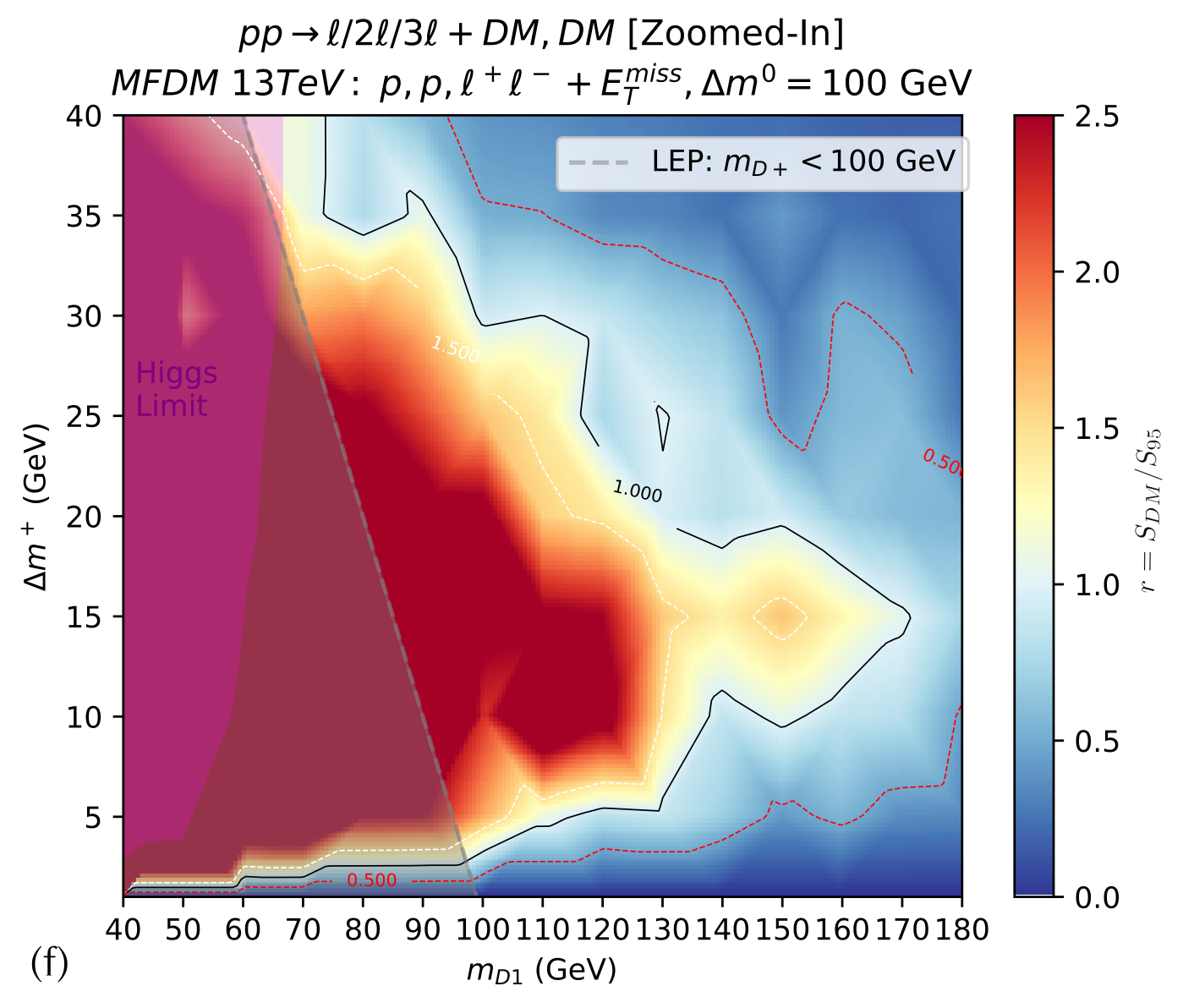}
	\caption{MFDM 13 TeV $r$-value contours as a function of $\Delta m^+$ and $m_{D1}$ for $\Delta m^0=1$(a)$,10$(c)$,100$(e) GeV and zoomed in (b),(d),(f).
		The magenta region and grey region indicate the current Higgs-to-invisible limit  \cite{Sirunyan_2019} of ~0.15 BR and LEP bounds on charginos for the fermionic DM case \cite{Pierce_2007} respectively.
	}
	\label{fig:DMD3plotsMFDM}
\end{figure*}
\clearpage

As $\Delta m^0$ is further increased to 10 GeV and 100 GeV in Fig.~\ref{fig:DMD3plotsMFDM}(b) and (c) respectively, the same $r$-value exclusion patterns are consistently observed.
This no-lose theorem appears with the 2-lepton $r$-value due to the following scenarios: for small $\Delta m^0$, the cross section is large for light $D_2$ production, but with suppressed coupling between $D_1-D'$ (see Fig.~\ref{fig:Decays13TeV}(c)). Then, for large $\Delta m^0$, this coupling is increased and (see Fig.~\ref{fig:Decays13TeV}(d)), while the heavy $D_2$ leads to suppressed production cross section. 

For the 3-lepton $r$-value, while $D_2$ production becomes suppressed for increasing $\Delta m^0$ as shown in Fig.~\ref{fig:XSecsTeV}, its decays $D_2\to W^+D^-$ and $D_2\to ZD'$ are enhanced, which can easily provide three leptons, including $\tau$-leptons required by the relevant signal regions. In addition, $D'D^\pm$ production cross section is not affected by the $\Delta m^0$ variation.

As $\Delta m^0$ is increased, the Higgs-to-invisible limit  \cite{Sirunyan_2019} of  $\sim 0.15$ BR excludes larger regions in Fig.~\ref{fig:DMD3plotsMFDM}(c)-(f), where $H\to D_1D_1$ becomes the Higgs' dominant decay channel (since this coupling is proportional to $Y_{DM}$), until $D_1$ is too heavy to be produced on-shell.
Other than this, the similarities in $r$-value plots for these three $\Delta m^0$ scenarios means we only need to consider the 2D plane ($\Delta m^+, m_{D1}$).
\begin{figure*}[!ht]
	\subfloat[]{
	\includegraphics[width=0.49\textwidth]{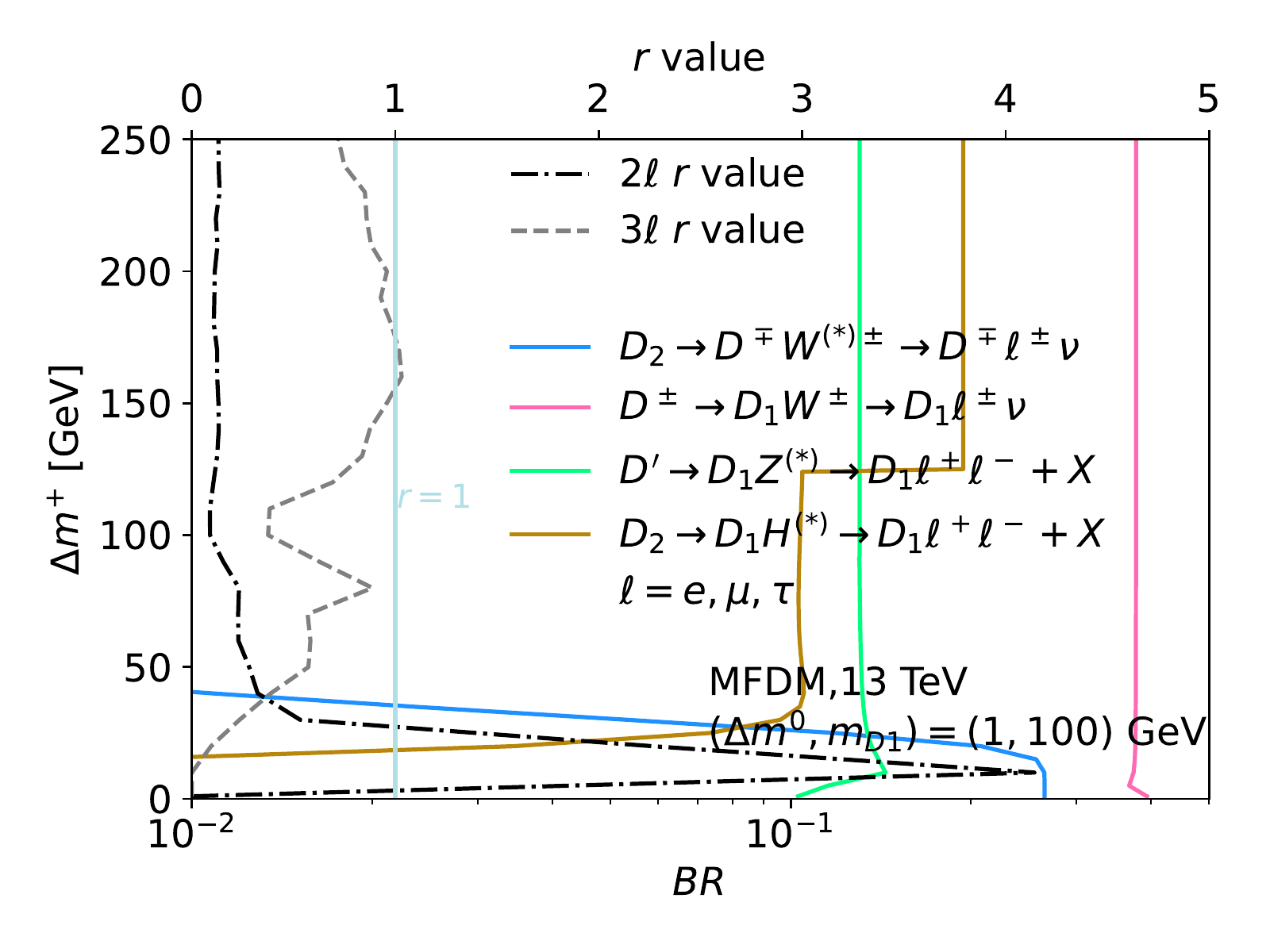}
	}
\subfloat[]{
		\includegraphics[width=0.49\textwidth]{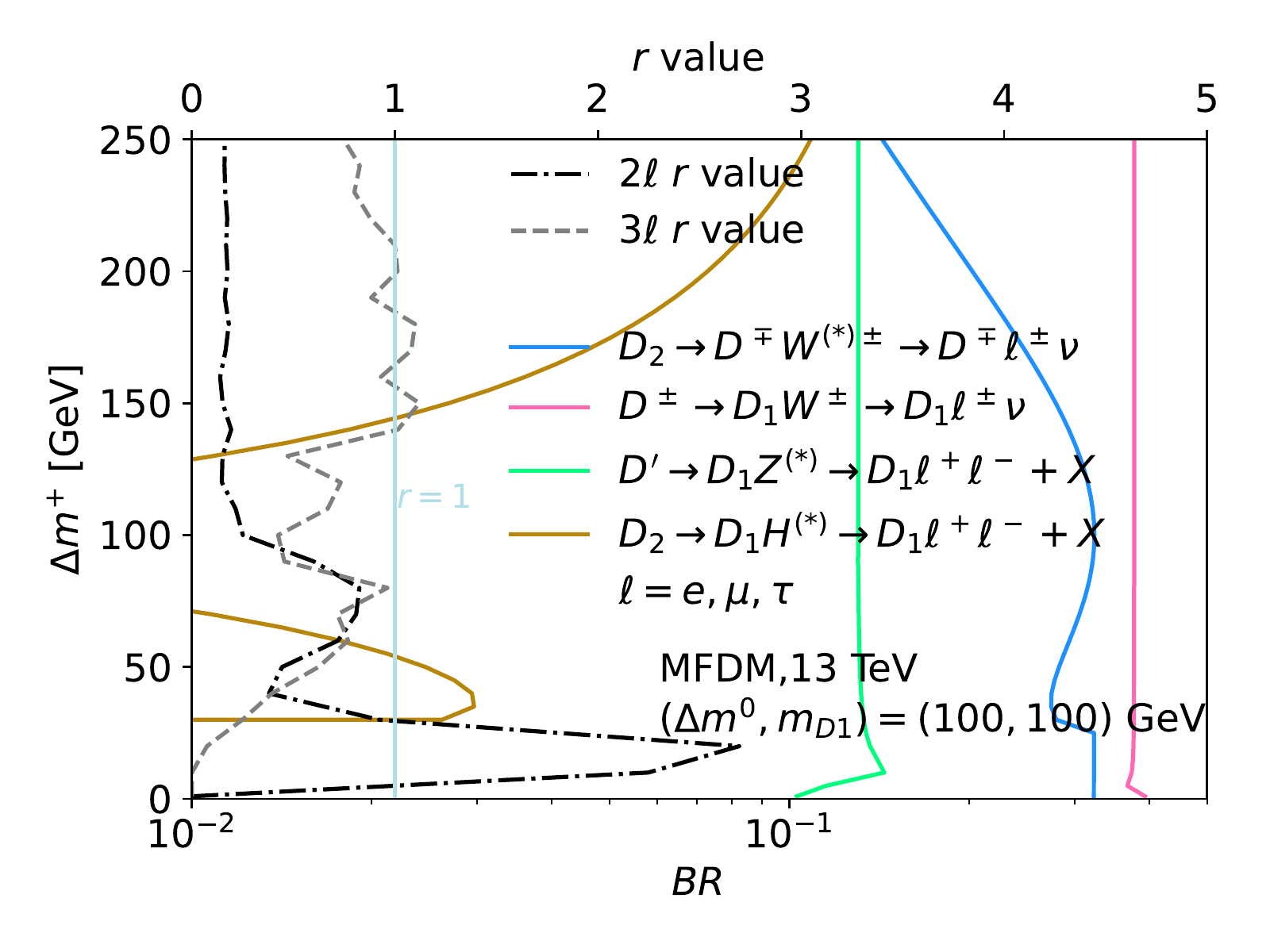}
		}
	\caption{Branching ratios (BR) as a function of the mass split $\Delta m^+$ for various MFDM decays, superimposed with the 2- and 3-lepton $r$-values, for $\Delta m^0=1$~GeV(a) and $\Delta m^0=100$~GeV(b). The DM mass is fixed to $m_{D_1}=100$~GeV, corresponding to a vertical slice of Fig.~\ref{fig:DMD3plotsMFDM}(a) and (e).} 
	\label{fig:D3BRsMFDM_2}
\end{figure*}
Fig.~\ref{fig:D3BRsMFDM_2}  visualises the relation of $r$-values and BR by superimposing  the decay BR from Fig.~\ref{fig:Decays13TeV}, rotated to match the contour plots in  Fig.~\ref{fig:DMD3plotsMFDM},  with the  2-(dot-dashed line) and 3-(dashed line) lepton $r$-values from a vertical slice of Fig.~\ref{fig:DMD3plotsMFDM}   at $m_{D_1}=100$~GeV as a function of $\Delta m^+$.

For $\Delta m^0=1$~GeV (Fig.~\ref{fig:D3BRsMFDM_2}(a)), the dominant contribution to the $r$-value switches from 2-lepton final states to 3-lepton final states around $\Delta m^+>45$~GeV, due to the change in dominant branching of $D_2\rightarrow D^\mp W^{*\pm}(\rightarrow{l}\nu)$ (blue line) to $D_2\rightarrow D_1 H^*(\rightarrow \tau^+ \tau^-)$ (brown line) as $\Delta m^+$ (and $m_{D2}$) increases. 
This corresponds to a phase space excluded specifically by the analysis {\tt cms\_sus\_16\_039}  \cite{cms_sus_16_039}, requiring three leptons with at least one $\tau$-lepton, which can be provided in significant quantities by the $H^*$ decay. Since the dominant productions at $\Delta m^0=1$ include $D_2D^+$ and $D_2D'$ (see Fig.~\ref{fig:XSecsTeV}(c)), the dominating contribution to the $r$-value changes from 
$D_2D^+\rightarrow {l}\nu D^+ {l}\nu D_1 +X$ (total of two charged leptons) to $D_2D^+\rightarrow {l}{l} D_1 {l}\nu D_1 + X$ and $D_2D'\rightarrow \tau^+\tau^- D_1 {l}{l} D_1$ (total of at least three charged leptons, including $\tau$-leptons). 
As the Higgs boson becomes on-shell at $\Delta m^+>130$~GeV, the $H\rightarrow W^+W^-$ channel opens, in addition to the real $H\rightarrow \tau^+\tau^-$, noticeably contributing further to the 3-lepton $r$-value line in Fig.~\ref{fig:D3BRsMFDM_2}(a).

For $\Delta m^0=100$~GeV in Fig.~\ref{fig:D3BRsMFDM_2}(b), while $D_2$ production is suppressed (see Fig.~\ref{fig:XSecsTeV}(d)), 
the co-dominant pair production $D'D^{\pm}$ also provides three leptons via $D'\rightarrow D_1Z$ (green line) and $D^\pm\rightarrow D_1W^\pm$ (pink line).
In addition, although $D_2$ production being suppressed, the $D_2$ decays  $D_2\rightarrow D^\pm W^\pm(\rightarrow{l}\nu)$ and $D_2\rightarrow D' Z(\rightarrow{l}{l})$ become strongly enhanced throughout (since $W$ and $Z$ are on-shell). Combined, they provide enough leptons, including $\tau$-leptons, to compensate for the suppression of $D_2$ production and maintain a significant contribution to the 3-lepton $r$-value  with increasing  $\Delta m^0$.
As $\Delta m^+$ increases, an increase in the $D_2\rightarrow D_1 H$ branching, which also decreases the $D_2\rightarrow D^\pm W^\pm(\rightarrow{l}\nu)$ branching, 
contributes with $\tau$-leptons to the 3-lepton $r$-value around $40<\Delta m^+<60$~GeV. 
As detailed in section~\ref{sec:signal}, the $D_2D_1H$ coupling falls to zero again as $\Delta m^+$ reaches $\Delta m^0=100$~GeV where the mixing angle $\theta$ is such that $\cos(2\theta)$ and therefore the $D_2D_1H$ coupling falls to zero (see Fig.~\ref{fig:couplings}(a) red line). This decrease does not occur for the $\Delta m^0=1$~GeV case (see Fig.~\ref{fig:couplings}(a) blue line), since $\Delta m^+ > \Delta m^0$. However, this reduced contribution in Fig.~\ref{fig:D3BRsMFDM_2}(b) is compensated by the increase in $D_2\to W^\pm D^\mp$ and $D_2\to Z D'$. Finally, there is  a boost in the  3-lepton $r$-value for $\Delta m^+>100$~GeV where the $D_2D_1H$ coupling becomes more enhanced. 

\subsection{Complementarity between the LHC and Non-collider Experiments}

Finally we comment on the cosmological aspects of these models, first presenting relic density constraints on the i2HDM, then  followed by relic density and direct detection constraints on the MFDM model.

The velocity-averaged annihilation cross section is dominated by DM scatterings to SM particles via the Z boson or the Higgs or via the quartic interaction $D_{1}D_{1}VV$. For the i2HDM, the relevant coupling that controls the annihilation cross section here is $\lambda_{345}$, described in section~\ref{sec:models}. For the MFDM, instead the annihilation cross section is controlled  by the Yukawa coupling $Y_{DM}$, Eq.~(\ref{eq:YDM}) and the $D_2-D_1$ mixing angle Eq.~(\ref{eq:tan2theta}).
Ignoring  direct detection constraints, the i2HDM can account for the relic density in three distinct regions of parameter space  \cite{Goudelis:2013uca}. The first is the so-called Higgs funnel region $m_{D_{1}}\simeq \frac{m_{h}}{2}$, the second around $ m_{D_{1}}\simeq 72$~GeV and finally the heavy mass region $m_{D_{1}} \geq 500$~GeV. The Higgs funnel region is essentially a resonance annihilation regulated by the width and the coupling $\lambda_{345}$ and constrains the coupling to $10^{-4}<\lambda_{345}<10^{-2} $ if relic density has to be exactly satisfied.  For the region around $m_{D1}\simeq 72$~GeV, the annihilation proceeds primarily through the $D_{1}D_{1}VV$ quartic interaction, via a pure gauge coupling. As $m_{D_{1}}$ approaches the WW mass threshold, this annihilation is over efficient, and results generically in under abundant dark matter. Finally,  there is a region of parameter space with $m_{D_{1}} > 500$~GeV with destructive interference between the quartic interaction and a t-channel diagram 
involving heavier i2HDM states.  Moving on to direct detection considerations, we realise that since  DM-nucleon scattering cross section scales as $\lambda_{345}^{2}$, almost all of the parameter space in the low mass region is highly constrained by Xenon100  \cite{XENON100:2012itz} {and XENON1T  \cite{Aprile:2017iyp}}. Note that the constraints on the invisible Higgs force the coupling $\lambda_{345} \leq 6\times 10^{-3}$.  Therefore the only viable region of parameter space consistent with relic density and  direct detection constraints remaining corresponds to the Higgs funnel region. 

In view of the above discussions, we set $\lambda_{345}\simeq 0$, such that the only region where relic density is satisfied is the high mass region via pure quartic gauge couplings. In principle the Higgs funnel region exists for non zero $\lambda_{345}$, and this region should be treated with caution. 
Meanwhile, for the MFDM model, DM scattering through the $Z$ boson
is suppressed due to a the $D_1-D'$ mass split. Therefore in MFDM,  direct detection constraints rely instead on DM scattering via Higgs exchange, which is proportional to the Yukawa coupling $Y_{DM}$ (see Eq.~(\ref{eq:YDM})) and $D_2-D_1$ mixing angle (see Eq.~(\ref{eq:tan2theta})).

\begin{figure}[!htb]
	\subfloat[]{
		\includegraphics[width=0.5\textwidth]{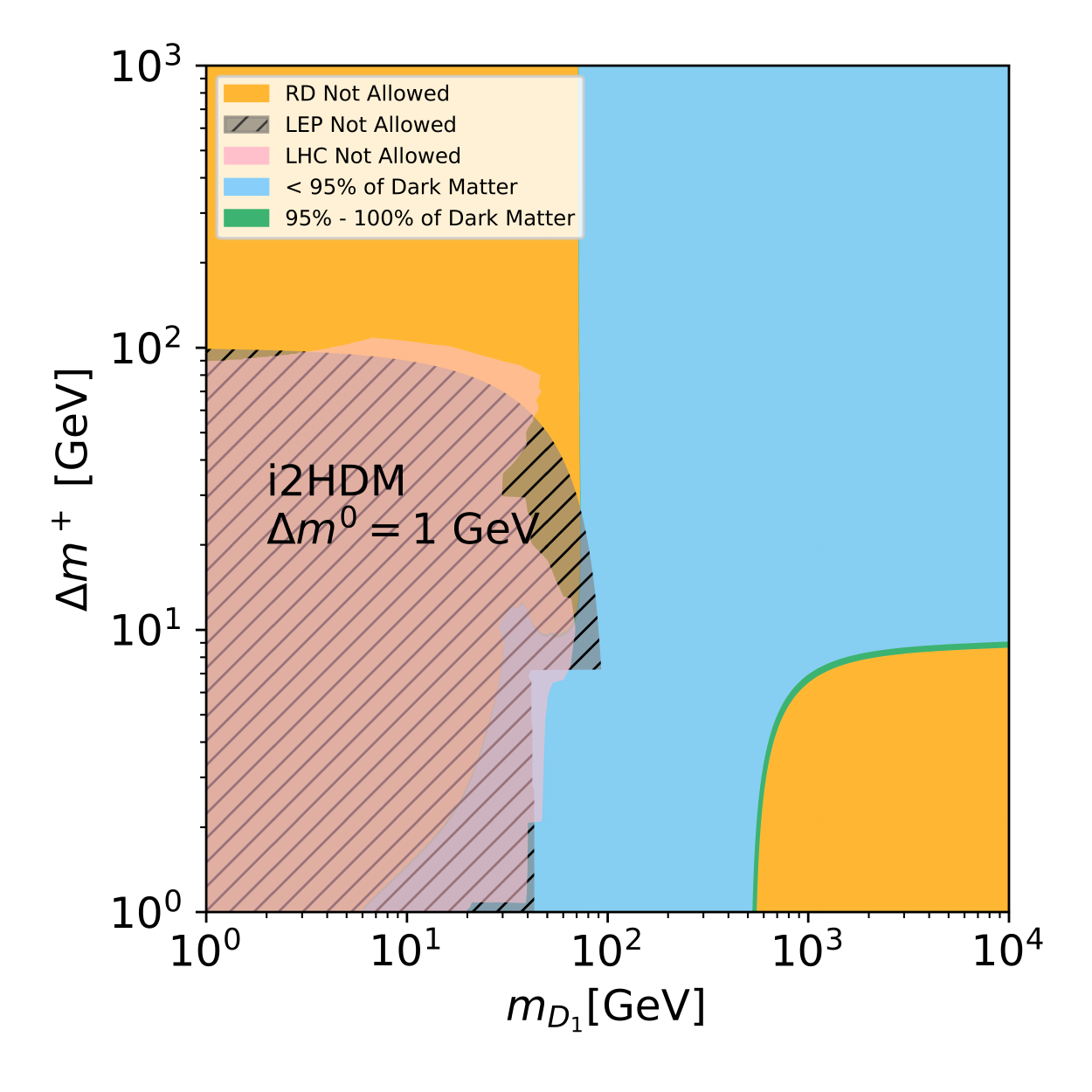}%
	}
	\subfloat[]{
		\includegraphics[width=0.5\textwidth]{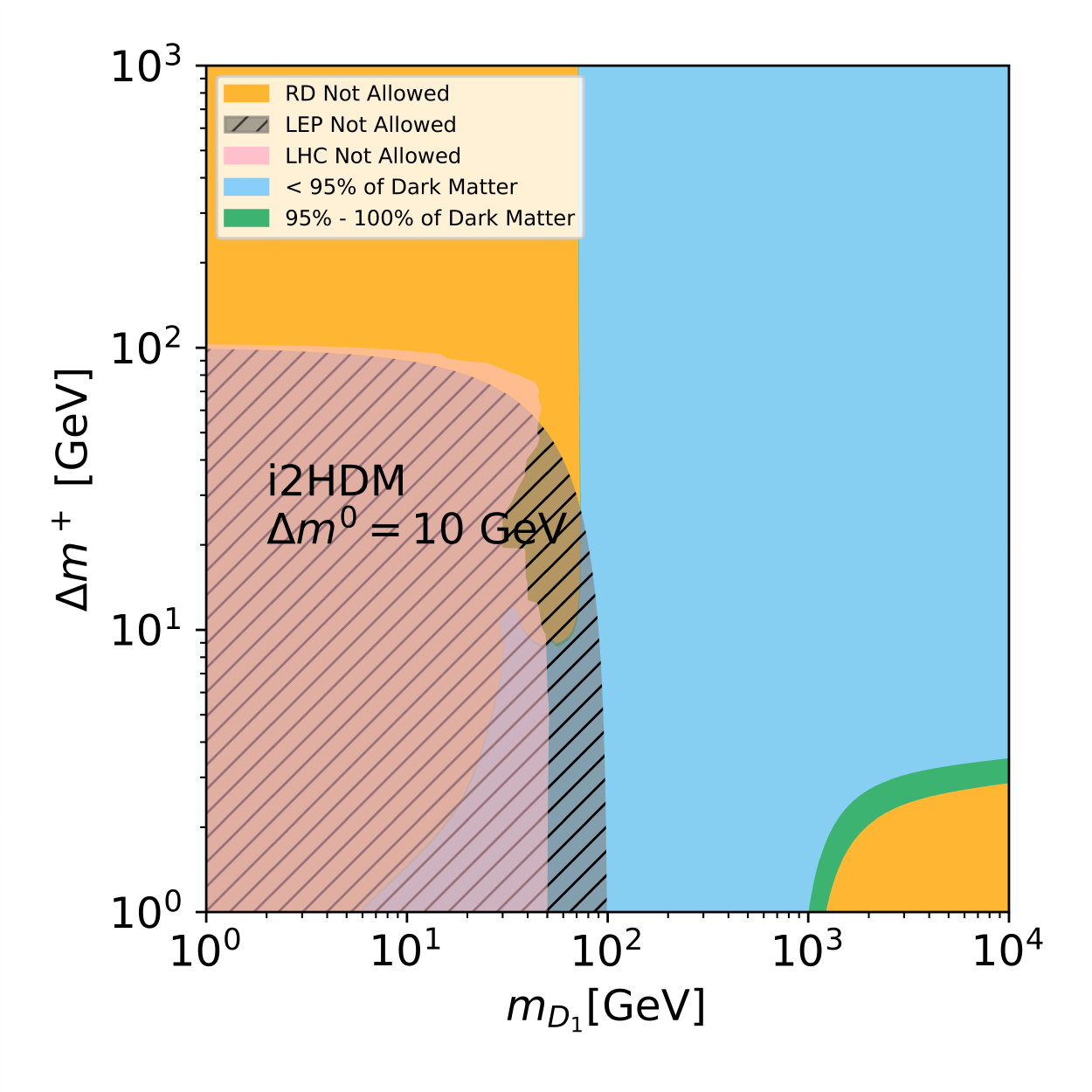}
	}\\
	\subfloat[]{
		\includegraphics[width=0.5\textwidth]{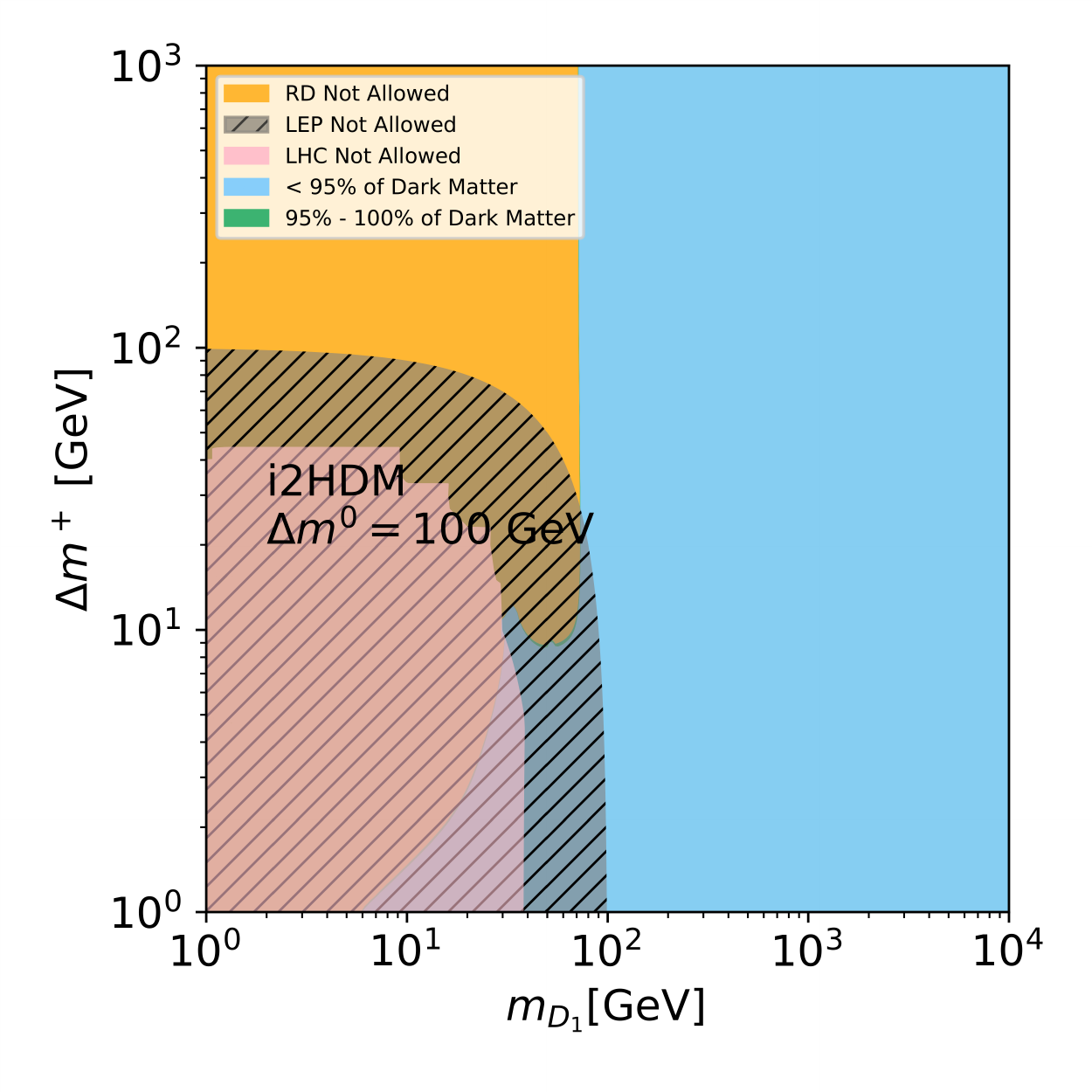}}%
	\caption{LHC potential to exclude the i2HDM parameter space complementing relic density, LEP (hatched region)\cite{Lundstr_m_2009}, \cite{Pierce_2007} and LHC limits. Plots (a),(b) and (c) show regions for $\Delta m^+=1,\ 10,\ 100$~GeV respectively.}
	\label{fig:LHC-complement-i2HDM}
\end{figure}

We now present the results from Fig.~\ref{fig:i2HDM-DeltaM} combined with the non-collider constraints on the i2HDM DM masses and mass splitting. Fig.~\ref{fig:LHC-complement-i2HDM} shows the excluded regions from relic density (RD: orange) LEP (grey) and LHC limits (pink). The allowed regions include predicted DM that explains <95\% of observed DM (blue) and DM that explains 95\%-100\% of observed DM (green). 
\clearpage

Since we assumed that $\lambda_{345}\sim 0$ in section~\ref{sec:models}, the $HD_1D_1$ vertex necessary for DM scattering is not permitted and  direct detection results therefore do not constrain this phase space for our scenario.
As discussed in the context of Fig.~\ref{fig:LHC-complement-i2HDM}, the LEP limits cover the majority of the LHC exclusion region, with the exception  of a small region with $m_{D_1}>40$~GeV, $2<\Delta m^+<8$~GeV for $\Delta m^0=1$ GeV, and a region around $\Delta m^+>60$~GeV, $m_{D_1}<50$~GeV for $\Delta m^0=1,10$~GeV. 
The relic density limit excludes regions around $m_{D_1}<70$~GeV before co-annihilation of $D_2-D_1$ through $Z$ and $D^+D_1$ through $W^+$ channels (proportional to the gauge weak coupling) open up, and $m_{D_1}>600$~GeV, $\Delta m^{+}<9$~GeV for $\Delta m^0=1$~GeV. This second region then reduces to $m_{D_1}>1000$~GeV, $\Delta m^{+}<3$~GeV for $\Delta m^0=10$~GeV and even further for  $\Delta m^0=100$~GeV. For large DM mass but small mass split, the relic density remains too large for relic density constraints. However, for large DM masses and mass split, direct annihilation through $WW$ and $ZZ$ opens, dropping the relic density.

\begin{figure}[!htb]
	        	\subfloat[]{
	\includegraphics[width=0.5\textwidth]{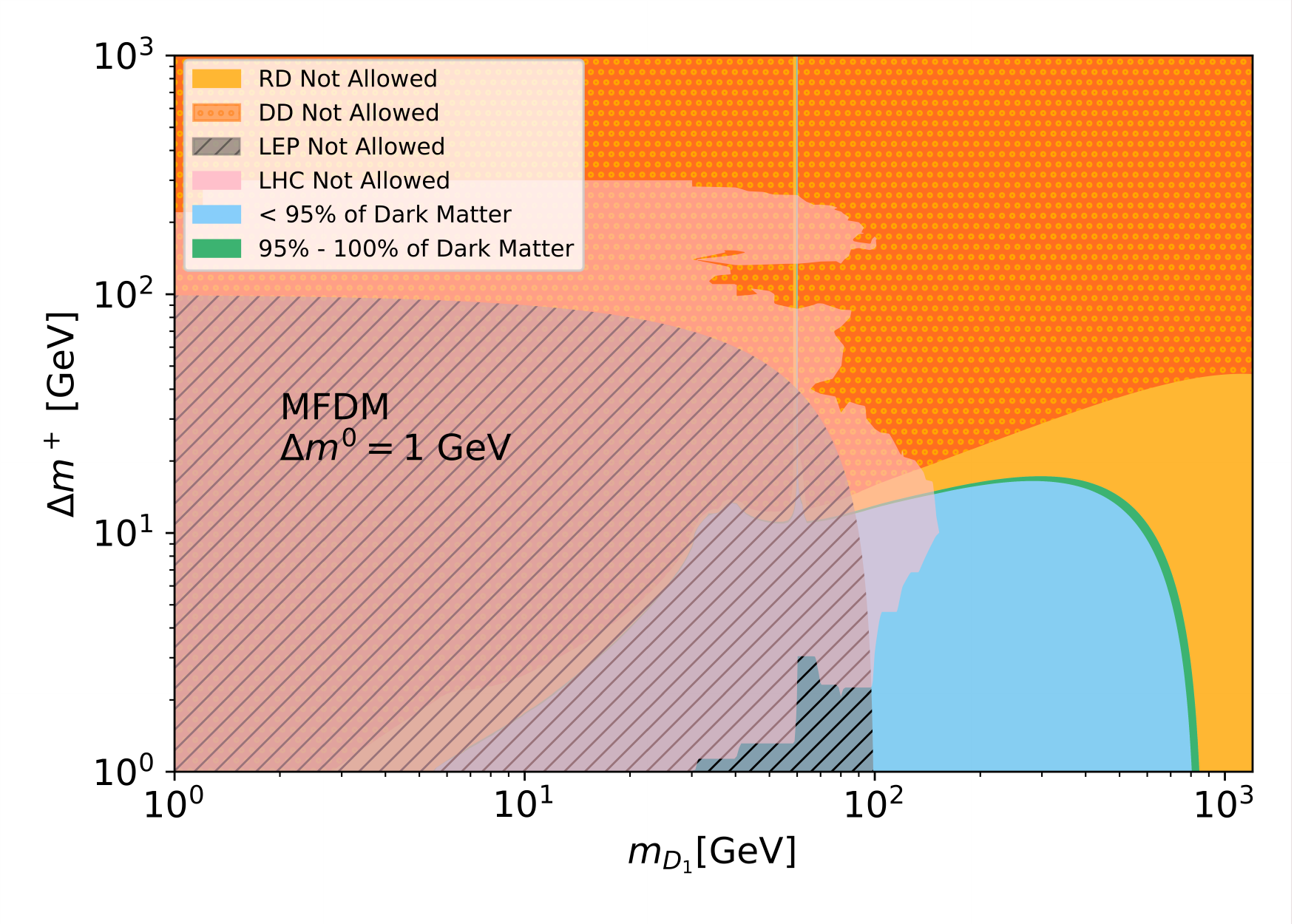}%
}
        	\subfloat[]{
	\includegraphics[width=0.5\textwidth]{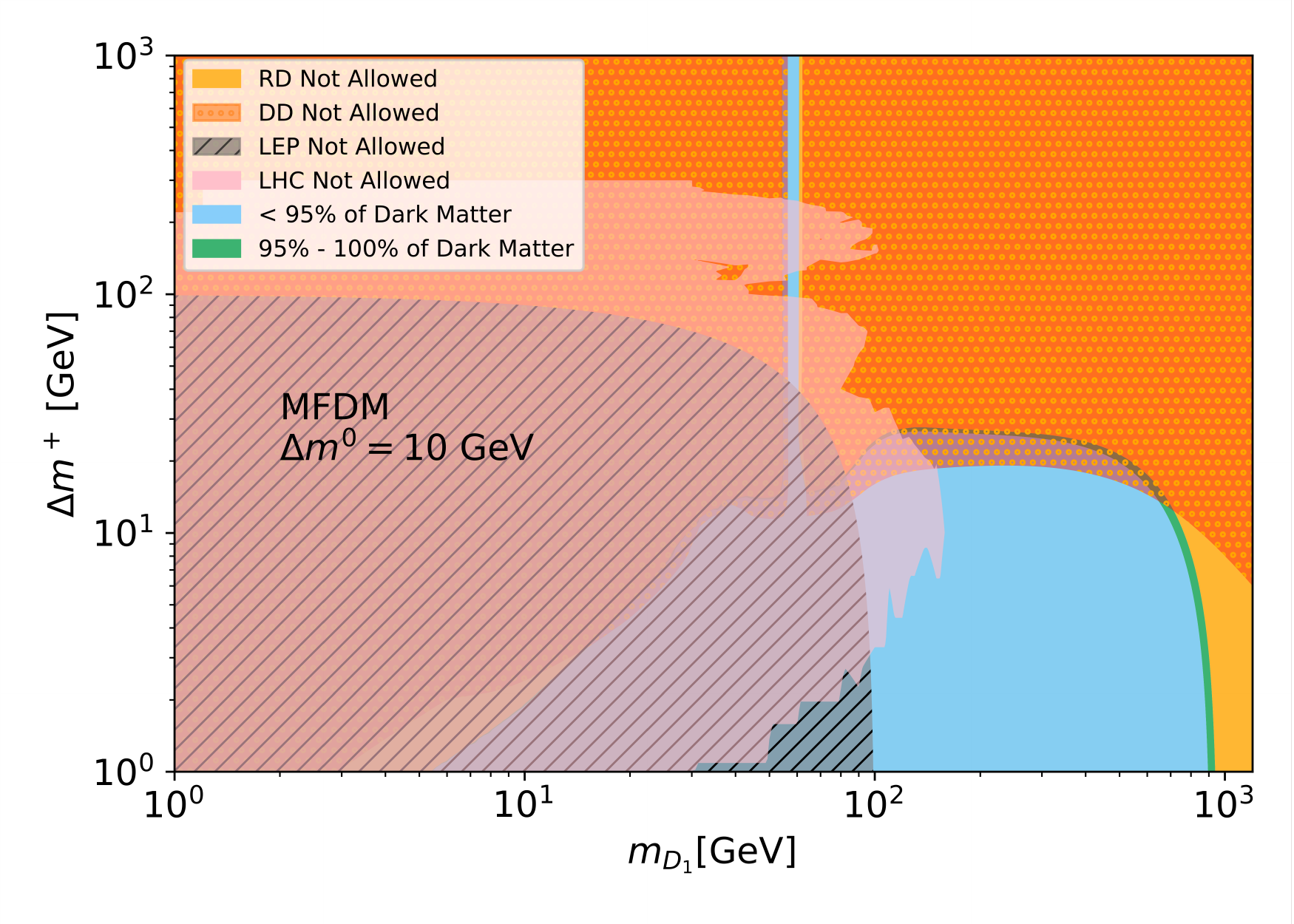}
}\\
        	\subfloat[]{
	\includegraphics[width=0.5\textwidth]{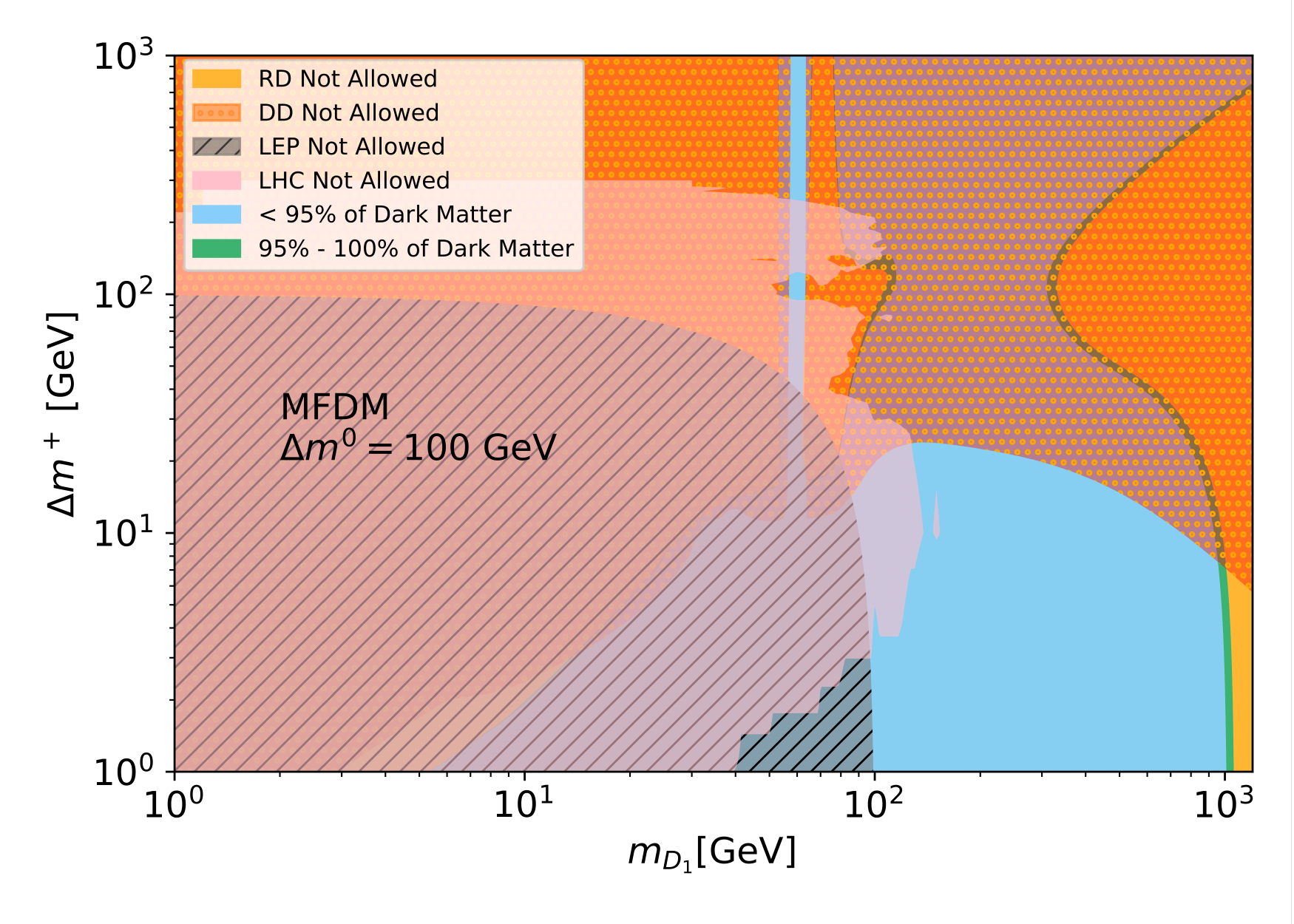}}
	\caption{LHC potential to exclude the MFDM model parameter space complementing non-collider constraints direct detection, relic density, LEP (hatched region)\cite{Pierce_2007} and LHC limits. Plots (a),(b) and (c) show regions for $\Delta m^+=1,\ 10,\ 100$~GeV respectively.}
	\label{fig:LHC-complement}
\end{figure}

In Fig.~\ref{fig:LHC-complement}, we now present the results from Fig.~\ref{fig:DMD3plotsMFDM} combined with the non-collider constraints on the MFDM model DM masses and mass splitting. Again, the allowed regions include predicted DM that explains <95\% of observed DM (blue) and DM that explains 95\%-100\% of observed DM (green). Relic density (orange),  LEP limits (grey) and LHC limits (pink) are also displayed presented in  Fig.~\ref{fig:LHC-complement}. For the MFDM model we now include regions excluded  from direct detection (DD: red) since we allow $D_1D_1H$ interactions for this model. 
As discussed in the context of Fig.~\ref{fig:DMD3plotsMFDM}, the new LHC limits extend the phase-space coverage significantly beyond the LEP limits, including a substantial portion of the Higgs-funnel region around $m_{D_1}=60$~GeV.

The  direct detection exclusion region from XENON1T  \cite{Aprile:2017iyp} 
limits an increasing range of $\Delta m^+$ with increasing $\Delta m^0$ as Yukawa coupling $Y_{DM}$ (see Eq.~(\ref{eq:YDM})) increases, since  direct detection in the MFDM model comes from DM scattering with Higgs. This is approximately within the region $\Delta m^+>10$~GeV and  $m_{D_1}<350$~GeV for $\Delta m^0=1$~GeV in Fig.~\ref{fig:LHC-complement}(a) and $\Delta m^+>6$~GeV, for $\Delta m^0=10$~GeV in Fig.~\ref{fig:LHC-complement}(b).  However, when $\Delta m^0$ is sufficiently large (Fig.~\ref{fig:LHC-complement}(c)) there is little change in this exclusion region, above $\Delta m^+>6$~GeV, as the $Y_{DM}$ coupling relevant for  direct detection converges towards $\sqrt{\Delta m^0}$.

In the region of $m_{D_1}>800$~GeV the relic density excludes a larger range of $m_{D_1}$ with increasing $\Delta m^+$ as the DM co-annihilation of $DD^+$ through $W$ and of $D_1D'$ through $Z$ becomes suppressed, enhancing the relic density above the exclusion limit for smaller values of $m_{D_1}$. 
However, starting with $m_{D_1}=100$~GeV, as the DM mass increases, the region of excluded $\Delta m^+$ decreases due to an enhancement of the $D_1D^-W^+$ and $D_1D'Z$ couplings proportional to $\cos\theta$.
The same couplings are relevant for DM self-annihilation through $D^\pm$ to $W^+W^-$ or through $D'$ to $ZZ$ respectively. 
This trend continues until the required relic density is too high as $m_{D_1}>300$~GeV, for which less DM annihilation suppression is needed.
This happens roughly within the regions $m_{D_1}>600$~GeV, $\Delta m^+<15$~GeV and $m_{D_1}>80$~GeV $\Delta m^+>15$~GeV for $\Delta m^0=1$~GeV in Fig.~\ref{fig:LHC-complement}(a). Then, for $\Delta m^0=10$~GeV, the lower limit for $m_{D_1}<600$~GeV increases to $\Delta m^+>20$~GeV, and the limit for smaller $\Delta m^+<10$~GeV changes to $m_{D_1}>900$~GeV in Fig.~\ref{fig:LHC-complement}(b). This excluded space reduces further for $\Delta m^0=100$~GeV in Fig.~\ref{fig:LHC-complement}(c) to $m_{D_1}>1050$~GeV within $\Delta m^+<30$~GeV.
This effect is due to increasing $\Delta m^0$ and is attributed to an enhancement of the  DM co-annihilation of $DD^+$ through $W$ and of $D_1D'$ through $Z$, with couplings proportional to $\cos\theta$ (see Fig.~\ref{fig:couplings}(b)).
\clearpage

\section{Conclusions}
\label{sec:conclusions}
	In this paper, we have explored 
	the full parameter space relevant to 2- and 3-lepton signatures at the LHC emerging from  two representative minimal consistent scenarios  with scalar and fermion DM: the  i2HDM and MFDM model, respectively.
	In  our analysis, we have suggested a 
	new parametrisation for both frameworks  
	in terms of the DM mass, $m_{D_1}$,  the mass difference between it  and its charged multiplet partner, $\Delta m^+ = m_{D^+}-m_{D_1}$, as well as  the mass difference between $D^+$ and $D_2$ (the next-to-lightest neutral $Z_2$-odd  particle), 
	$\Delta m^0 = m_{D_2}-m_{D^+}$.
	This parametrisation allowed us to  understand  and
	interpret  better the properties of the models under study and visualise a no-lose theorem covering   their full parameter spaces. This approach  
	is generic and quite model-independent since
	the mentioned mass differences are related to the couplings of the DM particle to the SM sector.
	
	Our numerical results have led to the newest and  most up-to-date 
	LHC limits on the i2HDM and MFDM model parameter spaces, coming from the complementary combination of the aforementioned 2- and 3-lepton signatures.
	In particular, in the case of the i2HDM, we have found new regions which the LHC can cover above  the LEP limits, for example: 
	a) for a small 	$\Delta m^0 \lesssim 1$~GeV, the 
	LHC excludes regions 
	with $45 \mbox{ GeV} \lesssim  m_{D1} \lesssim 55 \mbox{ GeV}$ and $\Delta m^+ \lesssim 8 \mbox{ GeV}$ as well as  
	with $10 \mbox{ GeV} \lesssim  m_{D1} \lesssim 50 \mbox{ GeV}$ and $60 \mbox{ GeV} \lesssim\Delta m^+ \lesssim 95 \mbox{ GeV}$;
	b)  for a larger  $\Delta m^0 \simeq 10$~GeV, the LHC excludes regions with $10 \mbox{ GeV} \lesssim  m_{D1} \lesssim 50 \mbox{ GeV}$ and $50 \mbox{ GeV} \lesssim\Delta m^+ \lesssim 85 \mbox{ GeV}$. Furthermore, in the case of the  MFDM model, the 
	LHC probes a notably  larger DM parameter space, in comparison, for example:
	a) for $\Delta m^+ \simeq 10\mbox{ GeV}$, the 
	LHC excludes DM masses up to 150--160 GeV, hence, 
	going well beyond not only the LEP limits but also the LHC ones in mono-jet;
	b) for a much  larger   $\Delta m^+ \simeq 150-200\mbox{ GeV}$, the 	LHC excludes DM masses up to 100 GeV, which is also  well beyond such LEP and LHC constraints. Therefore, the 
 parametrisation that we have suggested thus allows one to clearly establish a no-lose theorem for the MFDM model parameter space: the increase of $\Delta m^0$ leads to  the increase of the $W^\mp D_1 D^\pm$ coupling and respective increase of the $D_1 D^\pm$ production rate while  kinematically suppressing  the $D_2 D^\pm$ production rate. These two trends roughly compensate each other making the exclusion picture 
	in the $(m_{D_1}, \Delta m^+)$ plane quasi-independent of  $\Delta m^0$. 
	
	We have also found another important  complementarity between the 2- and 3-lepton signatures. The first one mainly covers  low values of  $\Delta m^+$  while the second one 
    is enhanced in the region of large $\Delta m^+$ values. To corroborate our quantitative results, 
     we have implemented and validated a 8 TeV ATLAS  multi-lepton analysis into the CheckMATE package and made available this  analysis publicly to the community.
    We have then created a map of the LHC efficiencies and cross-section limits for such 2- and
    3-lepton signatures for the simple parametrisation of the parameter  space 
    that we have suggested, which would then allow for a quick and easy model-independent reinterpretation of the LHC limits for analogous classes of models.
     
   Finally, we have produced combined constraints 
   from the LHC, DM relic density and  DM direct  search experiments 
   delineating the current status of the  i2HDM and MFDM model parameter spaces. These combined limits indicate yet another  important complementarity, the one between non-collider experiments and the LHC, both of which will continue to probe unknown territory of  DM parameter space, hopefully resulting in a DM signal discovery. In such a quest, we deem the multi-lepton signatures studied here to be of the utmost importance at the LHC. 

\section*{Acknowledgements}

AB and SM acknowledge support from the STFC Consolidated Grant ST/L000296/1 and are partially
financed through the NExT Institute. AB also acknowledges  support from a Soton-FAPESP
grant.  All authors
acknowledge the use of the IRIDIS High-Performance Computing Facility and associated
support services at the University of Southampton in completing this work.

\appendix

\newpage
\clearpage

\section{8 TeV Validation: i2HDM\label{app:i2HDM8}}
Appendix \ref{app:i2HDM8} details the validation of our CheckMATE recast for 8 TeV LHC exclusion limits for 2-lepton final states by comparing with the existing MadAnalysis implementation~\cite{recast:ATLAS-SUSY-2013-11}. The CheckMATE analysis code was written based on and validated with the original experimental results, which searched for direct production of charginos, neutralinos and sleptons in final states with two leptons and missing transverse momentum~\cite{Aad_2014} at the LHC. This was implemented using the CheckMATE's AnalysisManager in the current public build, and the SUSY analysis is available at \cite{ArranCheckMATESUSYCode}. The cutflows are given in tables~\ref{tab:ATLASSUSYcuts1} and~\ref{tab:ATLASSUSYcuts2}.

\begin{table}[!ht]
	\centering
	\begin{tabular}{l| c|}
		Global Cut&\\
		\hline
		\hline
		$E_T^{miss}$   &$>0$ GeV\\
		Base leptons&2\\
		$e^+e^-$ trigger&$97\%$\\
		$\mu^+\mu^-$ trigger&$89\%$\\
		$e\mu$ trigger&$75\%$\\
		Signal leptons&2\\
		Leading lepton $p_T$&$>35$ GeV\\
		sub-leading lepton $p_T$&$>20$ GeV\\
		$M_{\ell\ell}$&$>20$ GeV\\
		jets&0\\
		$|M_{\ell\ell}-M_Z|$&$>10$ GeV\\
	\end{tabular}
	\caption{Cutflow for all events in the 8 TeV ATLAS SUSY analysis for 2-lepton$+E_T^{miss}$ finals states, implemented in CheckMATE.}
	\label{tab:ATLASSUSYcuts1}
\end{table}

\begin{table}[!ht]
	\centering
	\begin{tabular}{l| l| l| l|l|l|l|l}
		SR&$m^{90}_{T2}$&$m^{120}_{T2}$&$m^{150}_{T2}$&$WWa$&$WWb$&$WWc$&Zjets\\
		\hline
		\hline
		$M_{\ell\ell}$	&		&	&  &$<120$&$<170$  &&	\\
		$p_T(\ell\ell)$	&		&	&  &$>80$    &&&$>80$	\\
		$E_T^{miss,rel}$	&		&	&  & $>80$ &   &&$>80$	\\
		$m_{T2}$        &$>90$	&$>120$	&$>150$&&$>90$&$>100$&
	\end{tabular}
	\caption{Cutflows used for the specific signal regions in the 8 TeV ATLAS SUSY analysis for 2-lepton$+E_T^{miss}$ finals states, implemented in CheckMATE.}
	\label{tab:ATLASSUSYcuts2}
\end{table}

The 8 TeV ATLAS analysis searching for invisible decays of a Higgs boson produced
in association with $Z$~\cite{Aad_2014}, previously recasted for MadAnalysis~\cite{recast:ATLAS-HIGG-2013-03} was also written for the CheckMATE analysis performed here, as it also looks for final states with two leptons and missing energy. The public code of the Higgs analysis is available at \cite{ArranCheckMATEHIGGSCode}.
The cutflow is given in table~\ref{tab:ATLASHIGGcuts}.
\begin{table}[!ht]
	\centering
	\begin{tabular}{l| c}
		Global Cut&\\
		\hline
		\hline
		Base leptons&2\\
		Lepton $p_T$&$>20$ GeV\\
		Z-window&$76<M_{\ell\ell}<106$ GeV\\
		$E_T^{miss}$&$>90$ GeV\\
		$d\phi(E_T^{miss}, p_T^{miss})$ &$< 0.2$\\
		$\Delta \phi(p_T(\ell\ell), E_T^{miss})$&$>2.6$\\
		$\Delta \phi(\ell, \ell)$&$<1.7$\\
		$|\frac{E_T^{miss} - p_T(\ell\ell)}{ p_T(\ell\ell) }|$&$>0.2$\\
		jets&0\\
	\end{tabular}
	\caption{Cutflow for all events in the 8 TeV ATLAS Higgs analysis for 2-lepton$+E_T^{miss}$ finals states, implemented in CheckMATE.}
	\label{tab:ATLASHIGGcuts}
\end{table}

The events used for the validations were generated with CalcHEP, with 100000 events produced for each 9 benchmark points, using the SLHA files provided from HepData\cite{hepdata} and 1 benchmark point for $HZ\rightarrow$ invisible, with $M_H=125.5$ GeV. Leptonic decays of $Z$ in the $HZ$ production, $\chi^\pm\chi 2$ production and $W$ in the $\chi^+\chi^-$, were also specified in CalcHEP to improve efficiency, which were then showered with CheckMATE's built-in PYTHIA8 module. Detector effects are also applied via CheckMATE with  a DELPHES module. Validation for the SUSY analysis is available at ~\cite{ArranCheckMATESUSY} and Higgs analysis available at ~\cite{ArranCheckMATEHIGGS}.

The motivation behind fixing $m_{D+}$ is because it is mostly only important for $D^+D^-$ production, which give an additional EW coupling factor compared to $D_2D_1$ production, only providing significant contributions to $r$-value at very light $m_{D+}$. The $ZD_1D_1$ production is also less dominant as the $ZZD_1D_1$ coupling is quadratic and therefore weak compared to other couplings. The lowest allowed LEP limit of $m_{D+}=85$ GeV is used and the higher value of $m_{D+}=150$ GeV for comparison.
\begin{figure*}[!ht]
	\centering
	\subfloat[]{
		\includegraphics[width=0.49\textwidth]{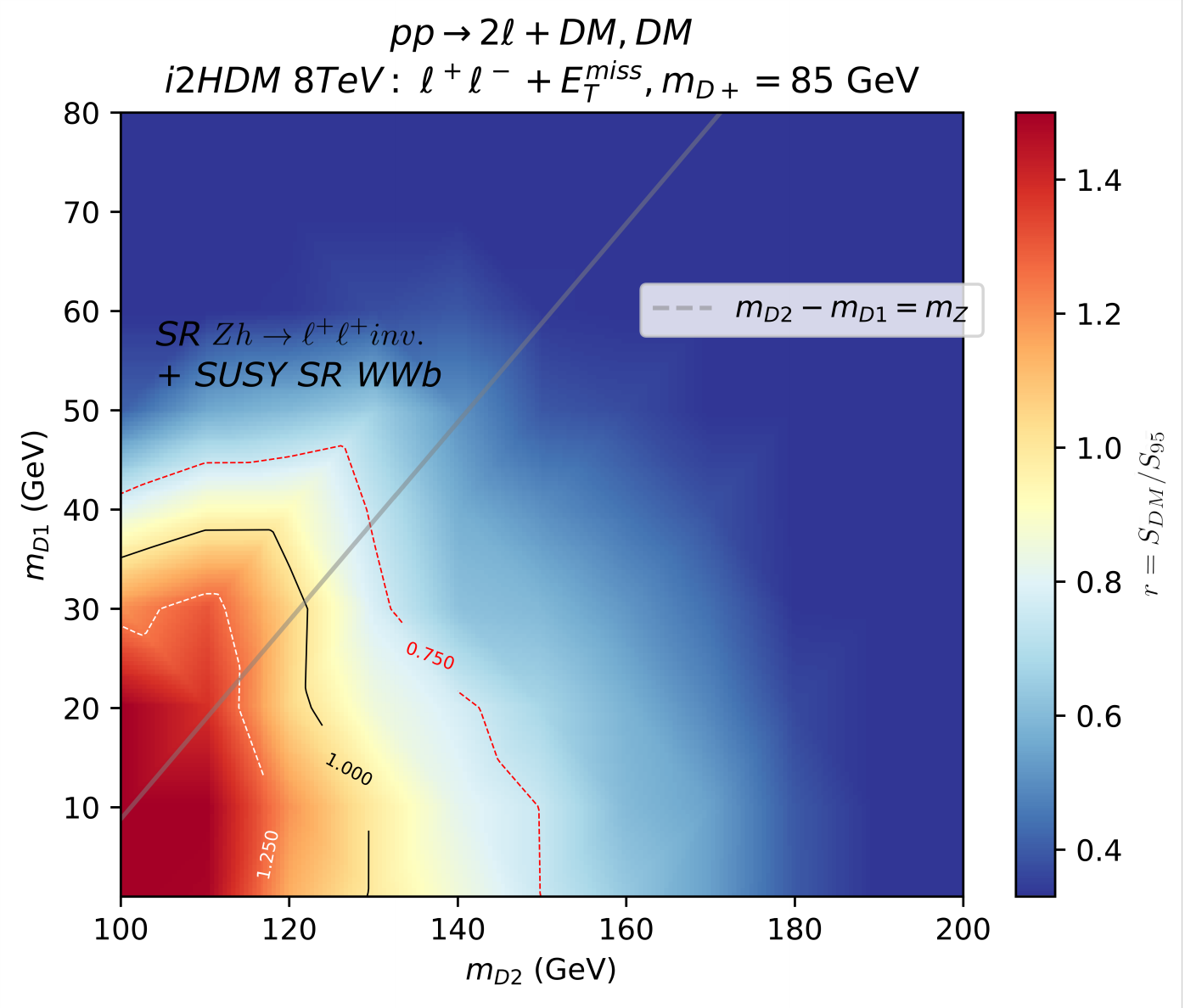}
	}
	\subfloat[]{
		\includegraphics[width=0.49\textwidth]{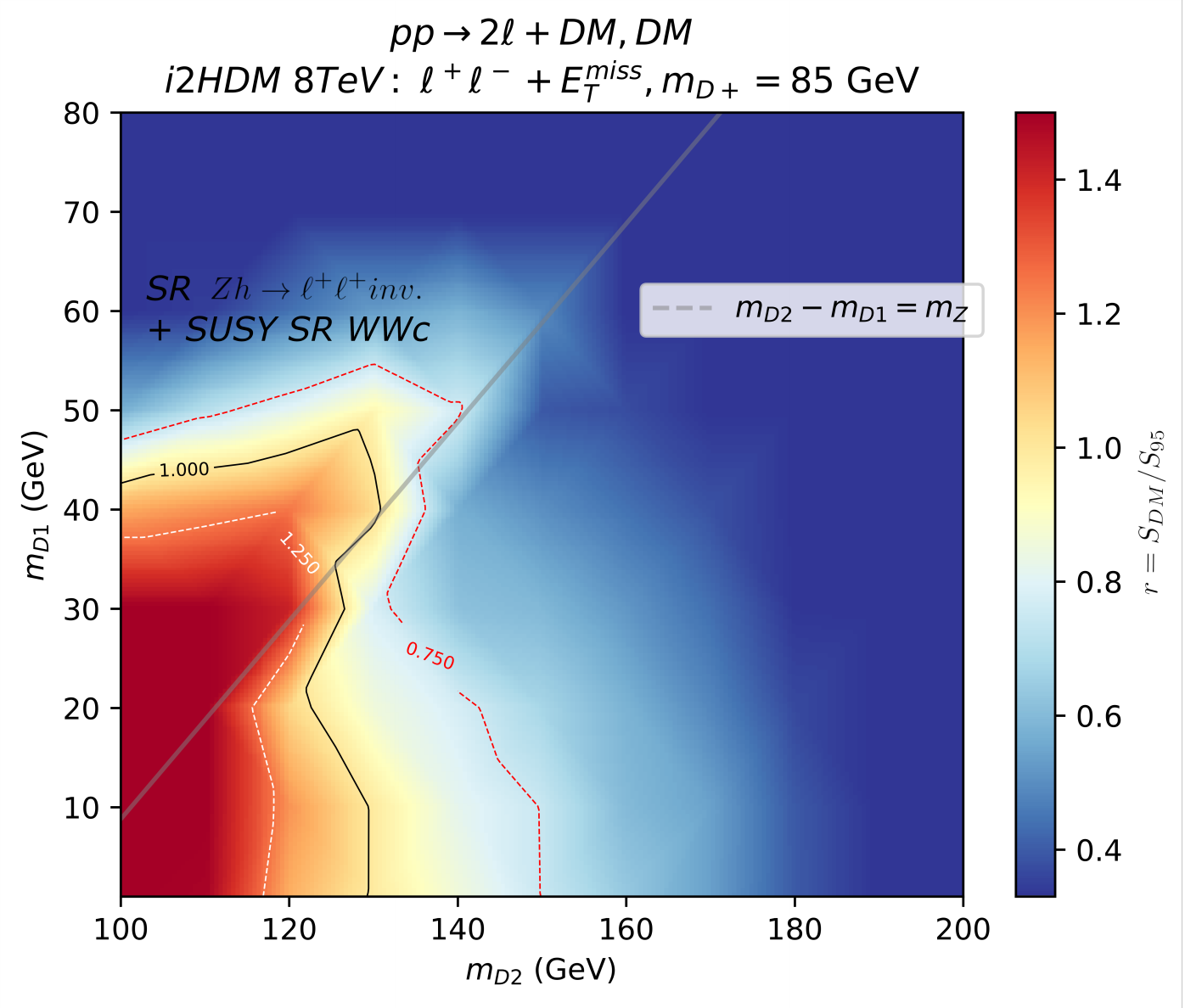}
	}
	\caption{$r$-value Exclusion plots as a function of $m_{D_2}$ in i2HDM at 8 TeV, $m_{D+}=85$ GeV, signal regions $WWb$+Higgs (a) and $WWc$+Higgs (b).}
	\label{fig:i2HDM8TeV85}
\end{figure*}

\begin{figure*}[!ht]
	\centering
	\subfloat[]{
		\includegraphics[width=0.49\textwidth]{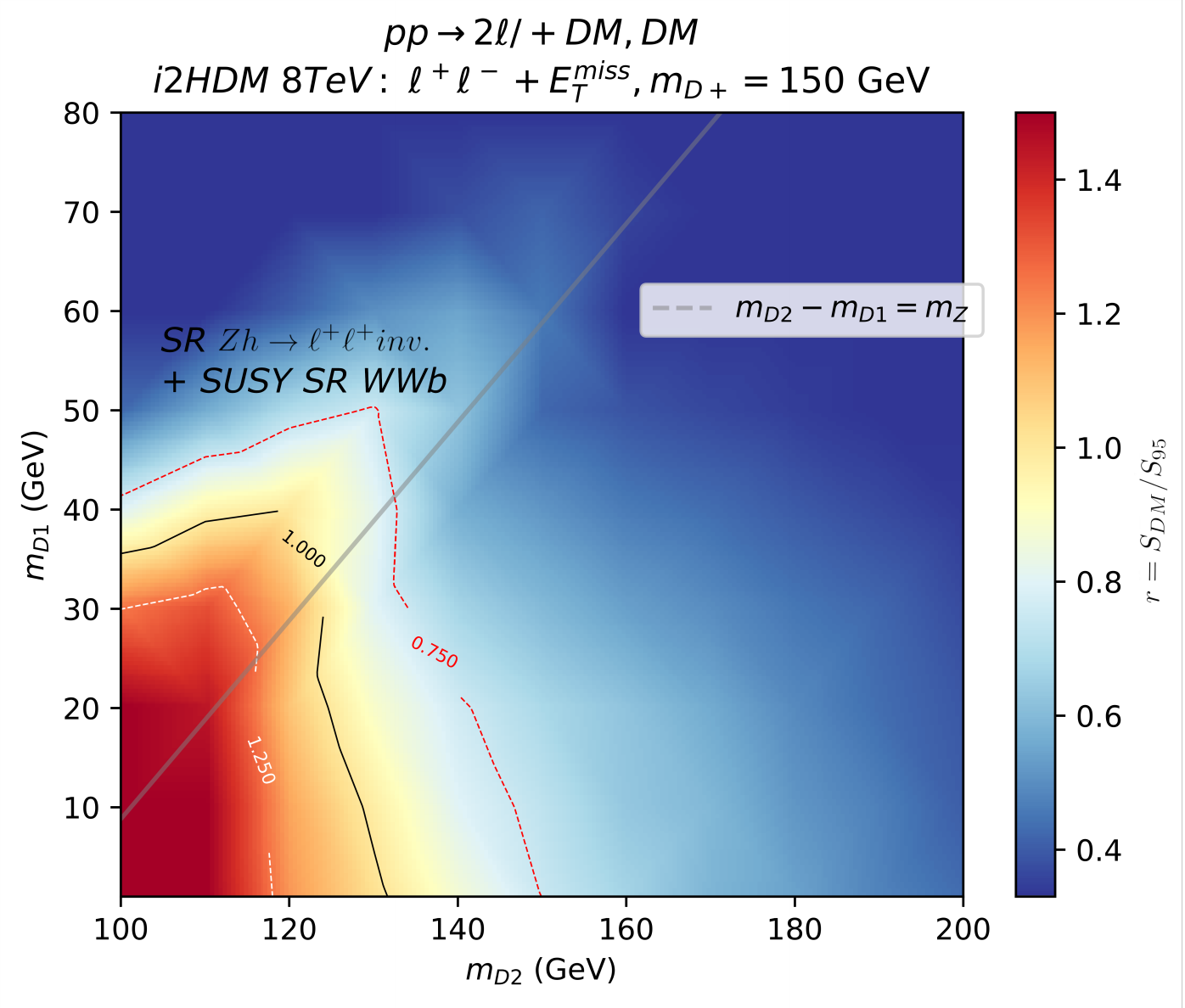}
	}
	\subfloat[]{
		\includegraphics[width=0.49\textwidth]{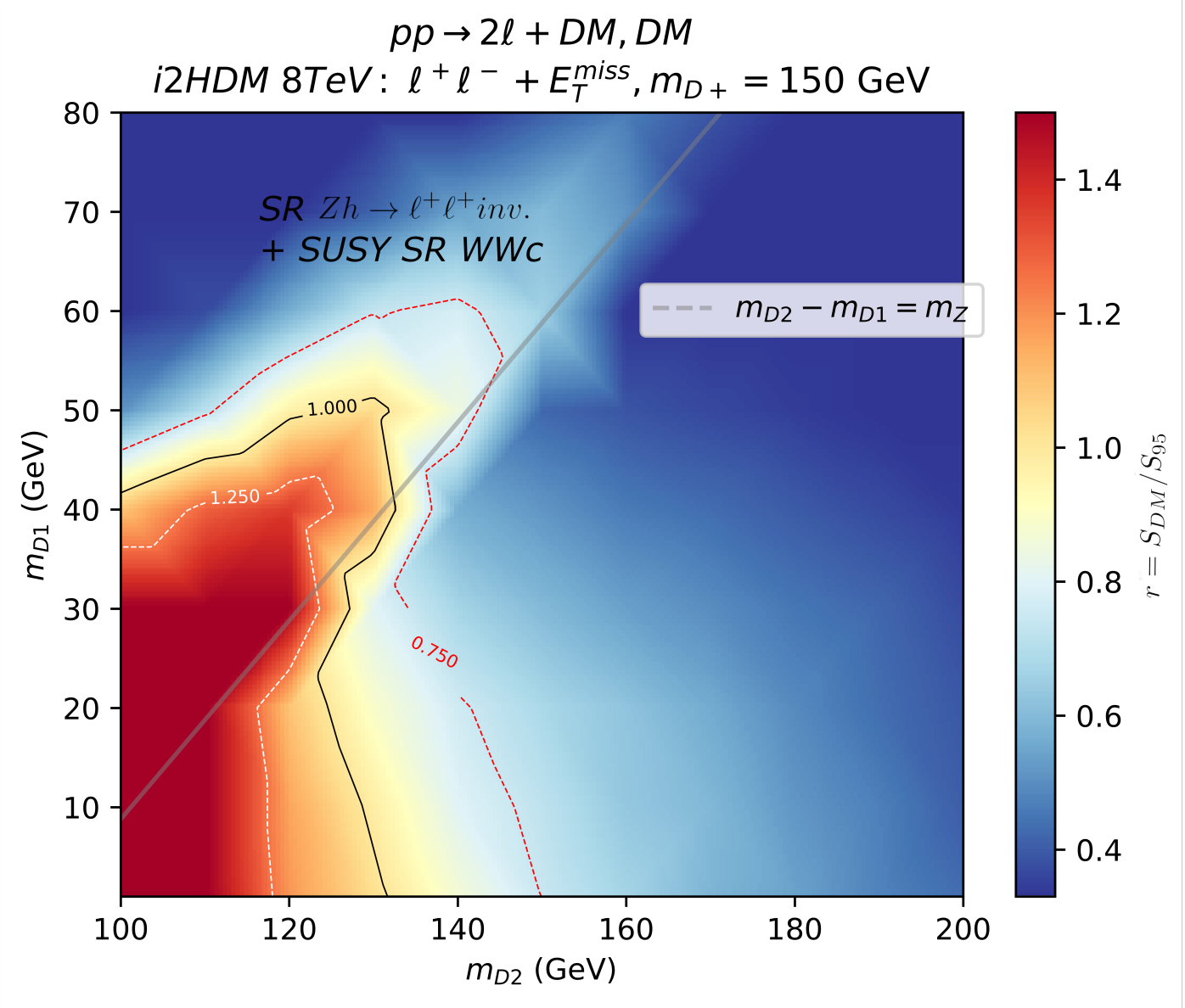}
	}
	\caption{$r$-value exclusion plots as a function of $m_{D_2}$ in i2HDM for 8 TeV, $m_{D+}=150$ GeV, signal regions $WWb$+Higgs (a) and $WWc$+Higgs (b).}
	\label{fig:i2HDM8TeV150}
\end{figure*}
The $r$-value contour plots for $m_{D+}=85$ GeV in Fig.~\ref{fig:i2HDM8TeV85}(a) for the $WWb$ signal region shows that the Run 1 ATLAS 2-lepton analysis excludes for the lightest DM mass of $m_{D1}\leq35$ GeV for $m_{D2}=100$ GeV, reaching a maximum of $\sim40$ GeV as $m_{D2}$ is increased. In the case of Fig.~\ref{fig:i2HDM8TeV85}(b), the $WWc$ signal region reports stronger limits, excluding the lightest DM mass of $m_{D1}\leq40$ GeV for $m_{D2}=100$ GeV, reaching a maximum of $\sim45$ GeV as $m_{D2}$ is increased. For $m_{D1}\rightarrow0$, both cases exclude up to a mass of $m_{D2}=130$ GeV.

By increasing $m_{D^+}$ to $150$ GeV in Fig.~\ref{fig:i2HDM8TeV150}(a), the $WWb$ signal region excludes up to $m_{D1}=35$ GeV at $m_{D2}=100$ GeV, to $m_{D1}=40$ GeV with increasing $m_{D2}$. In comparison, the $WWc$ signal region excludes masses from $m_{D1}=40$ GeV at $m_{D2}=100$ GeV, to $m_{D1}=50$ GeV as $m_{D2}$ is increased.

Constraints increase with larger $m_{D+}$ in both $WWb$ and $WWc$ signal regions, due to increased contributions from $D^+D^-$ production and from $D_2D^+$ production.

For larger $m_{D2}$, the signal events coming from $D_2D_1$ production is smaller
as the $Z$ decay from $D_2$ is replaced in favour of $W$ decay to $D^+$ and its decay to an additional $W^+$ and $D_1$, which produces much softer leptons than required by the signal region cuts.

As $m_{D2}$ is increased, $m_{D1}$ is further constrained when considering the phase space above $m_{D2}-m_{D1}=m_Z$, due to harder lepton production $Z$ decaying from $D_2$. Beyond this line, for a large enough $m_{D2}-m_{D1}$ mass splitting, the Z-veto required by the SUSY analysis is no longer fulfilled by the signal, as real Z decay emerges in the production. Instead, the Z-window required by the Higgs analysis accepts these events, and becomes the dominant signal region as the mass splitting is further increased, independent of $m_{D+}$.

From Fig.~\ref{fig:i2HDM8TeV85}(a) and ~\ref{fig:i2HDM8TeV150}(a) the $WWb$+Higgs analyses agree with the general shape in \cite{B_langer_2015} Fig. 1, for both $m_D+=85$ GeV and $150$ GeV, but with lower $r$-value overall. However, results with larger $r$-value are obtained when considering $WWc$+Higgs signal regions in Fig.~\ref{fig:i2HDM8TeV85}(b) and ~\ref{fig:i2HDM8TeV150}(b).
While \cite{B_langer_2015} considered both $WWb$ and the $HZ\rightarrow$invisible, they did not consider the $WWc$ signal region.  This is because, although the $WWc$ signal region gives a larger observed $r$ value than $WWb$, the expected $r$ value is lower than $WWb$ (deeming it the better channel by analysis tools). It is worth noting that the MadAnalysis validation for this signal region is overestimated compared to the experimental paper by a small amount, while the CheckMATE analysis implemented is closer to the experimental findings for survived number of MC events. contour exclusion limits for $WWc$, where a higher $m_T2$ cut on the leptons is implemented, shows higher observed $r$-value than the original paper's $WWb$ contour exclusion limits.

\clearpage
\section{i2HDM 13 TeV, 2- or 3-lepton Final states\label{app:i2HDM13}}
In appendix\ref{app:i2HDM13} the analysis in \cite{B_langer_2015} is then extended to higher centre of mass energies, using all available ATLAS and CMS 13 TeV analyses in CheckMATE. There is no equivalent to the Higgs $ZH\rightarrow \ell^+\ell^-$+ invisible recasted code as of writing, which is why the $r$-value beyond the $m_{D2}-m_{D1}=m_Z$ line is negligible.

\begin{figure*}[ht!]
	\centering
	\subfloat[]{
		\includegraphics[width=0.49\textwidth]{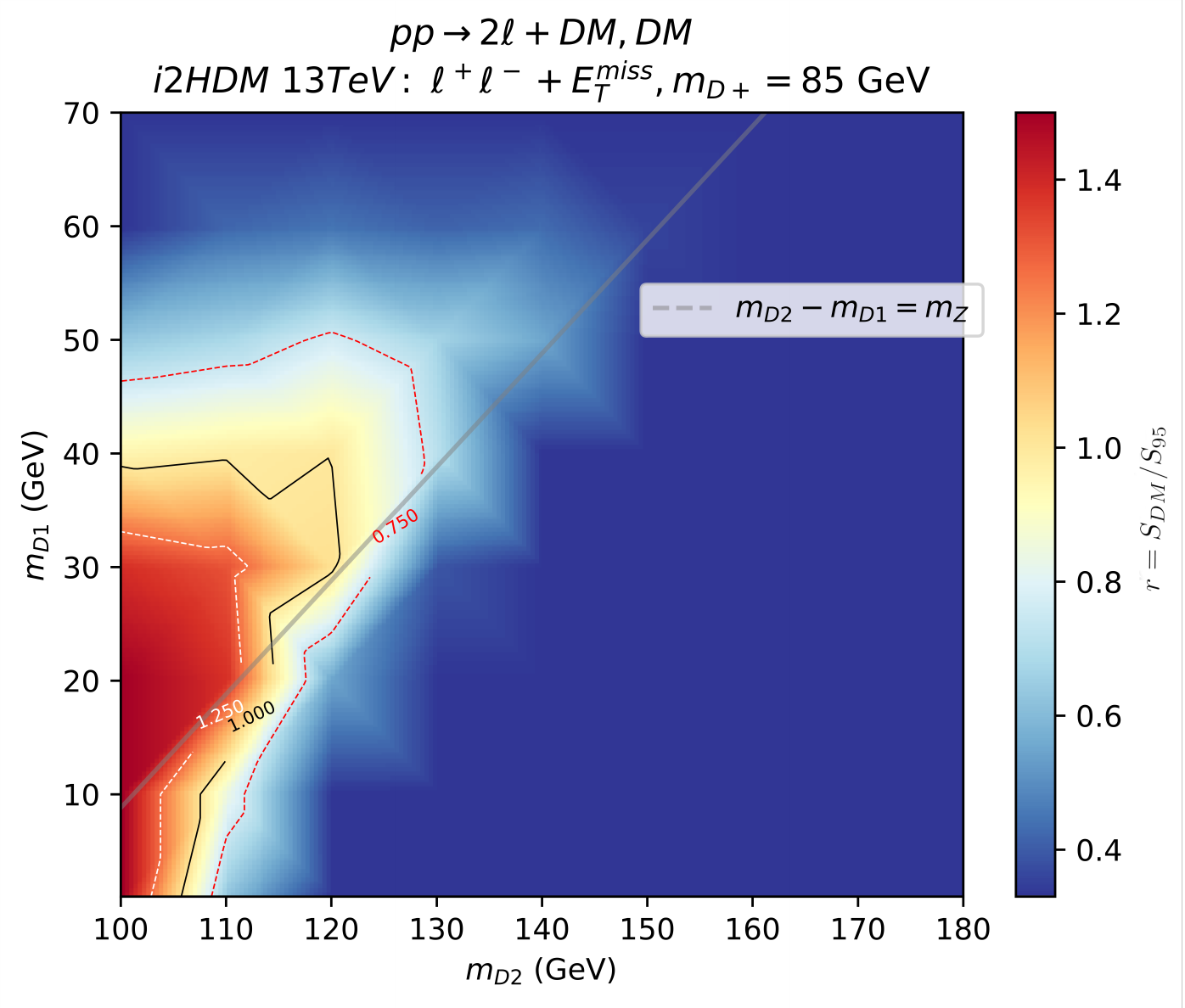}
	}
	\subfloat[]{
		\includegraphics[width=0.49\textwidth]{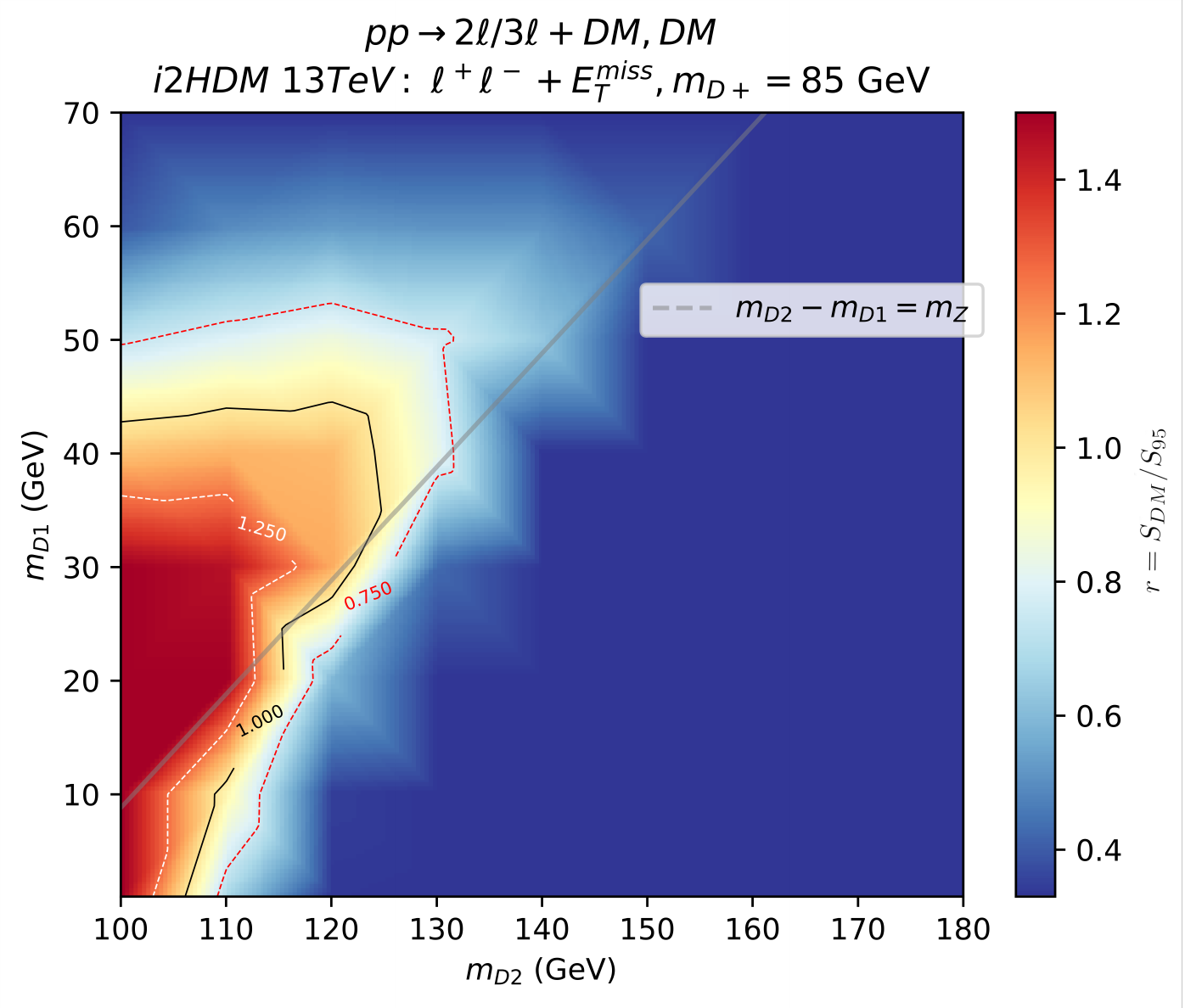}
	}
	\caption{$r$-value exclusion plots in i2HDM for 13 TeV, $m_{D+}=85$ GeV, using signal regions with the highest $r$-values for two leptons (a) and 2- or 3-lepton (b) final states.}
	\label{fig:i2HDM13TeV85}
\end{figure*}

Extended to 13 TeV, Fig.~\ref{fig:i2HDM13TeV85} shows that this does not necessarily improve results, due to scaling of vector boson backgrounds in this phase space. Also shown between Fig.~\ref{fig:i2HDM13TeV85}(a) and (b) is the improvement to $r$-value due to the inclusion of 3-lepton final states. This introduces 6 additional contributing diagrams where $D_2$,$D^+$ production (a) plays an important role. $D_2$ provides two leptons via $D_2\rightarrow Z(\ell^+\ell^-),D_1$ and $D^+$ gives one extra lepton via $D^+\rightarrow W^+(\ell^+\nu),D_1$ which contributes in heavier $D_1$ (and thus softer leptons) phase spaces than 2-lepton exclusive searches.

\begin{figure*}[ht!]
	\centering
	\subfloat[]{
		\includegraphics[width=0.49\textwidth]{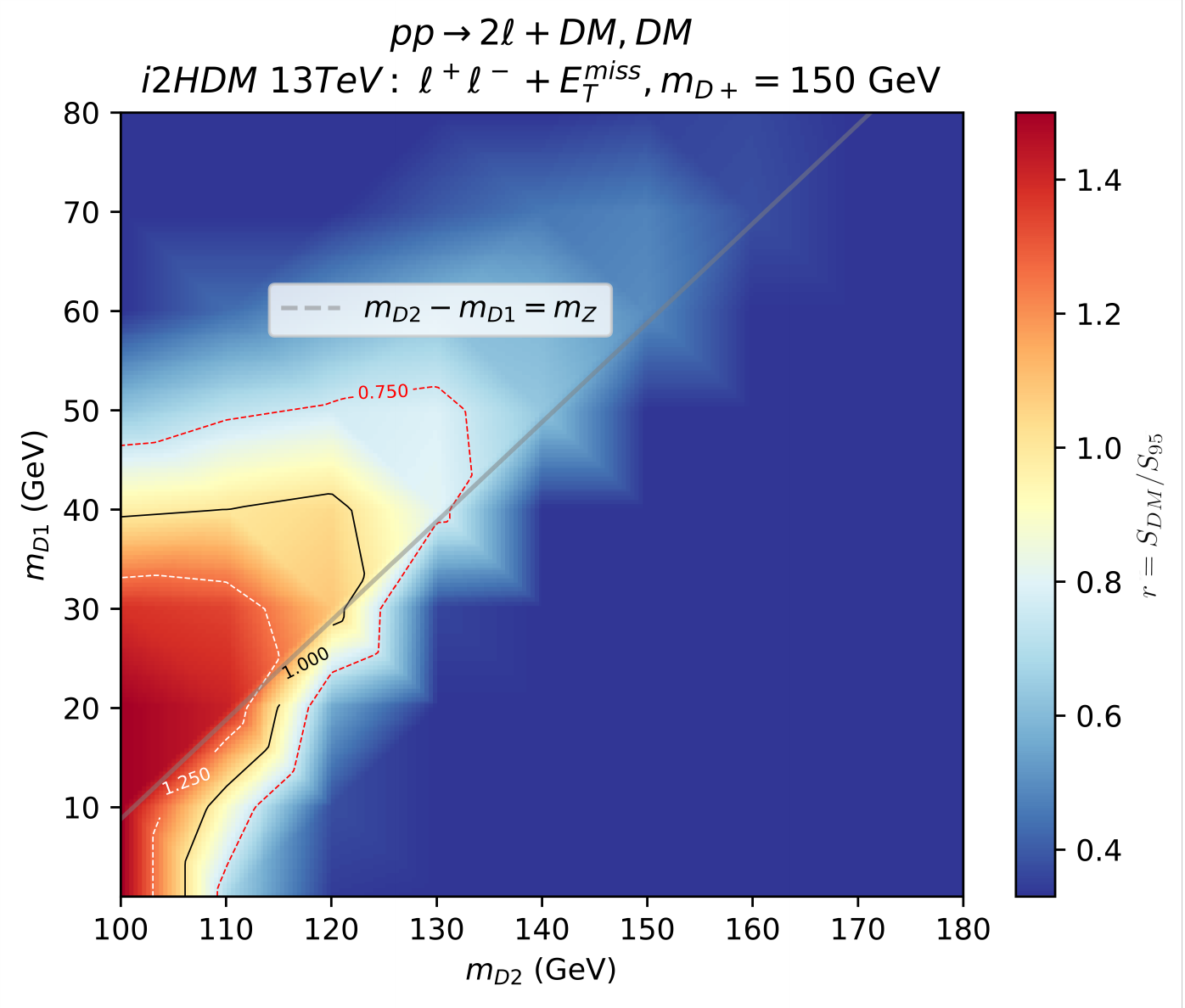}
	}
	\subfloat[]{
		\includegraphics[width=0.49\textwidth]{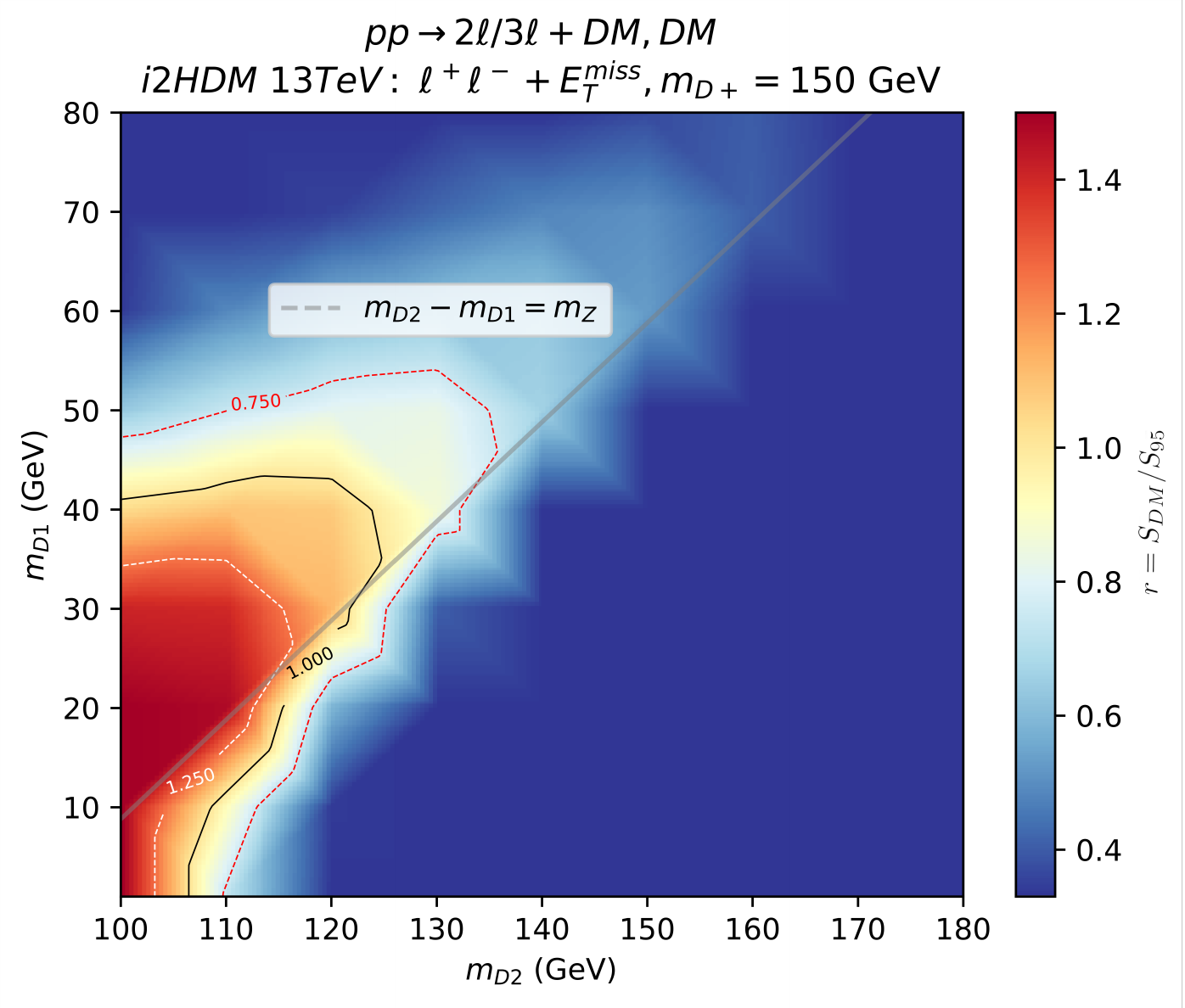}
	}
	\caption{$r$-value exclusion plots in i2HDM for 13 TeV, $m_{D+}=150$ GeV, using signal regions with the highest $r$-values for two leptons (a) and 2- or 3-lepton (b) final states.}
	\label{fig:i2HDM13TeV150}
\end{figure*}

If $D^+$ is heavy enough, it can decay via $D_2$, but this only occurs in Fig~\ref{fig:i2HDM13TeV150}.
The $m_{D1}$ extends from 40 GeV to 45 GeV limit, while $m_{D2}$ can extend from 120 GeV to 125 GeV.
\newpage
Shown in Fig.~\ref{fig:i2HDM13TeV150} is the improvement from increasing lepton multiplicity for the larger $m_{D+}$ value. Although less apparent than in Fig.~\ref{fig:i2HDM13TeV85}, there are extensions to exclusion limits from under $m_{D1}=40$ GeV to above the 40 GeV line and similarly small extensions to the $m_{D2}$ limit.

The diagrams with three leptons in the final state can now contribute to the relevant phase space, but as this chain of decays contains more steps than other contributions, the soft leptons it produces mostly do not pass the cuts here. Contrary to the 8 TeV case, there is not much increase in limits when moving to heavier $D_2$. In fact, for 2- and 3-lepton searches, Fig.~\ref{fig:i2HDM13TeV85}(b) has limits of $m_{D1}=44-45$ GeV for $m_{D2}=100$ GeV, while for $m_{D+}=150$ GeV figure~\ref{fig:i2HDM13TeV150}(b) only has limits reaching $m_{D1}=40-41$ GeV due to more processes that do not necessarily fulfil the signal criteria.

In principle one would combine 13 TeV result with the 8 TeV exclusion limits for a comprehensive picture of LHC limits. Although 8 TeV limits are stronger than 13 TeV limits for the phase space in \cite{B_langer_2015} and in this appendix, 8 TeV results do not improve within the phase pace of our 13 TeV results in new parametrisation, shown in section~\ref{sec:results}.
\clearpage
\section{MFDM 8 TeV\label{app:MFDM8}}

In appendix~\ref{app:MFDM8} 8 TeV $r$-value contour plots for MFDM are next discussed, starting with the same 8 TeV analyses that were used in the i2HDM case. 

\begin{figure*}[!ht]
	\centering
	\subfloat[]{
		\includegraphics[width=0.49\textwidth]{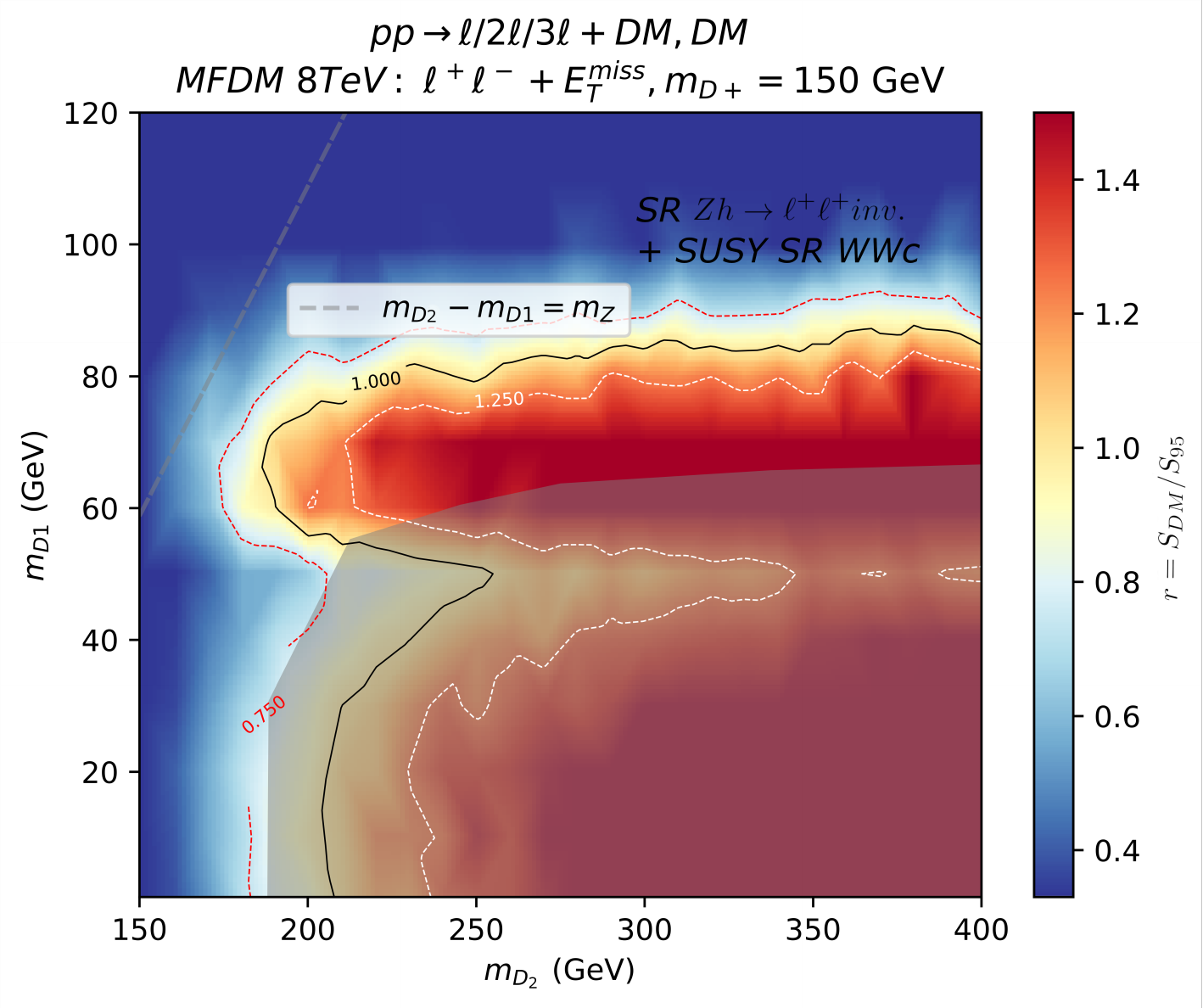}
	}
	\subfloat[]{
		\includegraphics[width=0.49\textwidth]{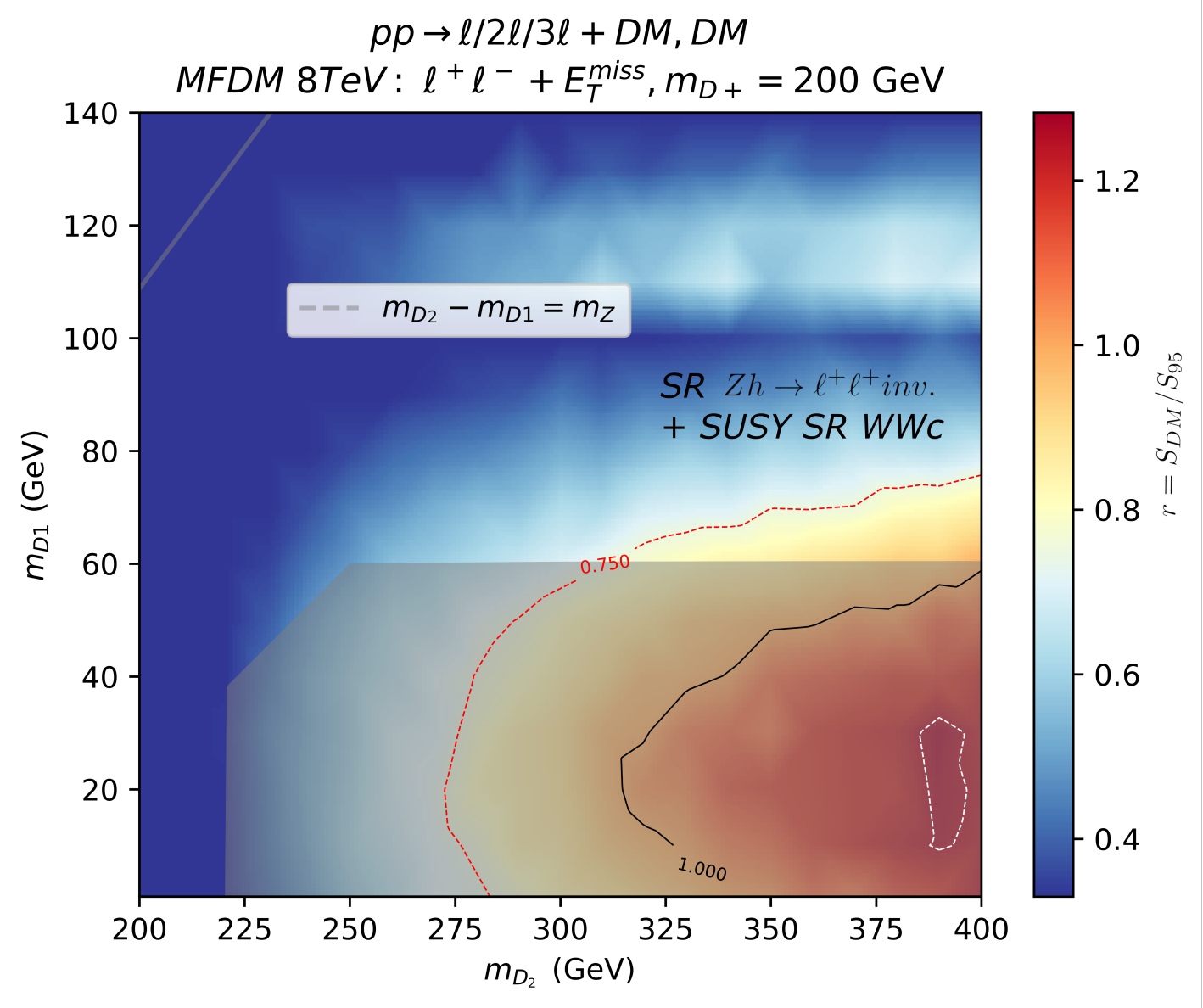}
	}
	\caption{The $r$-value exclusion plots in MFDM for 8 TeV, $m_{D+}=150$ GeV (a) and $m_{D+}=200$ GeV (b),  from the signal regions $WWc$+Higgs.}
	\label{fig:MFDM8TeVFixDP}
\end{figure*}
Plots in Fig.~\ref{fig:MFDM8TeVFixDP} show fixed $m_{D+}$ results, with scans in the $m_{D_1}$-$m_{D_2}$ plane. The shaded region shows the Higgs to invisible excluded region, covering a large phase space of the CheckMATE excluded region. In figure~\ref{fig:MFDM8TeVFixDP}(a), where $m_{D^+}=150$ GeV, the region above $m_{D_1}=60$ GeV has limits reaching to $m_{D_1}<85$ GeV excluded while $m_{D_2}>180$ excluded. As $m_{D_2}$ increases, this increases the split between $D_2$ and $D^+/D'$, increasing the Yukawa coupling in Eq.~(\ref{eq:YDM}). This facilitates more decays that would lead to leptonic final states, thus $r$-value is increased in the positive x-axis direction. In the y-axis, as DM mass is increased, the exclusion changes from {\tt atlas\_higgs\_2013\_03} analysis at low DM mass, to the {\tt atlas\_1403\_5294} analysis at higher DM mass. Looking at Tables~\ref{tab:ATLASSUSYcuts1},~\ref{tab:ATLASHIGGcuts}, this is due to the Z-window changing to a Z-veto with softer leptons being produced in association with harder DM for the same input energy.

Fig.~\ref{fig:MFDM8TeVFixDP}(b) shows $m_{D^+}=200$ GeV, where no exclusion outside of the Higgs to invisible limit of $m_{D_1}<60$ GeV excluded for $m_{D_2}>220$ GeV. This is because $m_{D^+}$ and $m_{D'}$ become too heavy to produce at these masses, so the sources of leptonic final states are suppressed.

\newpage
\begin{figure*}
	\centering
	\subfloat[]{
		\includegraphics[width=0.49\textwidth]{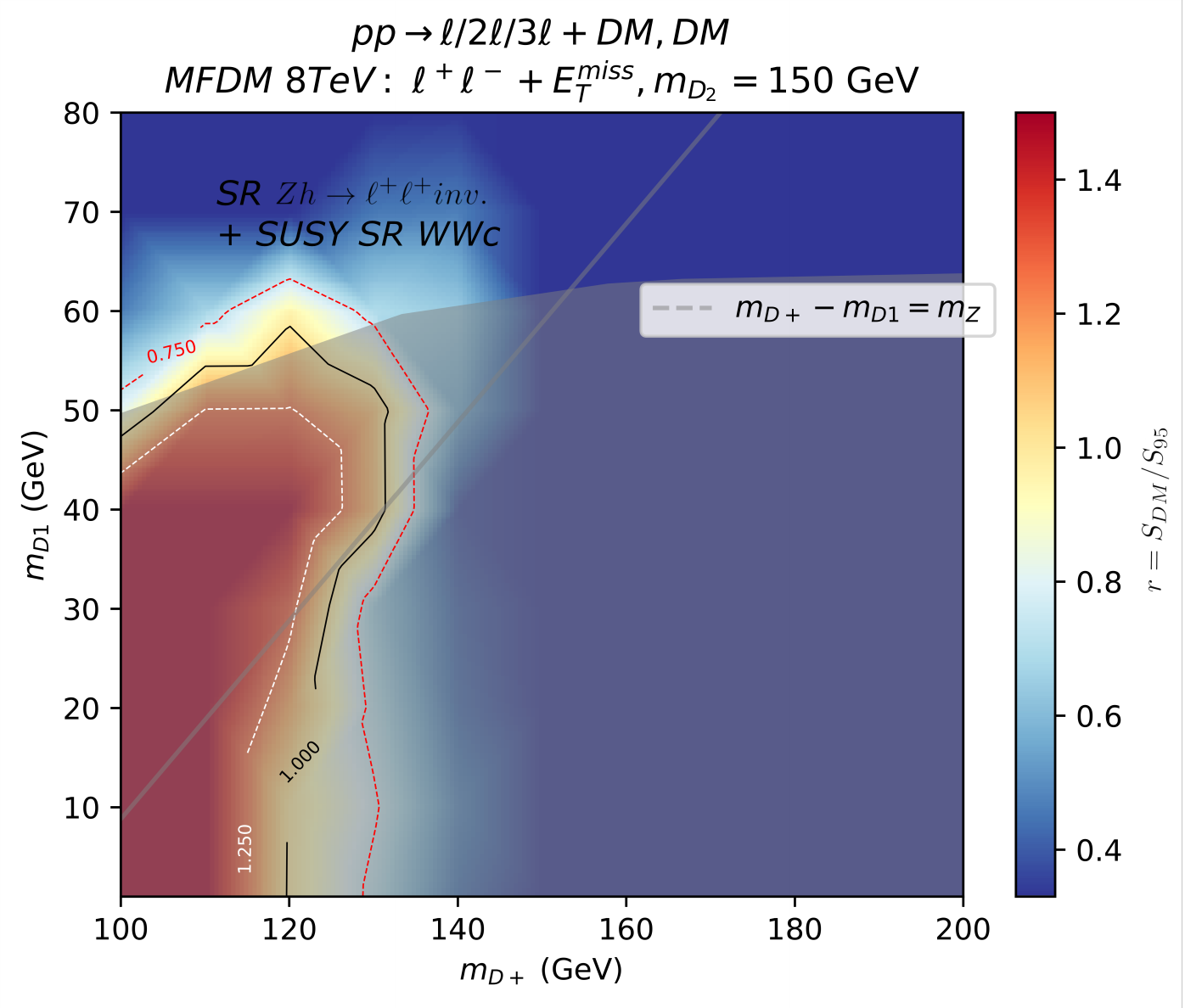}
	}
	\subfloat[]{
		\includegraphics[width=0.49\textwidth]{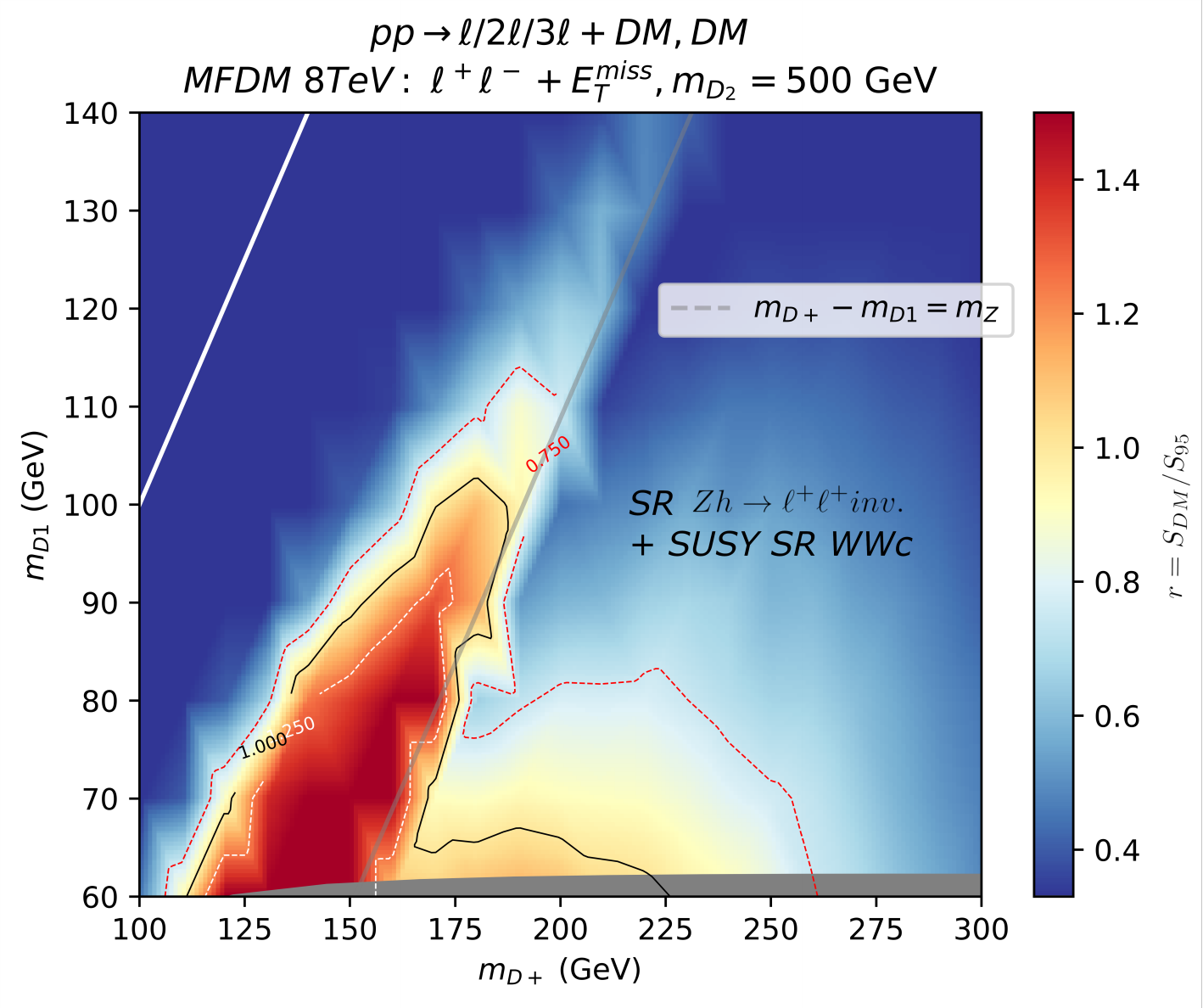}
	}
	\caption{$r$-value exclusion plots in MFDM for 8 TeV, $m_{D_2}=150$ GeV (a), $m_{D_2}=500$ GeV (b) from signal regions $WWc$+Higgs.}
	\label{fig:MFDM8TeVFixD3}
\end{figure*}
The plots in Fig.~\ref{fig:MFDM8TeVFixD3} for MFDM at 8 TeV with fixed $m_{D_2}$, showing the $m_{D_1}$-$m_{D^+}$ plane, are more analogous to those displayed for the i2HDM results shown previously. Figure~\ref{fig:MFDM8TeVFixD3}(a) with $m_{D_2}=150$ GeV is mostly excluded by Higgs to invisible limits, of $m_{D_1}<60$ GeV excluded.

On the other hand, Fig.~\ref{fig:MFDM8TeVFixD3}(b) limits, for $m_{D_2}=500$ GeV, extend much further. It excludes a peak at $m_{D_1}=100$ GeV, $m_{D^+}=180$ GeV, and follows along $M_Z$ in terms of the mas split between $m_{D^+}$ and $m_{D_1}$. This is when $D'$ can decay via real $Z$ boson to two leptons, and $D_1$ in the final state. It is also close to when this mass split equals $M_W$, facilitating two $D^\pm$ decays via real $W^\pm$ to a charged lepton and neutrino, along with $D_1$ in the final state.

In Fig.~\ref{fig:MFDM8TeVFixD3}(b) there is an additional region of large $r$-value at $160$ GeV $<m_{D^+}<225$ GeV, with a limit of $m_{D_1}=65$ GeV excluded, coming from the {\tt atlas\_higgs\_2013\_03} analysis with harder leptonic decays, and lighter $D_1$.

\newpage
\clearpage
\section{Numerical Overlaid Plots\label{app:plots}}

\begin{figure*}[!ht]
	\subfloat[]{
		\includegraphics[width=0.49\textwidth]{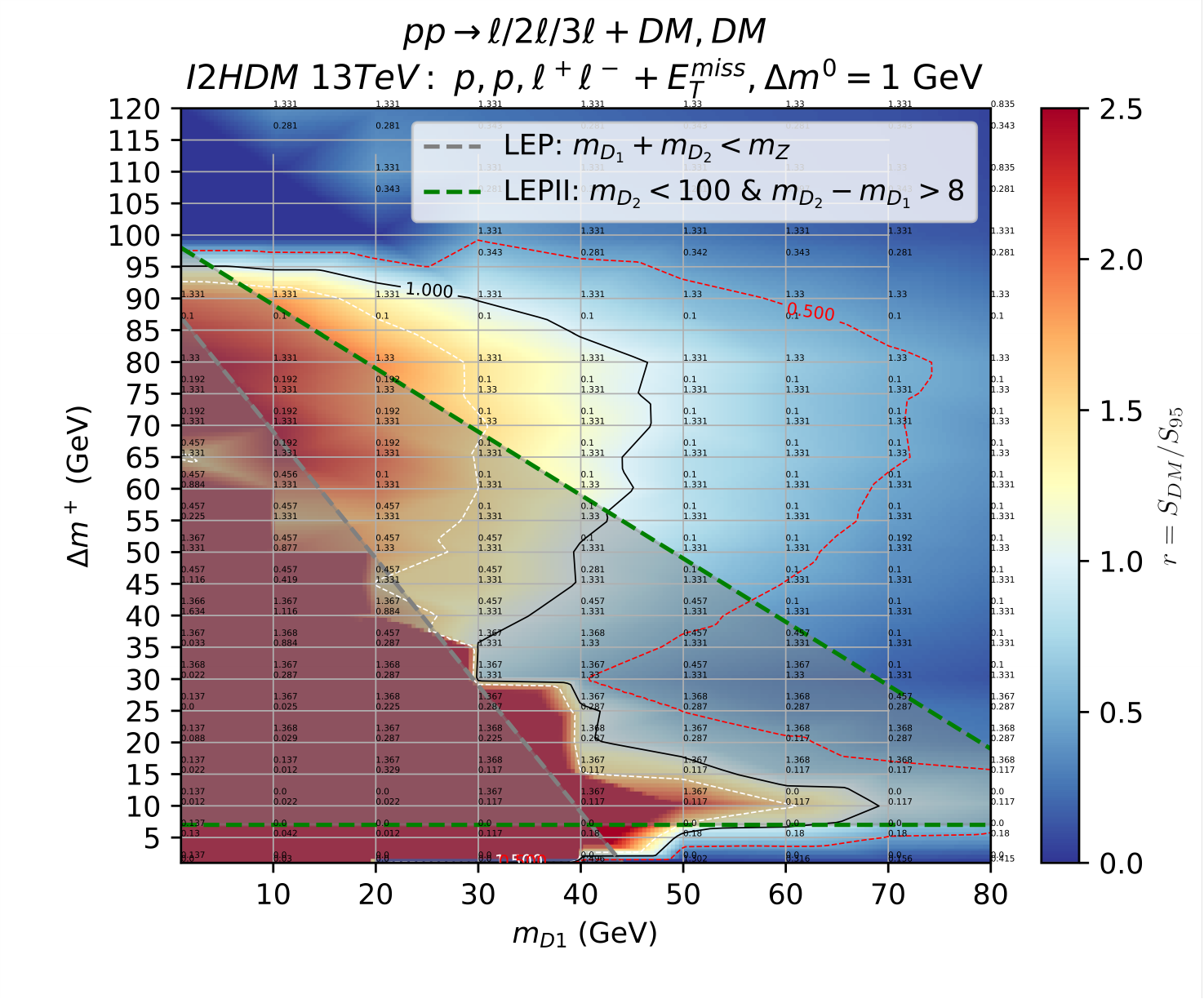}
	}
	\subfloat[]{
		\includegraphics[width=0.49\textwidth]{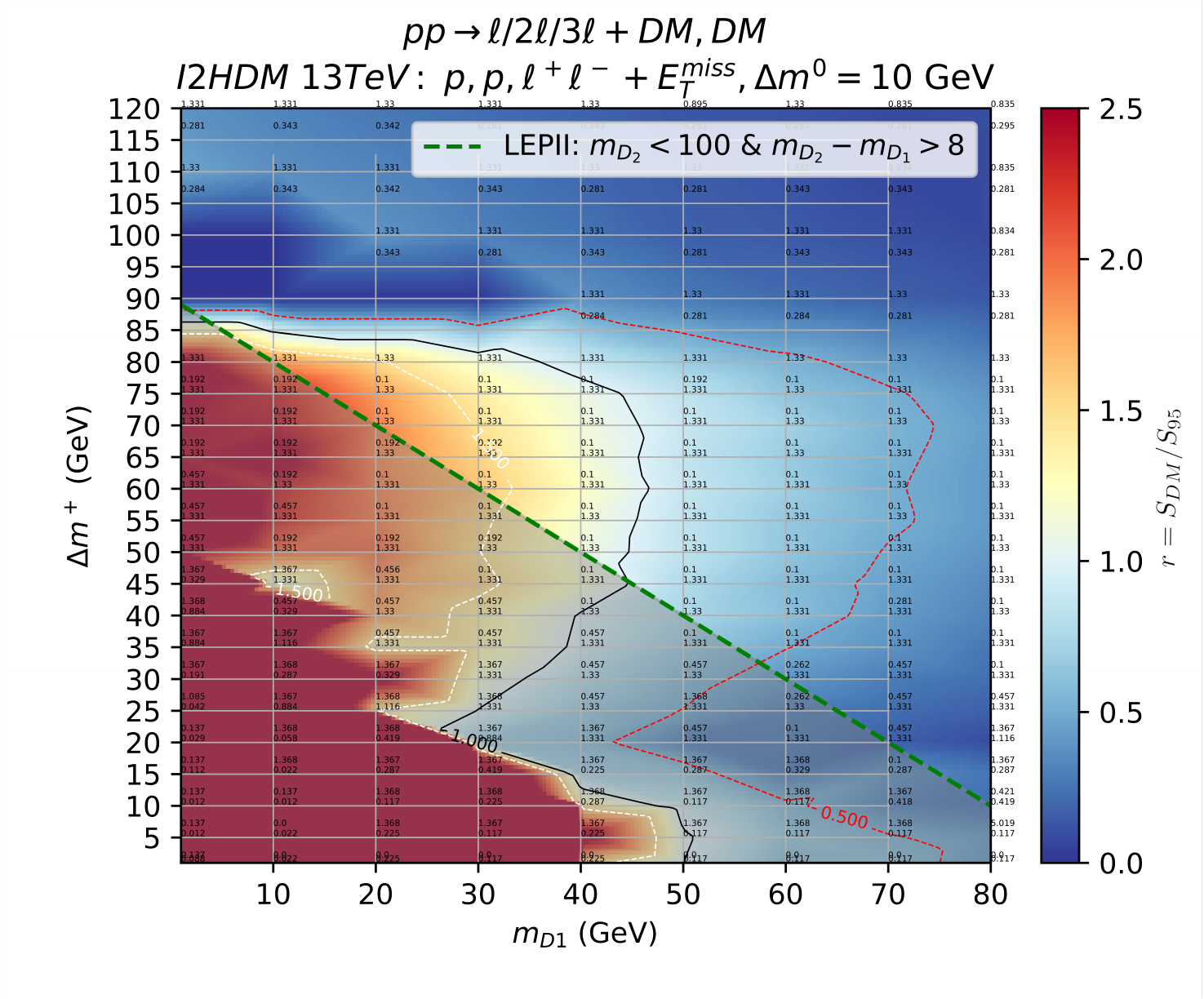}
	}\\
	\subfloat[]{
		\includegraphics[width=0.49\textwidth]{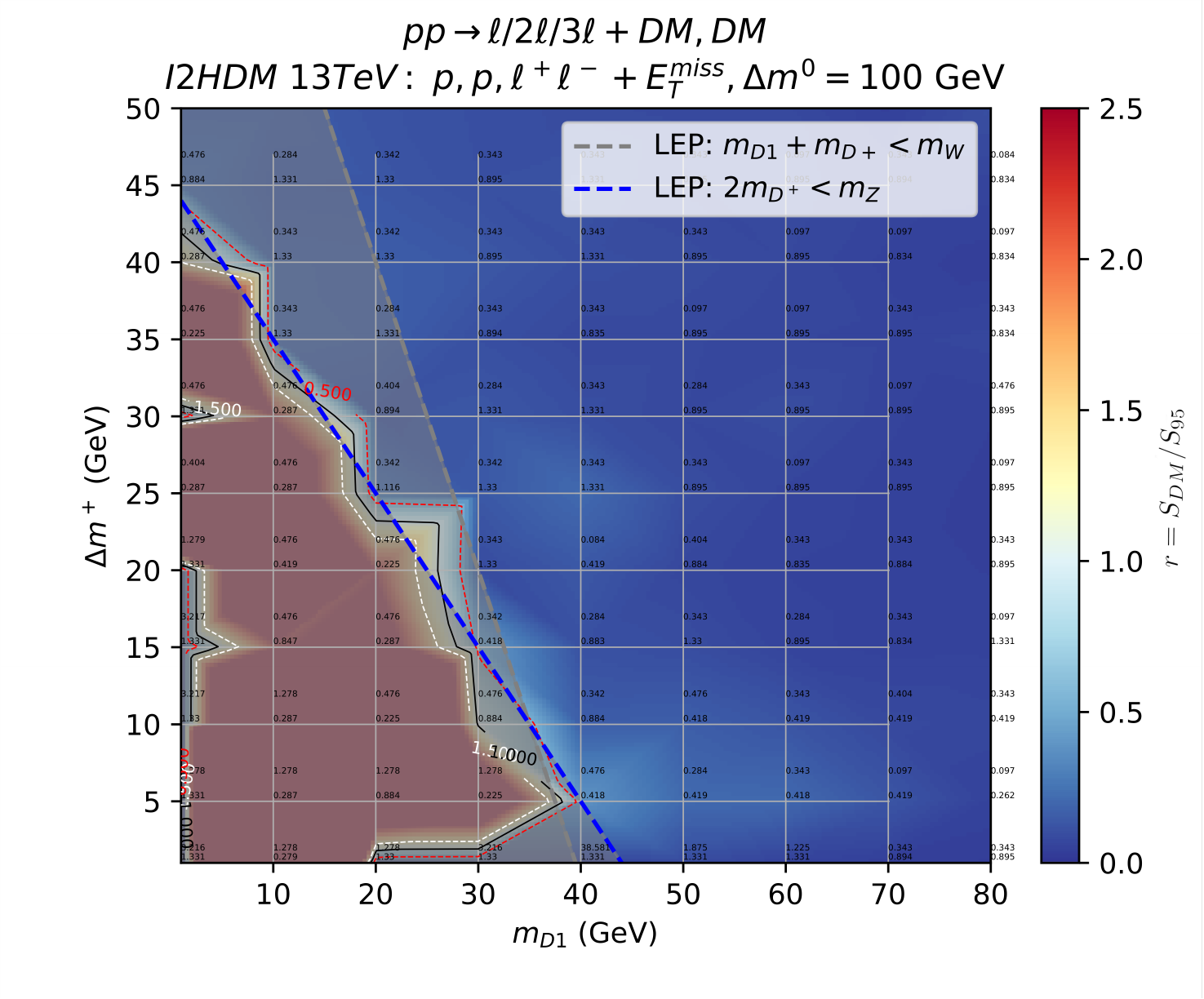}}
	\caption{i2HDM 13 TeV $r$-value exclusion plots, overlaid with total cross-section yields from 2-lepton(top number) and 3-lepton (bottom number) contributions to $r$-value. This is with the parametrisation in terms of  $\Delta m^+$ $\Delta m^0$. Plot (a) shows the case where $\Delta m^0=1$ GeV, while plot (b) shows $\Delta m^0=10$ GeV and plot (c) shows $\Delta m^0=100$ GeV. These are overlaid with limits from LEP I, LEP II experiments~\cite{Lundstr_m_2009,Pierce_2007}.}
	\label{fig:i2HDM-DeltaM_Numb}
\end{figure*}

\begin{figure*}[!ht]
	\centering
		\includegraphics[width=0.49\textwidth]{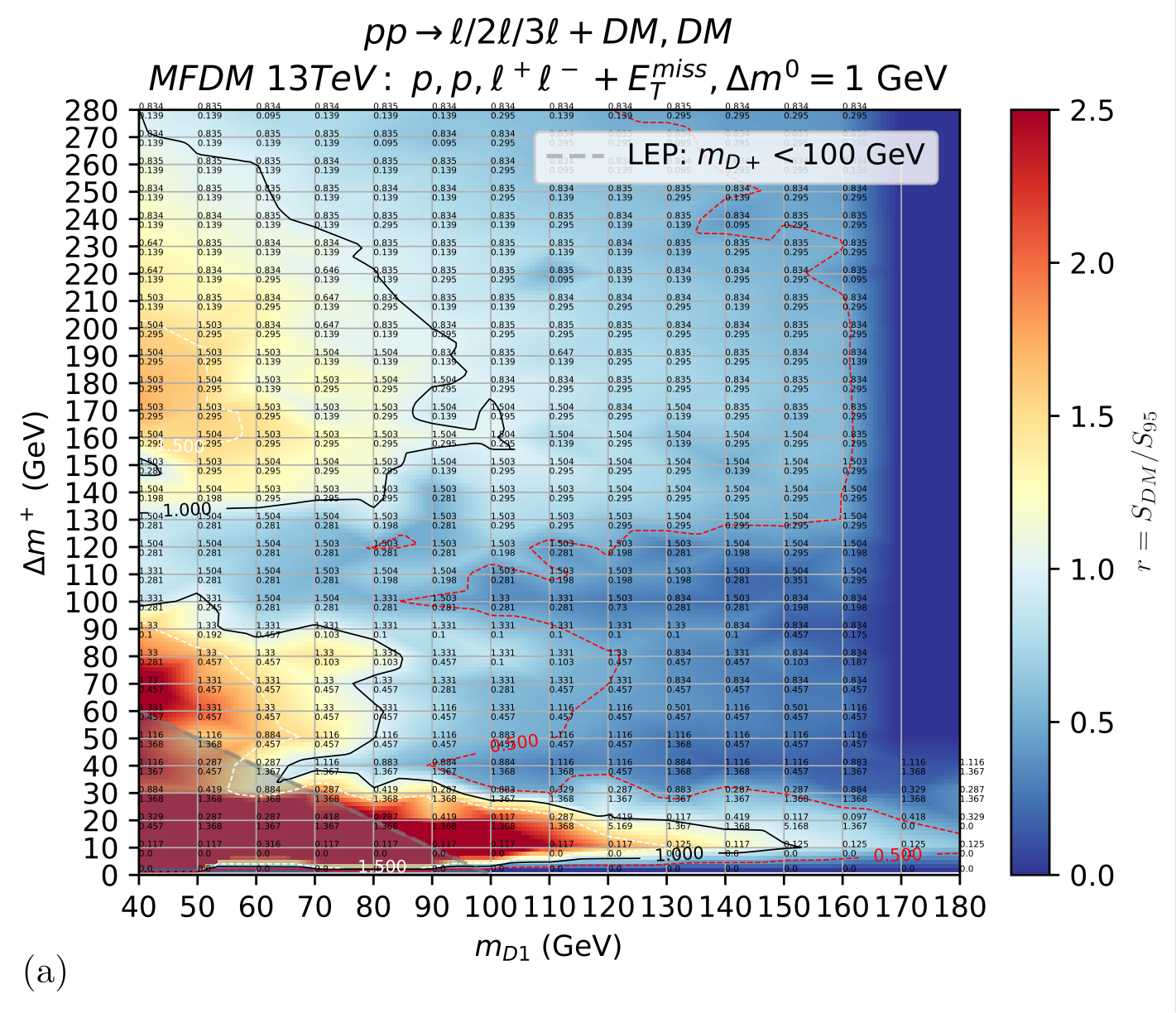}
		\includegraphics[width=0.49\textwidth]{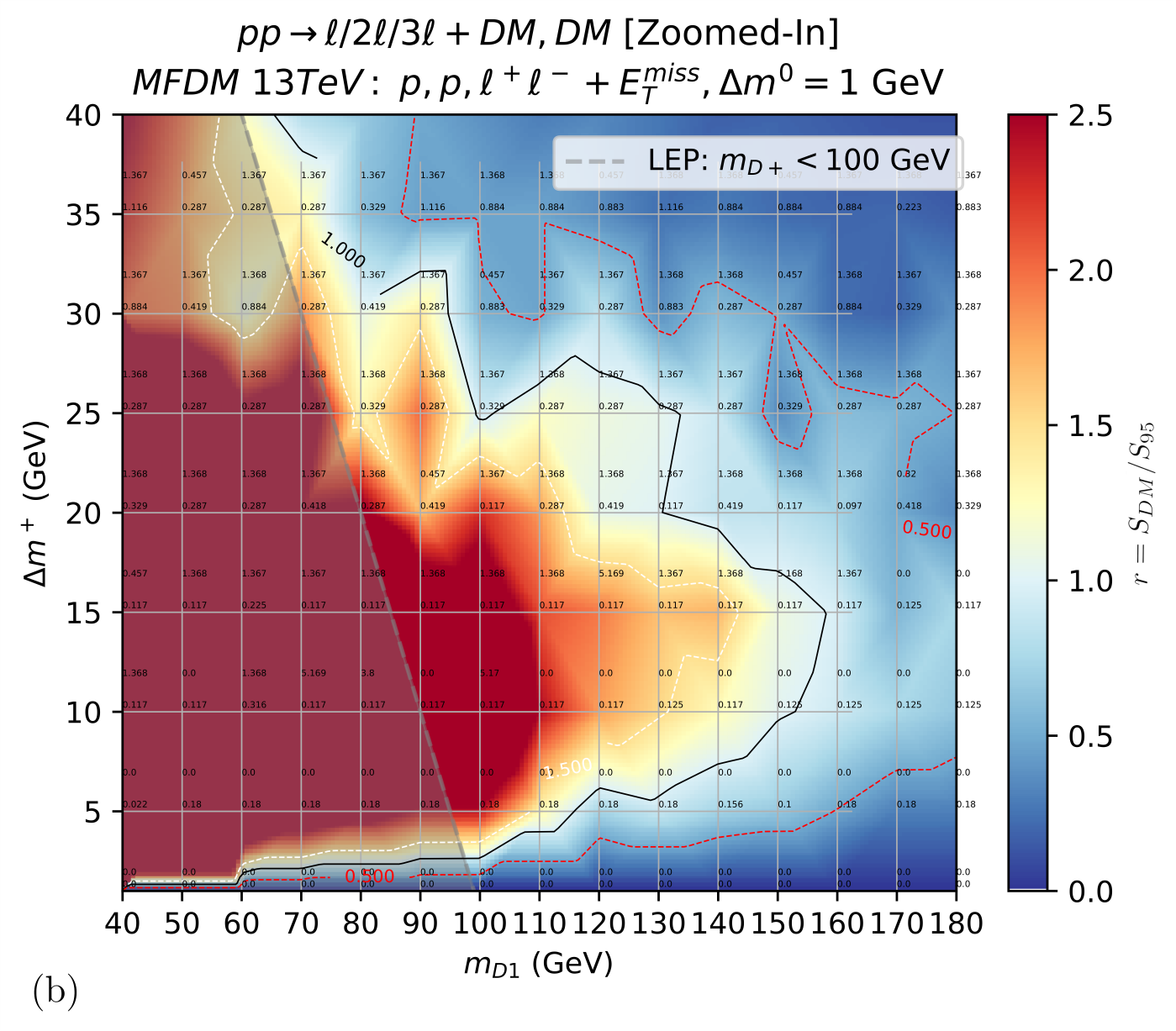}
\\
		\includegraphics[width=0.49\textwidth]{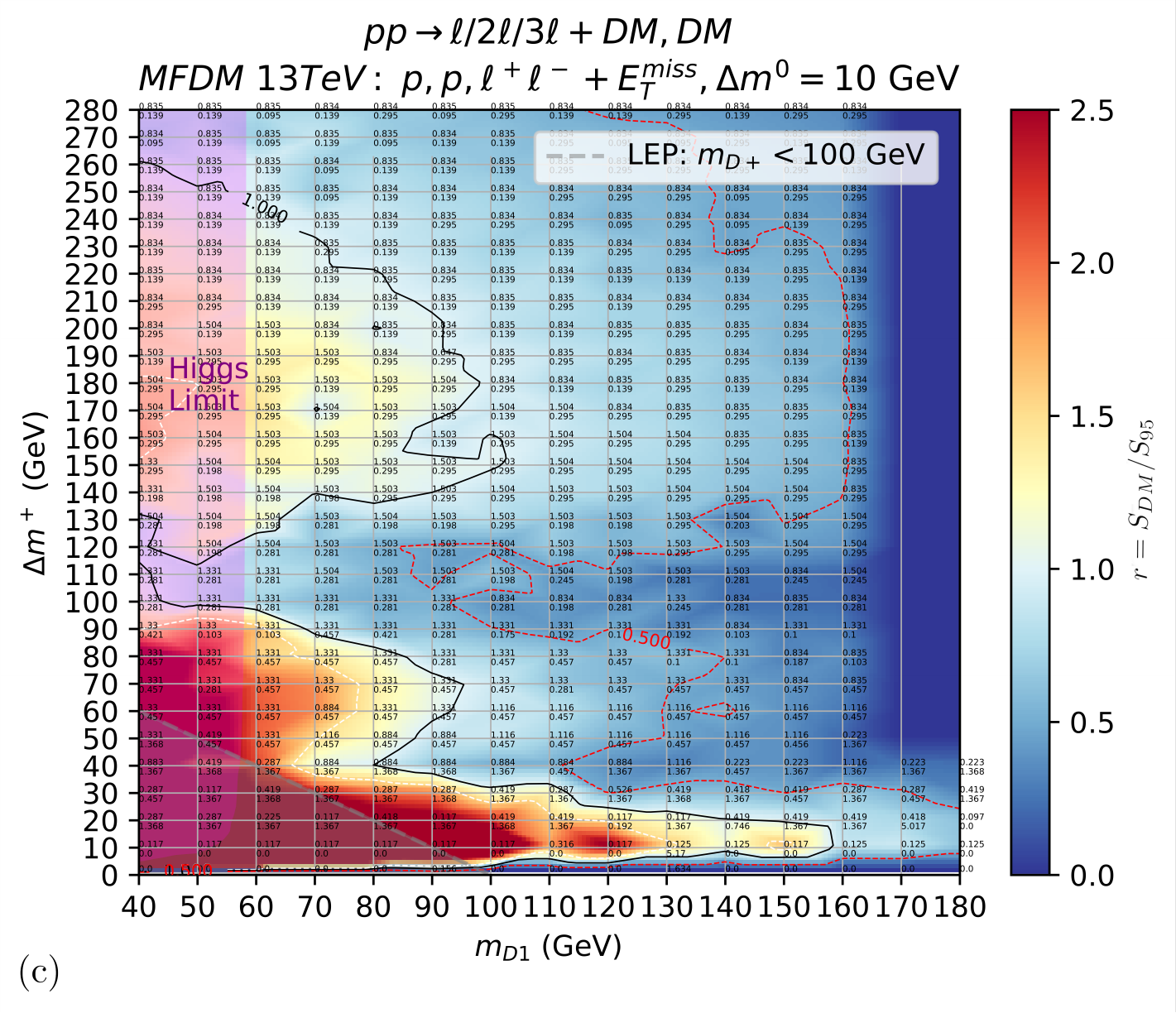}
		\includegraphics[width=0.49\textwidth]{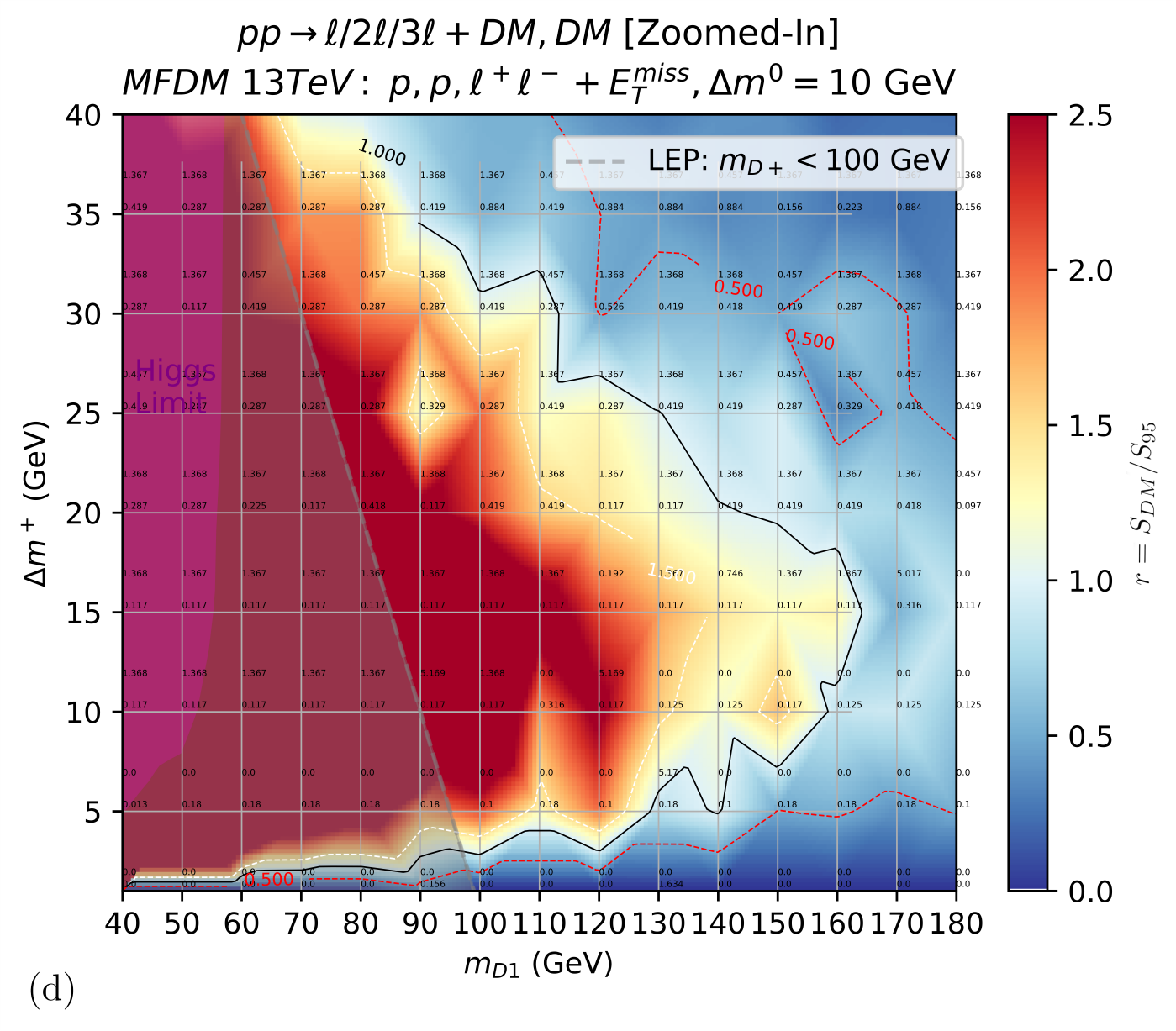}
	\\
		\includegraphics[width=0.49\textwidth]{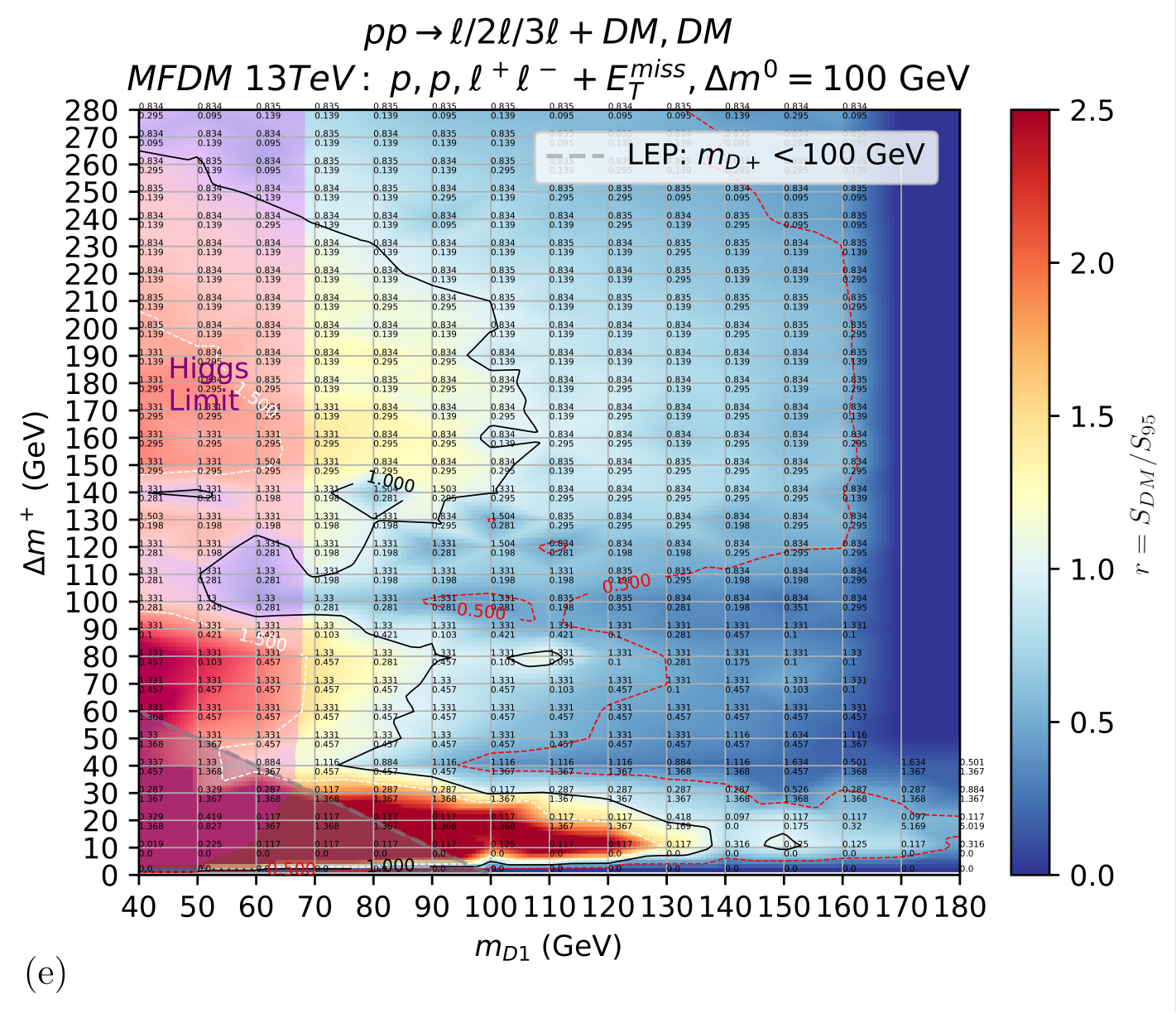}
		\includegraphics[width=0.49\textwidth]{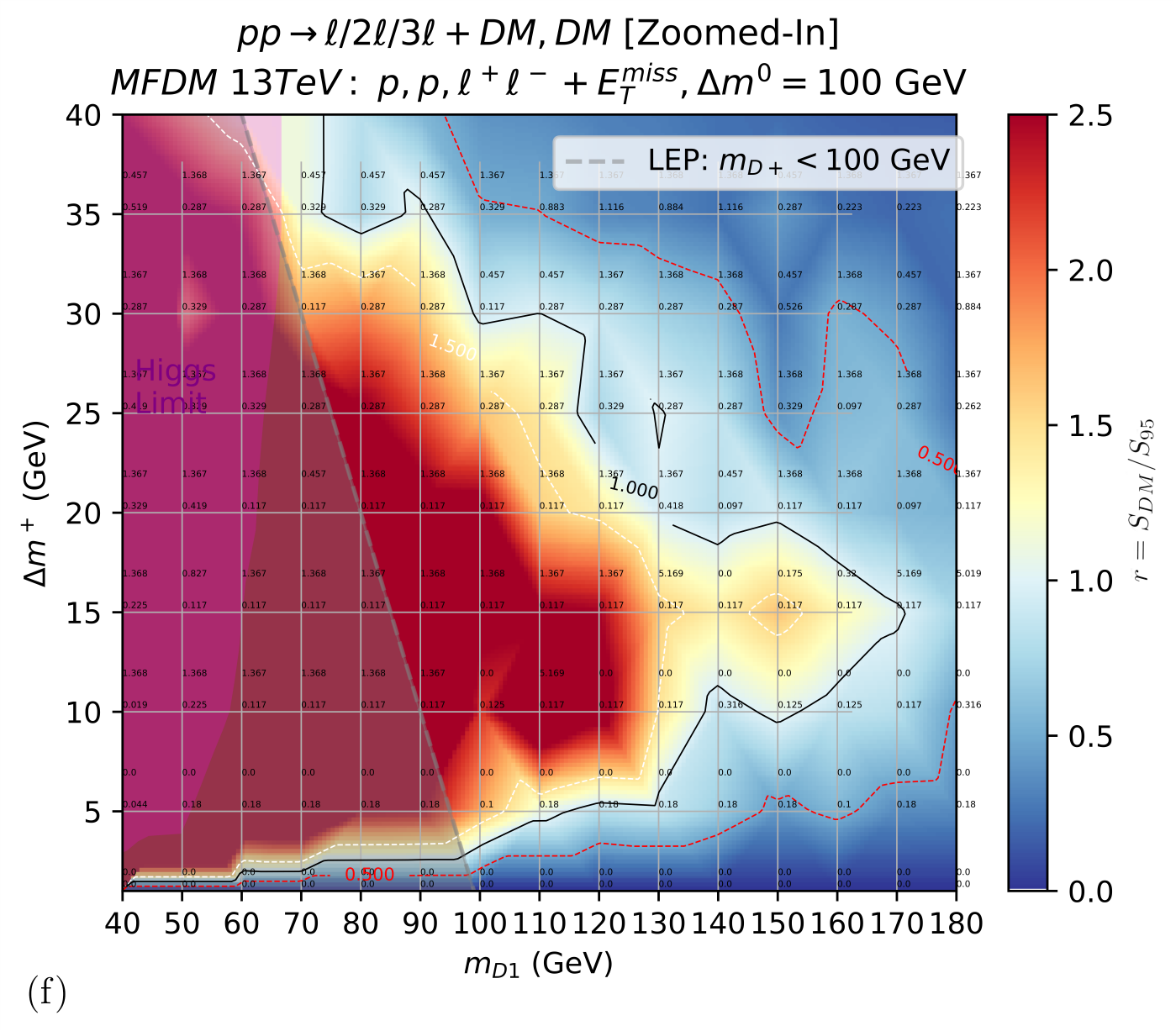}
	\caption{MFDM 13 TeV $r$-value exclusion plots overlaid with cross-section yields from 2-lepton(top number) and 3-lepton (bottom number) contributions, for parameter space  $\Delta m^+$-$m_{D1}$ for $\Delta m^0=1$(a)-(b), $10$(c)-(d), $100$(e)-(f) GeV.  The magenta region and grey region indicate the current Higgs-to-invisible limit\cite{Sirunyan_2019} of ~0.15 branching fraction and LEP bounds on charginos for the fermionic DM case\cite{Pierce_2007} respectively.}
	\label{fig:DMD3plotsMFDM_Numb}
\end{figure*}

In appendix~\ref{app:plots} we present the exclusion plots overlaid with cross sections in $fb$ for 2 or 3-leptonic final states. We first present the results for i2HDM in Fig.\ref{fig:i2HDM-DeltaM_Numb} and  MFDM in Fig.\ref{fig:DMD3plotsMFDM_Numb}.

\clearpage

\clearpage

\section{Sample Exclusion Formulae\label{app:excl-form}}
In appendix~\ref{app:excl-form} we describe the samples and formulae used to understand the tables presented in Appendix \ref{app:i2HDM-exc},\ref{app:MFDM-exc}. The input samples are separated between A,B and C, in both i2HDM and MFDM cases to improve efficiency.

For i2HDM, set A produces 50,000 events of $D^+D^-$ and $D_1D_2$ pairs, while specifying the decays $D^\pm\rightarrow\ell^\pm,\nu,D_1$ (via $W$)and $D_2\rightarrow\ell^+,\ell^-,D_1$ (via $Z$). Set B produces 150,000 events of $D^\pm D_2$, while specifying $D_2\rightarrow\ell^+,\ell^-,D_1$ (via $Z$) to obtain at least two leptons. Set C produces 100,000 events of the $D_1D_1Z$ production, while specifying leptonic $Z$ decay.

For MFDM, set A produces 50,000 total events, of $D^+D^-$ and $D_1D'$ pairs, while specifying decays of $D^\pm\rightarrow\ell^\pm,\nu,D_1$ (via $W$) and decays $D'\rightarrow\ell^+,\ell^-,D_1$ (via $Z$). Set B produces 150,000 $D'D_2$ events, without specifying decays. Set C produces 100,000 events of $D^\pm D'$ and $D^\pm D_2$ pairs, while requiring $D'\rightarrow\ell^+,\ell^-,D_1$ decays,
$D_2\rightarrow\ell^\pm,\nu,D^\pm$ (via $W$) or $D_2\rightarrow\ell^-,\ell^+,D'$ decays (via $Z$). This last set means $D'$ or $D^\pm$ coming from $D_2$ can also decay hadronically, to fulfil more 2-lepton and 3-lepton thresholds of the analyses.

The exclusion limits between input samples, for the same signal region of a given analysis, can be related by the equation
\begin{equation}
\sigma^{95}_A\epsilon_A=\sigma^{95}_B\epsilon_B=\sigma^{95}_C\epsilon_C
\end{equation}
allowing for exclusion $r$ values to be calculated, also noting that
\begin{equation}
r=r_A+r_B+r_C=\frac{\sigma^{DM}_A}{\sigma^{95}_A}+\frac{\sigma^{DM}_B}{\sigma^{95}_B}+\frac{\sigma^{DM}_C}{\sigma^{95}_C}.
\end{equation}

\clearpage

\section{i2HDM Cross-Section Limits\label{app:i2HDM-exc}}
In appendix~\ref{app:i2HDM-exc} we present the tables for i2HDM cross-section limits for 2 or 3-lepton final state samples A, B and C. Table~\ref{tab:i2hdm_exc_1},\ref{tab:i2hdm_exc_2},\ref{tab:i2hdm_exc_3} show the results, with $1<m_{D1}<80$ GeV, $1<\Delta m^+<120$ GeV, for $\Delta m^0=1,\ 10,\ 100$ GeV respectively.
\begin{table}[!ht]
	\begin{center}
		\resizebox{\columnwidth}{!}{%
			\begin{tabular}{|l|l|l||l|l|l|l|l|l||l|l|l|l|l|l|}
				$m_{D1}$ &  $\Delta m^+$ & $\Delta m^0$ & 2$\ell$ $\sigma^{95}_A$ (fb) &$\frac{100}{\sqrt{N_{MC}}}$ & 2$\ell$ $\sigma^{95}_B$ (fb)&$\frac{100}{\sqrt{N_{MC}}}$ & 2$\ell$ $\sigma^{95}_C$ (fb)&$\frac{100}{\sqrt{N_{MC}}}$ & 3$\ell$ $\sigma^{95}_A$ (fb)&$\frac{100}{\sqrt{N_{MC}}}$ & 3$\ell$ $\sigma^{95}_B$ (fb)&$\frac{100}{\sqrt{N_{MC}}}$ & 3$\ell$ $\sigma^{95}_C$ (fb)&$\frac{100}{\sqrt{N_{MC}}}$ \\
				\hline \hline
1	&	5	&	1	&	3.26 $\times10^3$	&	71	&	-	&	100	&	6.51$\times10^4$	&	71	&	-	&	-	&	1.21$\times10^3$	&	24	&	-	&	-	\\
1	&	10	&	1	&	97.0	&	41	&	-	&	100	&	-	&	-	&	-	&	-	&	1.21$\times10^3$	&	24	&	-	&	100	\\

1	&	20	&	1	&	1.47$\times10^3$	&	58	&	6.63$\times10^3$	&	71	&	-	&	100	&	-	&	-	&	933	&	21	&	-	&	-	\\

1	&	40	&	1	&	1.02$\times10^5$	&	35	&	8.17$\times10^4$	&	58	&	8.17$\times10^4$	&	71	&	-	&	-	&	1.2$\times10^3$	&	8	&	-	&	-	\\

1	&	60	&	1	&	8.84$\times10^3$	&	45	&	5.3$\times10^3$	&	20	&	2.94$\times10^4$	&	58	&	-	&	-	&	220	&	6	&	-	&	100	\\

1	&	80	&	1	&	783	&	11	&	326	&	4	&	1.15$\times10^3$	&	9	&	-	&	-	&	93.0	&	6	&	-	&	-	\\

10	&	5	&	1	&	698	&	58	&	3.14$\times10^3$	&	71	&	-	&	100	&	-	&	-	&	-	&	-	&	-	&	-	\\
10	&	10	&	1	&	161	&	38	&	674	&	45	&	-	&	-	&	-	&	-	&	-	&	-	&	-	&	-	\\

10	&	20	&	1	&	287	&	45	&	-	&	100	&	1.43$\times10^4$	&	71	&	-	&	-	&	1.87$\times10^3$	&	30	&	-	&	100	\\

10	&	40	&	1	&	1.40$\times10^4$	&	50	&	1.29$\times10^4$	&	28	&	2.23$\times10^4$	&	45	&	-	&	-	&	531	&	5	&	6.82$\times10^4$	&	71	\\

10	&	60	&	1	&	4.44$\times10^3$	&	26	&	507	&	5	&	604	&	7	&	-	&	-	&	165	&	5	&	-	&	-	\\

10	&	80	&	1	&	150	&	5	&	248	&	4	&	630	&	7	&	-	&	-	&	80.0	&	5	&	-	&	-	\\

10	&	120	&	1	&	281	&	6	&	1.32$\times10^3$	&	8	&	411	&	6	&	-	&	-	&	62.0	&	4	&	-	&	-	\\

20	&	5	&	1	&	97.0	&	41	&	877	&	71	&	-	&	-	&	-	&	-	&	-	&	-	&	-	&	-	\\
20	&	10	&	1	&	140	&	35	&	562	&	41	&	-	&	-	&	-	&	-	&	-	&	-	&	-	&	-	\\

20	&	20	&	1	&	4.78$\times10^3$	&	58	&	1.08$\times10^4$	&	50	&	-	&	-	&	-	&	-	&	9.32$\times10^3$	&	21	&	-	&	-	\\

20	&	40	&	1	&	6.31$\times10^3$	&	38	&	6.02$\times10^3$	&	21	&	1.76$\times10^4$	&	45	&	-	&	-	&	366	&	7	&	-	&	-	\\

20	&	60	&	1	&	247	&	6	&	377	&	4	&	438	&	6	&	-	&	-	&	148	&	5	&	-	&	-	\\

20	&	80	&	1	&	91.0	&	4	&	230	&	3	&	534	&	6	&	-	&	-	&	62.0	&	5	&	-	&	-	\\

20	&	120	&	1	&	247	&	6	&	1.50$\times10^3$	&	9	&	321	&	5	&	-	&	100	&	58.0	&	4	&	9.40$\times10^3$	&	58	\\

40	&	5	&	1	&	3.00$\times10^3$	&	58	&	2.25$\times10^3$	&	29	&	-	&	-	&	-	&	-	&	-	&	-	&	-	&	-	\\
40	&	10	&	1	&	344	&	24	&	390	&	15	&	-	&	-	&	-	&	-	&	-	&	-	&	-	&	-	\\

40	&	20	&	1	&	797	&	24	&	1.39$\times10^3$	&	18	&	-	&	100	&	-	&	-	&	2.18$\times10^3$	&	10	&	-	&	-	\\

40	&	40	&	1	&	315	&	7	&	710	&	6	&	327	&	5	&	-	&	-	&	387	&	4	&	-	&	-	\\

40	&	60	&	1	&	99.0	&	4	&	300	&	4	&	302	&	5	&	835	&	41	&	114	&	9	&	2.00$\times10^3$	&	45	\\

40	&	80	&	1	&	56.0	&	3	&	203	&	3	&	325	&	5	&	313	&	25	&	57.0	&	6	&	2.00$\times10^3$	&	45	\\

40	&	100	&	1	&	171	&	5	&	1.06$\times10^3$	&	7	&	341	&	5	&	-	&	100	&	76.0	&	4	&	1.41$\times10^4$	&	71	\\

40	&	120	&	1	&	184	&	5	&	1.22$\times10^3$	&	8	&	229	&	4	&	6.91$\times10^3$	&	70	&	46.0	&	3	&	1.41$\times10^4$	&	71	\\

60	&	5	&	1	&	900	&	32	&	519	&	14	&	-	&	-	&	-	&	-	&	-	&	-	&	-	&	-	\\
60	&	10	&	1	&	292	&	22	&	219	&	11	&	-	&	-	&	-	&	-	&	-	&	-	&	-	&	-	\\

60	&	20	&	1	&	344	&	24	&	798	&	21	&	-	&	-	&	-	&	-	&	1.90$\times10^3$	&	10	&	-	&	-	\\

60	&	40	&	1	&	228	&	6	&	585	&	5	&	236	&	4	&	-	&	100	&	240	&	6	&	-	&	-	\\

60	&	60	&	1	&	79.0	&	3	&	255	&	4	&	224	&	4	&	836	&	41	&	71.0	&	7	&	900	&	30	\\

60	&	80	&	1	&	45.0	&	3	&	182	&	3	&	248	&	4	&	627	&	35	&	44.0	&	5	&	2.48$\times10^3$	&	50	\\

60	&	100	&	1	&	139	&	5	&	1.03$\times10^3$	&	7	&	252	&	4	&	952	&	24	&	188	&	6	&	561	&	13	\\

60	&	120	&	1	&	168	&	5	&	1.43$\times10^3$	&	8	&	202	&	4	&	-	&	100	&	45.0	&	3	&	1.41$\times10^4$	&	71	\\

80	&	5	&	1	&	563	&	25	&	267	&	10	&	-	&	-	&	-	&	-	&	-	&	-	&	-	&	-	\\
80	&	10	&	1	&	266	&	21	&	151	&	9	&	-	&	-	&	-	&	-	&	-	&	-	&	-	&	-	\\

80	&	20	&	1	&	319	&	15	&	414	&	10	&	3.95$\times10^3$	&	38	&	-	&	-	&	1.31$\times10^3$	&	8	&	-	&	-	\\

80	&	40	&	1	&	188	&	5	&	536	&	5	&	198	&	4	&	1.25$\times10^3$	&	50	&	255	&	13	&	1.10$\times10^3$	&	33	\\

80	&	60	&	1	&	68.0	&	3	&	256	&	4	&	198	&	4	&	251	&	22	&	83.0	&	7	&	1.24$\times10^3$	&	35	\\

80	&	80	&	1	&	39.0	&	2	&	176	&	3	&	223	&	4	&	264	&	23	&	44.0	&	5	&	830	&	29	\\

80	&	100	&	1	&	140	&	5	&	1.20$\times10^3$	&	8	&	232	&	4	&	7.03$\times10^3$	&	71	&	64.0	&	4	&	-	&	100	\\

80	&	120	&	1	&	271	&	8	&	680	&	7	&	220	&	5	&	439	&	16	&	128	&	5	&	489	&	12	\\

		\end{tabular}%
	}
\caption{i2HDM table of limits on the cross sections for A,B and C samples, for $1<m_{D1}<80$ GeV, $1<\Delta m^+<120$ GeV, $\Delta m^0=1$ GeV.}
\label{tab:i2hdm_exc_1}
\end{center}
\end{table}

\clearpage

\begin{table}[!ht]
	\begin{center}
		\resizebox{\columnwidth}{!}{%
			\begin{tabular}{|l|l|l||l|l|l|l|l|l||l|l|l|l|l|l|}
				$m_{D1}$ &  $\Delta m^+$ & $\Delta m^0$ & 2$\ell$ $\sigma^{95}_A$ (fb) &$\frac{100}{\sqrt{N_{MC}}}$ & 2$\ell$ $\sigma^{95}_B$ (fb)&$\frac{100}{\sqrt{N_{MC}}}$ & 2$\ell$ $\sigma^{95}_C$ (fb)&$\frac{100}{\sqrt{N_{MC}}}$ & 3$\ell$ $\sigma^{95}_A$ (fb)&$\frac{100}{\sqrt{N_{MC}}}$ & 3$\ell$ $\sigma^{95}_B$ (fb)&$\frac{100}{\sqrt{N_{MC}}}$ & 3$\ell$ $\sigma^{95}_C$ (fb)&$\frac{100}{\sqrt{N_{MC}}}$ \\
				\hline \hline
1	&	5	&	10	&	110	&	43	&	877	&	71	&	-	&	-	&	-	&	-	&	1.03$\times10^3$	&	22	&	-	&	-	\\
1	&	10	&	10	&	257	&	66	&	877	&	71	&	-	&	-	&	-	&	-	&	684	&	18	&	-	&	-	\\

1	&	20	&	10	&	358	&	50	&	-	&	100	&	1.43$\times10^4$	&	71	&	-	&	-	&	554	&	16	&	-	&	-	\\

1	&	40	&	10	&	1.10$\times10^4$	&	50	&	9.46$\times10^3$	&	27	&	-	&	100	&	2.28$\times10^4$	&	58	&	618	&	5	&	-	&	-	\\

1	&	60	&	10	&	4.75$\times10^3$	&	27	&	562	&	5	&	1.11$\times10^3$	&	9	&	-	&	100	&	170	&	5	&	-	&	-	\\

1	&	80	&	10	&	168	&	5	&	267	&	4	&	1.10$\times10^3$	&	9	&	-	&	-	&	83.0	&	5	&	-	&	-	\\

1	&	120	&	10	&	296	&	7	&	1.52$\times10^3$	&	9	&	407	&	6	&	-	&	98	&	70.0	&	4	&	-	&	-	\\

10	&	5	&	10	&	176	&	40	&	-	&	-	&	-	&	-	&	-	&	-	&	-	&	-	&	-	&	-	\\
10	&	10	&	10	&	182	&	56	&	585	&	58	&	-	&	-	&	-	&	-	&	-	&	-	&	-	&	-	\\

10	&	20	&	10	&	358	&	50	&	1.43$\times10^4$	&	58	&	1.43$\times10^4$	&	71	&	-	&	-	&	1.28$\times10^4$	&	25	&	-	&	-	\\

10	&	40	&	10	&	5.48$\times10^3$	&	58	&	3.79$\times10^3$	&	28	&	-	&	100	&	-	&	-	&	403	&	4	&	-	&	-	\\

10	&	60	&	10	&	536	&	9	&	409	&	5	&	710	&	7	&	-	&	-	&	144	&	5	&	-	&	-	\\

10	&	80	&	10	&	121	&	4	&	267	&	4	&	714	&	7	&	-	&	-	&	74.0	&	5	&	-	&	-	\\

10	&	120	&	10	&	264	&	6	&	1.57$\times10^3$	&	9	&	336	&	5	&	856	&	22	&	180	&	6	&	1.01$\times10^3$	&	17	\\

20	&	5	&	10	&	3.07$\times10^3$	&	52	&	1.69$\times10^4$	&	71	&	-	&	-	&	-	&	-	&	-	&	-	&	-	&	-	\\
20	&	10	&	10	&	1.41$\times10^3$	&	49	&	8.77$\times10^3$	&	71	&	-	&	-	&	-	&	-	&	3.42$\times10^4$	&	41	&	-	&	-	\\

20	&	20	&	10	&	4.19$\times10^3$	&	45	&	6.97$\times10^3$	&	33	&	1.04$\times10^4$	&	50	&	-	&	-	&	1.48$\times10^3$	&	8	&	-	&	-	\\

20	&	40	&	10	&	1.17$\times10^3$	&	13	&	842	&	6	&	461	&	6	&	-	&	-	&	297	&	7	&	-	&	-	\\

20	&	60	&	10	&	161	&	5	&	296	&	4	&	554	&	6	&	716	&	38	&	106	&	8	&	2.50$\times10^3$	&	50	\\

20	&	80	&	10	&	82.0	&	4	&	236	&	3	&	542	&	6	&	455	&	30	&	69.0	&	7	&	2.50$\times10^3$	&	50	\\
20	&	100	&	10	&	263	&	6	&	1.42$\times10^3$	&	8	&	398	&	5	&	1.28$\times10^3$	&	27	&	200	&	6	&	1.14$\times10^3$	&	18	\\

20	&	120	&	10	&	222	&	6	&	1.43$\times10^3$	&	8	&	285	&	5	&	951	&	24	&	153	&	5	&	856	&	16	\\

40	&	5	&	10	&	1.02$\times10^3$	&	30	&	544	&	13	&	-	&	-	&	-	&	-	&	-	&	-	&	-	&	-	\\
40	&	10	&	10	&	929	&	25	&	1.05$\times10^3$	&	16	&	9.50$\times10^3$	&	58	&	-	&	-	&	3.31$\times10^3$	&	13	&	-	&	-	\\

40	&	20	&	10	&	950	&	12	&	1.75$\times10^3$	&	9	&	290	&	5	&	-	&	100	&	865	&	6	&	-	&	100	\\

40	&	40	&	10	&	212	&	6	&	445	&	5	&	298	&	5	&	-	&	-	&	197	&	5	&	-	&	-	\\

40	&	60	&	10	&	78.0	&	3	&	226	&	3	&	379	&	5	&	456	&	30	&	70.0	&	7	&	4.98$\times10^3$	&	71	\\

40	&	80	&	10	&	60.0	&	3	&	214	&	3	&	328	&	5	&	313	&	25	&	67.0	&	7	&	2.00$\times10^3$	&	45	\\

40	&	100	&	10	&	201	&	5	&	1.30$\times10^3$	&	8	&	266	&	4	&	1.14$\times10^3$	&	26	&	205	&	6	&	797	&	15	\\

40	&	120	&	10	&	180	&	5	&	1.38$\times10^3$	&	8	&	205	&	4	&	569	&	18	&	147	&	5	&	553	&	13	\\

60	&	5	&	10	&	877	&	39	&	297	&	13	&	-	&	-	&	-	&	-	&	-	&	-	&	-	&	-	\\
60	&	10	&	10	&	831	&	38	&	337	&	14	&	-	&	-	&	-	&	-	&	2.77$\times10^3$	&	12	&	-	&	-	\\

60	&	20	&	10	&	578	&	9	&	1.43$\times10^3$	&	8	&	218	&	4	&	-	&	-	&	741	&	6	&	-	&	-	\\

60	&	40	&	10	&	126	&	4	&	367	&	4	&	218	&	4	&	1.00$\times10^3$	&	45	&	190	&	11	&	1.43$\times10^3$	&	38	\\

60	&	60	&	10	&	63.0	&	3	&	220	&	3	&	239	&	4	&	386	&	28	&	71.0	&	7	&	2.47$\times10^3$	&	50	\\

60	&	80	&	10	&	50.0	&	3	&	208	&	3	&	269	&	5	&	455	&	30	&	58.0	&	6	&	901	&	30	\\

60	&	100	&	10	&	185	&	5	&	1.51$\times10^3$	&	9	&	224	&	4	&	772	&	21	&	159	&	6	&	659	&	14	\\

60	&	120	&	10	&	165	&	5	&	1.19$\times10^3$	&	8	&	170	&	4	&	-	&	100	&	43.0	&	3	&	1.41$\times10^4$	&	71	\\

80	&	5	&	10	&	650	&	33	&	161	&	10	&	-	&	-	&	-	&	-	&	-	&	-	&	-	&	-	\\
80	&	10	&	10	&	509	&	16	&	476	&	9	&	2.00$\times10^4$	&	71	&	-	&	100	&	1.25$\times10^5$	&	41	&	2.42$\times10^5$	&	71	\\

80	&	20	&	10	&	786	&	12	&	1.07$\times10^3$	&	8	&	9.13$\times10^3$	&	29	&	-	&	100	&	717	&	6	&	-	&	-	\\

80	&	40	&	10	&	116	&	4	&	331	&	4	&	188	&	4	&	418	&	29	&	125	&	9	&	1.65$\times10^3$	&	41	\\

80	&	60	&	10	&	56.0	&	3	&	202	&	3	&	200	&	4	&	264	&	23	&	57.0	&	6	&	1.41$\times10^3$	&	38	\\

80	&	80	&	10	&	47.0	&	3	&	199	&	3	&	217	&	4	&	313	&	25	&	50.0	&	6	&	1.24$\times10^3$	&	35	\\

80	&	100	&	10	&	273	&	8	&	1.07$\times10^3$	&	9	&	254	&	6	&	6.94$\times10^3$	&	70	&	55.0	&	4	&	9.38$\times10^3$	&	58	\\

80	&	120	&	10	&	202	&	7	&	759	&	8	&	220	&	5	&	-	&	99	&	31.0	&	3	&	-	&	100
		\end{tabular}%
}
\caption{i2HDM table of limits on the cross sections for A,B and C samples, for $1<m_{D1}<80$ GeV, $1<\Delta m^+<120$ GeV, $\Delta m^0=10$ GeV.}
\label{tab:i2hdm_exc_2}
\end{center}
\end{table}

\clearpage

\begin{table}[!ht]
	\begin{center}
		\resizebox{\columnwidth}{!}{%
			\begin{tabular}{|l|l|l||l|l|l|l|l|l||l|l|l|l|l|l|}
				$m_{D1}$ &  $\Delta m^+$ & $\Delta m^0$ & 2$\ell$ $\sigma^{95}_A$ (fb) &$\frac{100}{\sqrt{N_{MC}}}$ & 2$\ell$ $\sigma^{95}_B$ (fb)&$\frac{100}{\sqrt{N_{MC}}}$ & 2$\ell$ $\sigma^{95}_C$ (fb)&$\frac{100}{\sqrt{N_{MC}}}$ & 3$\ell$ $\sigma^{95}_A$ (fb)&$\frac{100}{\sqrt{N_{MC}}}$ & 3$\ell$ $\sigma^{95}_B$ (fb)&$\frac{100}{\sqrt{N_{MC}}}$ & 3$\ell$ $\sigma^{95}_C$ (fb)&$\frac{100}{\sqrt{N_{MC}}}$ \\
				\hline \hline
1	&	5	&	100	&	-	&	-	&	2.38$\times10^3$	&	11	&	628	&	7	&	-	&	-	&	694	&	4	&	-	&	-	\\
1	&	10	&	100	&	-	&	-	&	2.27$\times10^3$	&	11	&	591	&	7	&	-	&	-	&	451	&	5	&	6.39$\times10^4$	&	71	\\
1	&	20	&	100	&	-	&	-	&	1.88$\times10^3$	&	10	&	450	&	6	&	-	&	-	&	504	&	3	&	6.43$\times10^4$	&	45	\\
1	&	30	&	100	&	-	&	-	&	1.64$\times10^3$	&	9	&	386	&	5	&	-	&	-	&	153	&	5	&	-	&	100	\\
1	&	40	&	100	&	3.59$\times10^3$	&	50	&	6.14$\times10^3$	&	38	&	4.78$\times10^3$	&	41	&	-	&	-	&	149	&	5	&	1.59$\times10^4$	&	58	\\
10	&	5	&	100	&	2.05$\times10^3$	&	38	&	7.17$\times10^3$	&	41	&	-	&	100	&	-	&	-	&	385	&	4	&	-	&	100	\\
10	&	10	&	100	&	4.78$\times10^3$	&	58	&	7.17$\times10^3$	&	41	&	-	&	100	&	-	&	-	&	367	&	4	&	-	&	-	\\
10	&	20	&	100	&	4.19$\times10^3$	&	45	&	1.05$\times10^4$	&	41	&	1.39$\times10^4$	&	58	&	-	&	-	&	178	&	5	&	-	&	-	\\
10	&	40	&	100	&	4.11$\times10^3$	&	25	&	1.28$\times10^3$	&	8	&	279	&	5	&	-	&	99	&	302	&	8	&	856	&	16	\\
20	&	5	&	100	&	7.36$\times10^3$	&	41	&	1.66$\times10^4$	&	35	&	2.21$\times10^4$	&	50	&	-	&	-	&	402	&	5	&	6.39$\times10^4$	&	71	\\
20	&	10	&	100	&	2.81$\times10^3$	&	50	&	-	&	100	&	-	&	-	&	-	&	-	&	307	&	4	&	-	&	100	\\
20	&	20	&	100	&	2.81$\times10^3$	&	50	&	-	&	-	&	-	&	-	&	-	&	-	&	140	&	4	&	-	&	100	\\
20	&	40	&	100	&	1.85$\times10^3$	&	17	&	1.01$\times10^3$	&	7	&	228	&	4	&	3.45$\times10^3$	&	49	&	196	&	7	&	728	&	16	\\
40	&	5	&	100	&	2.99$\times10^3$	&	38	&	8.97$\times10^3$	&	38	&	1.40$\times10^4$	&	58	&	6.43$\times10^5$	&	58	&	3.60$\times10^4$	&	8	&	3.94$\times10^4$	&	10	\\
40	&	10	&	100	&	2.45$\times10^3$	&	24	&	1.47$\times10^4$	&	33	&	1.47$\times10^4$	&	41	&	-	&	100	&	131	&	4	&	-	&	-	\\
40	&	20	&	100	&	1.20$\times10^3$	&	24	&	1.05$\times10^4$	&	41	&	-	&	100	&	6.92$\times10^3$	&	70	&	233	&	7	&	498	&	13	\\
40	&	40	&	100	&	1.14$\times10^3$	&	13	&	811	&	6	&	195	&	4	&	8.05$\times10^3$	&	69	&	197	&	6	&	511	&	12	\\
60	&	5	&	100	&	585	&	17	&	5.23$\times10^3$	&	29	&	2.09$\times10^4$	&	71	&	-	&	98	&	1.24$\times10^3$	&	8	&	2.11$\times10^3$	&	13	\\
60	&	10	&	100	&	708	&	18	&	6.28$\times10^3$	&	32	&	1.40$\times10^4$	&	58	&	4.06$\times10^3$	&	49	&	269	&	7	&	535	&	13	\\
60	&	20	&	100	&	2.44$\times10^3$	&	24	&	645	&	7	&	251	&	5	&	3.32$\times10^3$	&	48	&	207	&	7	&	536	&	14	\\
60	&	40	&	100	&	851	&	14	&	737	&	7	&	183	&	5	&	1.55$\times10^3$	&	56	&	120	&	9	&	375	&	20	\\
80	&	5	&	100	&	789	&	25	&	7.85$\times10^3$	&	45	&	8.73$\times10^3$	&	58	&	-	&	95	&	370	&	8	&	612	&	13	\\
80	&	10	&	100	&	512	&	16	&	6.98$\times10^3$	&	33	&	1.05$\times10^4$	&	50	&	1.10$\times10^3$	&	48	&	236	&	13	&	609	&	25	\\
80	&	20	&	100	&	1.46$\times10^3$	&	18	&	576	&	7	&	196	&	5	&	2.22$\times10^3$	&	68	&	138	&	10	&	513	&	23	\\
80	&	40	&	100	&	791	&	14	&	569	&	7	&	189	&	5	&	1.45$\times10^3$	&	29	&	157	&	6	&	519	&	12		
		\end{tabular}%
}
\caption{i2HDM table of limits on the cross sections for A,B and C samples, for $1<m_{D1}<80$ GeV, $1<\Delta m^+<120$ GeV, $\Delta m^0=100$ GeV.}
\label{tab:i2hdm_exc_3}
\end{center}
\end{table}

\newpage

\section{MFDM Cross-Section Limits
\label{app:MFDM-exc}}
In appendix~\ref{app:MFDM-exc} we present the tables for MFDM cross-section limits for 2- and 3-lepton final state samples A, B and C. For $40<m_{D1}<80$ GeV, $1<\Delta m^+<280$ GeV,  tables~\ref{tab:mfdm_exc_1},\ref{tab:mfdm_exc_2},\ref{tab:mfdm_exc_3} show the results for $\Delta m^0<1$ GeV, tables~\ref{tab:mfdm_exc_4},\ref{tab:mfdm_exc_5},\ref{tab:mfdm_exc_6} show the results for $\Delta m^0<10$ GeV and tables ~\ref{tab:mfdm_exc_7}\ref{tab:mfdm_exc_8},\ref{tab:mfdm_exc_9} show the results for $\Delta m^0<100$ GeV
\begin{table}[!ht]
	\begin{center}
		\resizebox{\columnwidth}{!}{%
		\begin{tabular}{|l|l|l||l|l|l|l|l|l||l|l|l|l|l|l|}
			$m_{D1}$ &  $\Delta m^+$ & $\Delta m^0$ & 2$\ell$ $\sigma^{95}_A$ (fb) &$\frac{100}{\sqrt{N_{MC}}}$ & 2$\ell$ $\sigma^{95}_B$ (fb)&$\frac{100}{\sqrt{N_{MC}}}$ & 2$\ell$ $\sigma^{95}_C$ (fb)&$\frac{100}{\sqrt{N_{MC}}}$ & 3$\ell$ $\sigma^{95}_A$ (fb)&$\frac{100}{\sqrt{N_{MC}}}$ & 3$\ell$ $\sigma^{95}_B$ (fb)&$\frac{100}{\sqrt{N_{MC}}}$ & 3$\ell$ $\sigma^{95}_C$ (fb)&$\frac{100}{\sqrt{N_{MC}}}$ \\
			\hline \hline
40	&	10	&	1	&	975	&	41	&	2.93$\times10^3$	&	41	&	2.11$\times10^3$	&	42	&	-	&	-	&	-	&	-	&	-	&	-	\\
40	&	20	&	1	&	2.05$\times10^3$	&	35	&	-	&	100	&	3.59$\times10^3$	&	33	&	-	&	-	&	-	&	100	&	8.74$\times10^3$	&	44	\\
40	&	40	&	1	&	2.33$\times10^3$	&	20	&	8.37$\times10^4$	&	71	&	4.53$\times10^3$	&	20	&	-	&	-	&	2.05$\times10^4$	&	32	&	547	&	6	\\
40	&	60	&	1	&	3.91$\times10^3$	&	24	&	1.81$\times10^4$	&	30	&	728	&	7	&	-	&	-	&	1.37$\times10^4$	&	45	&	190	&	6	\\
40	&	80	&	1	&	2.66$\times10^3$	&	20	&	7.13$\times10^3$	&	19	&	460	&	6	&	-	&	-	&	4.69$\times10^3$	&	33	&	139	&	7	\\
40	&	100	&	1	&	665	&	10	&	9.98$\times10^4$	&	71	&	3.02$\times10^3$	&	15	&	-	&	-	&	-	&	-	&	177	&	8	\\
40	&	120	&	1	&	275	&	6	&	-	&	-	&	3.94$\times10^3$	&	16	&	-	&	-	&	-	&	-	&	93	&	6	\\
40	&	140	&	1	&	182	&	5	&	1.50$\times10^4$	&	26	&	2.45$\times10^3$	&	13	&	-	&	100	&	5.93$\times10^3$	&	45	&	45	&	5	\\
40	&	160	&	1	&	113	&	4	&	1.61$\times10^4$	&	27	&	1.63$\times10^3$	&	10	&	-	&	-	&	1.77$\times10^3$	&	20	&	27	&	3	\\
40	&	180	&	1	&	90	&	3	&	1.25$\times10^4$	&	24	&	1.45$\times10^3$	&	10	&	-	&	-	&	1.06$\times10^3$	&	15	&	18	&	2	\\
40	&	200	&	1	&	73	&	3	&	9.02$\times10^3$	&	20	&	1.14$\times10^3$	&	9	&	-	&	-	&	703	&	13	&	14	&	2	\\
40	&	220	&	1	&	59	&	4	&	7.46$\times10^3$	&	28	&	1.06$\times10^3$	&	13	&	-	&	100	&	510	&	16	&	12	&	3	\\
40	&	240	&	1	&	549	&	11	&	2.36$\times10^3$	&	14	&	90	&	3	&	-	&	100	&	454	&	15	&	10	&	3	\\
40	&	260	&	1	&	605	&	12	&	1.81$\times10^3$	&	12	&	65	&	3	&	-	&	100	&	258	&	11	&	9	&	3	\\
40	&	280	&	1	&	426	&	10	&	1.25$\times10^3$	&	10	&	52	&	2	&	-	&	-	&	348	&	13	&	8	&	2	\\
60	&	10	&	1	&	5.26$\times10^3$	&	58	&	1.18$\times10^4$	&	50	&	1.91$\times10^3$	&	25	&	-	&	-	&	-	&	-	&	-	&	-	\\
60	&	20	&	1	&	552	&	20	&	4.30$\times10^3$	&	32	&	1.70$\times10^3$	&	24	&	-	&	-	&	6.84$\times10^4$	&	58	&	1.76$\times10^4$	&	36	\\
60	&	40	&	1	&	956	&	26	&	1.08$\times10^4$	&	50	&	741	&	16	&	-	&	-	&	2.56$\times10^4$	&	35	&	498	&	6	\\
60	&	60	&	1	&	4.75$\times10^3$	&	27	&	1.66$\times10^4$	&	29	&	638	&	7	&	-	&	-	&	1.71$\times10^4$	&	50	&	158	&	6	\\
60	&	80	&	1	&	2.66$\times10^3$	&	20	&	9.98$\times10^3$	&	22	&	408	&	6	&	-	&	100	&	5.27$\times10^3$	&	28	&	117	&	5	\\
60	&	100	&	1	&	611	&	9	&	1.13$\times10^5$	&	71	&	1.98$\times10^4$	&	36	&	-	&	-	&	-	&	-	&	154	&	7	\\
60	&	120	&	1	&	233	&	6	&	-	&	100	&	2.81$\times10^3$	&	14	&	-	&	-	&	-	&	100	&	87	&	6	\\
60	&	140	&	1	&	135	&	4	&	1.61$\times10^4$	&	27	&	3.14$\times10^3$	&	14	&	-	&	-	&	2.33$\times10^3$	&	23	&	32	&	3	\\
60	&	160	&	1	&	98	&	4	&	1.41$\times10^4$	&	25	&	1.70$\times10^3$	&	11	&	-	&	-	&	1.27$\times10^3$	&	17	&	20	&	3	\\
60	&	180	&	1	&	82	&	3	&	6.83$\times10^3$	&	17	&	1.16$\times10^3$	&	9	&	-	&	100	&	1.39$\times10^3$	&	26	&	19	&	4	\\
60	&	200	&	1	&	642	&	12	&	3.05$\times10^3$	&	16	&	143	&	4	&	-	&	100	&	642	&	12	&	13	&	2	\\
60	&	220	&	1	&	605	&	12	&	2.28$\times10^3$	&	13	&	98	&	3	&	-	&	-	&	515	&	11	&	11	&	2	\\
60	&	240	&	1	&	522	&	11	&	1.57$\times10^3$	&	11	&	68	&	3	&	-	&	-	&	444	&	15	&	9	&	3	\\
60	&	260	&	1	&	421	&	10	&	1.29$\times10^3$	&	10	&	57	&	3	&	-	&	-	&	279	&	12	&	8	&	2	\\
60	&	280	&	1	&	469	&	11	&	1.07$\times10^3$	&	9	&	45	&	2	&	-	&	100	&	346	&	16	&	8	&	3	\\
80	&	10	&	1	&	1.46$\times10^3$	&	50	&	1.60$\times10^3$	&	30	&	387	&	18	&	-	&	-	&	-	&	-	&	-	&	-	\\
80	&	20	&	1	&	494	&	19	&	1.79$\times10^3$	&	20	&	1.87$\times10^3$	&	26	&	-	&	-	&	6.84$\times10^4$	&	58	&	9.72$\times10^3$	&	27	\\
80	&	40	&	1	&	736	&	13	&	1.32$\times10^4$	&	32	&	1.06$\times10^3$	&	11	&	3.42$\times10^4$	&	71	&	1.46$\times10^4$	&	27	&	462	&	6	\\
80	&	60	&	1	&	7.39$\times10^3$	&	33	&	9.07$\times10^3$	&	21	&	670	&	7	&	-	&	-	&	4.57$\times10^3$	&	26	&	155	&	6	\\
80	&	80	&	1	&	2.02$\times10^3$	&	17	&	6.24$\times10^3$	&	18	&	310	&	5	&	-	&	-	&	5.15$\times10^3$	&	58	&	73	&	8	\\
80	&	100	&	1	&	508	&	9	&	6.65$\times10^4$	&	58	&	1.68$\times10^3$	&	11	&	-	&	-	&	-	&	-	&	118	&	6	\\
80	&	120	&	1	&	228	&	6	&	-	&	100	&	3.54$\times10^3$	&	15	&	-	&	-	&	6.03$\times10^3$	&	38	&	85	&	6	\\
80	&	140	&	1	&	124	&	4	&	1.41$\times10^4$	&	25	&	2.37$\times10^3$	&	13	&	-	&	-	&	2.01$\times10^3$	&	21	&	26	&	3	\\
80	&	160	&	1	&	95	&	4	&	1.61$\times10^4$	&	27	&	1.45$\times10^3$	&	10	&	-	&	-	&	852	&	14	&	19	&	3	\\
80	&	180	&	1	&	75	&	3	&	9.02$\times10^3$	&	20	&	1.02$\times10^3$	&	8	&	-	&	-	&	764	&	13	&	14	&	2	\\
80	&	200	&	1	&	535	&	11	&	2.72$\times10^3$	&	15	&	123	&	4	&	-	&	-	&	746	&	19	&	12	&	3	\\
80	&	220	&	1	&	579	&	12	&	2.91$\times10^3$	&	15	&	86	&	3	&	-	&	-	&	348	&	13	&	10	&	3	\\
80	&	240	&	1	&	421	&	10	&	1.46$\times10^3$	&	11	&	61	&	3	&	-	&	100	&	346	&	9	&	9	&	2	\\
80	&	260	&	1	&	444	&	10	&	1.36$\times10^3$	&	10	&	49	&	2	&	-	&	-	&	252	&	11	&	9	&	2	\\
80	&	280	&	1	&	405	&	10	&	907	&	9	&	43	&	2	&	-	&	-	&	225	&	10	&	7	&	2	
					\end{tabular}%
				}
\caption{MFDM table of limits on the cross sections for A,B and C samples, for $40<m_{D1}<80$ GeV, $1<\Delta m^+<280$ GeV, $\Delta m^0=100$ GeV.}
\label{tab:mfdm_exc_1}
\end{center}
\end{table}
\clearpage

\begin{table}[!ht]
	\begin{center}
				\resizebox{\columnwidth}{!}{%
		\begin{tabular}{|l|l|l||l|l|l|l|l|l||l|l|l|l|l|l|}
			$m_{D1}$ &  $\Delta m^+$ & $\Delta m^0$ & 2$\ell$ $\sigma^{95}_A$ (fb) &$\frac{100}{\sqrt{N_{MC}}}$ & 2$\ell$ $\sigma^{95}_B$ (fb)&$\frac{100}{\sqrt{N_{MC}}}$ & 2$\ell$ $\sigma^{95}_C$ (fb)&$\frac{100}{\sqrt{N_{MC}}}$ & 3$\ell$ $\sigma^{95}_A$ (fb)&$\frac{100}{\sqrt{N_{MC}}}$ & 3$\ell$ $\sigma^{95}_B$ (fb)&$\frac{100}{\sqrt{N_{MC}}}$ & 3$\ell$ $\sigma^{95}_C$ (fb)&$\frac{100}{\sqrt{N_{MC}}}$ \\
			\hline \hline
100	&	10	&	1	&	731	&	35	&	1.46$\times10^3$	&	29	&	407	&	19	&	-	&	-	&	-	&	-	&	-	&	-	\\
100	&	20	&	1	&	974	&	41	&	2.51$\times10^3$	&	38	&	431	&	19	&	-	&	-	&	2.93$\times10^4$	&	38	&	1.32$\times10^4$	&	31	\\

100	&	40	&	1	&	713	&	13	&	1.10$\times10^4$	&	29	&	1.07$\times10^3$	&	11	&	-	&	-	&	2.56$\times10^4$	&	35	&	422	&	6	\\

100	&	60	&	1	&	3.33$\times10^3$	&	22	&	1.11$\times10^4$	&	24	&	497	&	6	&	-	&	-	&	7.61$\times10^3$	&	33	&	164	&	6	\\

100	&	80	&	1	&	2.38$\times10^3$	&	19	&	6.24$\times10^3$	&	18	&	328	&	5	&	-	&	100	&	1.16$\times10^3$	&	28	&	89	&	9	\\

100	&	100	&	1	&	459	&	8	&	3.99$\times10^4$	&	45	&	1.46$\times10^3$	&	10	&	-	&	-	&	-	&	100	&	118	&	6	\\

100	&	120	&	1	&	174	&	5	&	-	&	-	&	3.57$\times10^3$	&	15	&	-	&	100	&	-	&	100	&	40	&	5	\\

100	&	140	&	1	&	111	&	4	&	1.33$\times10^4$	&	24	&	1.73$\times10^3$	&	11	&	-	&	-	&	1.43$\times10^3$	&	18	&	25	&	3	\\

100	&	160	&	1	&	85	&	3	&	1.61$\times10^4$	&	27	&	1.45$\times10^3$	&	10	&	-	&	-	&	727	&	13	&	16	&	2	\\

100	&	180	&	1	&	542	&	11	&	5.22$\times10^3$	&	20	&	138	&	4	&	-	&	100	&	515	&	11	&	13	&	2	\\

100	&	200	&	1	&	673	&	13	&	2.61$\times10^3$	&	14	&	103	&	4	&	-	&	100	&	477	&	10	&	10	&	2	\\

100	&	220	&	1	&	485	&	11	&	2.02$\times10^3$	&	13	&	77	&	3	&	-	&	100	&	410	&	10	&	9	&	2	\\

100	&	240	&	1	&	474	&	11	&	1.38$\times10^3$	&	10	&	54	&	3	&	-	&	-	&	278	&	12	&	9	&	2	\\

100	&	260	&	1	&	401	&	10	&	963	&	9	&	46	&	2	&	-	&	-	&	301	&	8	&	7	&	2	\\

100	&	280	&	1	&	435	&	10	&	869	&	8	&	39	&	2	&	-	&	100	&	270	&	8	&	7	&	2	\\

120	&	10	&	1	&	1.46$\times10^3$	&	50	&	2.92$\times10^3$	&	41	&	428	&	19	&	-	&	-	&	-	&	-	&	-	&	-	\\
120	&	20	&	1	&	338	&	13	&	2.03$\times10^3$	&	18	&	880	&	14	&	-	&	-	&	2.58$\times10^5$	&	58	&	5.03$\times10^4$	&	31	\\

120	&	40	&	1	&	846	&	12	&	9.30$\times10^3$	&	24	&	695	&	8	&	-	&	100	&	1.37$\times10^4$	&	45	&	307	&	8	\\

120	&	60	&	1	&	765	&	12	&	1.86$\times10^4$	&	33	&	775	&	8	&	-	&	-	&	7.61$\times10^3$	&	33	&	131	&	5	\\

120	&	80	&	1	&	1.39$\times10^3$	&	14	&	8.31$\times10^3$	&	20	&	323	&	5	&	-	&	-	&	3.26$\times10^3$	&	22	&	83	&	4	\\

120	&	100	&	1	&	356	&	7	&	1.13$\times10^5$	&	71	&	6.54$\times10^3$	&	21	&	-	&	-	&	1.37$\times10^4$	&	35	&	114	&	4	\\

120	&	120	&	1	&	174	&	5	&	5.64$\times10^4$	&	50	&	2.46$\times10^3$	&	13	&	-	&	-	&	4.24$\times10^3$	&	38	&	40	&	5	\\

120	&	140	&	1	&	110	&	4	&	1.41$\times10^4$	&	25	&	1.91$\times10^3$	&	11	&	-	&	-	&	1.17$\times10^3$	&	16	&	21	&	3\\	

120	&	160	&	1	&	80	&	3	&	8.35$\times10^3$	&	19	&	1.69$\times10^3$	&	11	&	-	&	-	&	963	&	15	&	15	&	2\\	

120	&	180	&	1	&	480	&	11	&	4.04$\times10^3$	&	18	&	129	&	4	&	-	&	-	&	583	&	11	&	12	&	2\\	

120	&	200	&	1	&	542	&	11	&	2.24$\times10^3$	&	13	&	79	&	3	&	-	&	-	&	457	&	10	&	10	&	2\\	

120	&	220	&	1	&	435	&	10	&	1.49$\times10^3$	&	11	&	62	&	3	&	-	&	-	&	354	&	13	&	9	&	3\\	

120	&	240	&	1	&	528	&	11	&	985	&	9	&	49	&	2	&	-	&	100	&	258	&	11	&	8	&	2\\	

120	&	260	&	1	&	379	&	10	&	1.03$\times10^3$	&	9	&	40	&	2	&	3.48$\times10^3$	&	71	&	230	&	10	&	8	&	2\\	

120	&	280	&	1	&	444	&	10	&	802	&	8	&	35	&	2	&	-	&	100	&	211	&	10	&	7	&	2\\

140	&	10	&	1	&	1.95$\times10^3$	&	58	&	1.35$\times10^3$	&	28	&	429	&	19	&	-	&	-	&	-	&	-	&	-	&	-\\	
140	&	20	&	1	&	287	&	12	&	2.03$\times10^3$	&	18	&	544	&	11	&	-	&	-	&	2.93$\times10^4$	&	38	&	5.92$\times10^3$	&	21\\	

140	&	40	&	1	&	570	&	10	&	1.05$\times10^4$	&	25	&	731	&	8	&	3.42$\times10^4$	&	71	&	2.28$\times10^4$	&	33	&	400	&	5\\	

140	&	60	&	1	&	641	&	11	&	1.2$\times10^4$	&	27	&	801	&	8	&	-	&	-	&	4.57$\times10^3$	&	26	&	121	&	5\\	

140	&	80	&	1	&	1.58$\times10^3$	&	15	&	5.50$\times10^3$	&	17	&	292	&	5	&	-	&	-	&	3.61$\times10^3$	&	23	&	96	&	5\\	

140	&	100	&	1	&	367	&	7	&	4.51$\times10^4$	&	45	&	7.89$\times10^3$	&	23	&	-	&	-	&	1.41$\times10^4$	&	58	&	107	&	6	\\

140	&	120	&	1	&	159	&	5	&	7.52$\times10^4$	&	58	&	2.90$\times10^3$	&	14	&	-	&	100	&	5.93$\times10^3$	&	45	&	42	&	5	\\

140	&	140	&	1	&	104	&	4	&	1.50$\times10^4$	&	26	&	2.28$\times10^3$	&	12	&	-	&	-	&	1.27$\times10^3$	&	17	&	20	&	3	\\

140	&	160	&	1	&	73	&	3	&	8.35$\times10^3$	&	19	&	1.30$\times10^3$	&	9	&	-	&	-	&	852	&	14	&	15	&	2	\\

140	&	180	&	1	&	413	&	10	&	2.72$\times10^3$	&	15	&	102	&	3	&	-	&	-	&	561	&	11	&	11	&	2	\\

140	&	200	&	1	&	549	&	11	&	2.05$\times10^3$	&	13	&	75	&	3	&	-	&	100	&	443	&	10	&	10	&	2	\\

140	&	220	&	1	&	444	&	10	&	1.24$\times10^3$	&	10	&	56	&	3	&	-	&	100	&	389	&	9	&	9	&	2	\\

140	&	240	&	1	&	426	&	10	&	1.17$\times10^3$	&	10	&	44	&	2	&	-	&	100	&	1.42$\times10^3$	&	32	&	9	&	3	\\

140	&	260	&	1	&	357	&	9	&	797	&	8	&	39	&	2	&	-	&	-	&	312	&	8	&	7	&	2	\\

140	&	280	&	1	&	383	&	10	&	659	&	7	&	34	&	2	&	-	&	100	&	277	&	8	&	7	&	1	
				\end{tabular}%
}
\caption{MFDM table of limits on the cross sections for A,B and C samples, for $100<m_{D1}<140$ GeV, $1<\Delta m^+<280$ GeV, $\Delta m^0=1$ GeV.}
\label{tab:mfdm_exc_2}
\end{center}
\end{table}
\clearpage

\begin{table}[!ht]
	\begin{center}
		\resizebox{\columnwidth}{!}{%
			\begin{tabular}{|l|l|l||l|l|l|l|l|l||l|l|l|l|l|l|}
				$m_{D1}$ &  $\Delta m^+$ & $\Delta m^0$ & 2$\ell$ $\sigma^{95}_A$ (fb) &$\frac{100}{\sqrt{N_{MC}}}$ & 2$\ell$ $\sigma^{95}_B$ (fb)&$\frac{100}{\sqrt{N_{MC}}}$ & 2$\ell$ $\sigma^{95}_C$ (fb)&$\frac{100}{\sqrt{N_{MC}}}$ & 3$\ell$ $\sigma^{95}_A$ (fb)&$\frac{100}{\sqrt{N_{MC}}}$ & 3$\ell$ $\sigma^{95}_B$ (fb)&$\frac{100}{\sqrt{N_{MC}}}$ & 3$\ell$ $\sigma^{95}_C$ (fb)&$\frac{100}{\sqrt{N_{MC}}}$ \\
				\hline \hline
160	&	10	&	1	&	418	&	26	&	1.34$\times10^3$	&	27	&	416	&	18	&	-	&	-	&	-	&	-	&	-	&	-	\\
160	&	20	&	1	&	375	&	28	&	1.04$\times10^3$	&	27	&	445	&	21	&	-	&	-	&	2.56$\times10^4$	&	35	&	8.84$\times10^3$	&	25	\\

160	&	40	&	1	&	496	&	11	&	4.57$\times10^3$	&	19	&	614	&	8	&	-	&	-	&	1.71$\times10^4$	&	29	&	415	&	6	\\

160	&	60	&	1	&	672	&	11	&	8.37$\times10^3$	&	22	&	606	&	7	&	-	&	100	&	4.57$\times10^3$	&	26	&	107	&	5	\\

160	&	80	&	1	&	1.67$\times10^3$	&	20	&	7.36$\times10^3$	&	24	&	304	&	6	&	-	&	100	&	5.60$\times10^3$	&	45	&	106	&	8	\\

160	&	100	&	1	&	907	&	15	&	-	&	100	&	925	&	11	&	-	&	100	&	9.89$\times10^3$	&	58	&	72	&	6	\\

160	&	120	&	1	&	139	&	4	&	7.52$\times10^4$	&	58	&	3.01$\times10^3$	&	14	&	-	&	100	&	3.71$\times10^3$	&	35	&	39	&	4	\\

160	&	140	&	1	&	93	&	4	&	1.19$\times10^4$	&	23	&	1.74$\times10^3$	&	11	&	-	&	-	&	1.27$\times10^3$	&	17	&	19	&	3	\\

160	&	160	&	1	&	409	&	10	&	5.22$\times10^3$	&	20	&	154	&	4	&	-	&	100	&	607	&	12	&	13	&	2	\\

160	&	180	&	1	&	449	&	10	&	2.61$\times10^3$	&	14	&	96	&	3	&	-	&	100	&	439	&	10	&	11	&	2	\\

160	&	200	&	1	&	397	&	10	&	1.55$\times10^3$	&	11	&	68	&	3	&	-	&	-	&	418	&	10	&	9	&	2	\\

160	&	220	&	1	&	417	&	10	&	1.28$\times10^3$	&	10	&	51	&	2	&	-	&	100	&	418	&	17	&	10	&	3	\\

160	&	240	&	1	&	444	&	10	&	1.03$\times10^3$	&	9	&	42	&	2	&	-	&	-	&	290	&	8	&	7	&	2	\\

160	&	260	&	1	&	366	&	9	&	772	&	8	&	37	&	2	&	-	&	100	&	211	&	10	&	7	&	2	\\

160	&	280	&	1	&	417	&	10	&	792	&	8	&	31	&	2	&	-	&	100	&	221	&	7	&	6	&	1	\\

180	&	10	&	1	&	570	&	30	&	723	&	20	&	537	&	21	&	-	&	-	&	-	&	-	&	-	&	-	\\
180	&	20	&	1	&	913	&	24	&	1.20$\times10^3$	&	16	&	687	&	14	&	-	&	-	&	-	&	-	&	-	&	-	\\

180	&	40	&	1	&	537	&	10	&	5.58$\times10^3$	&	18	&	442	&	6	&	-	&	-	&	2.05$\times10^4$	&	32	&	440	&	6
					\end{tabular}%
				}
\caption{MFDM table of limits on the cross sections for A,B and C samples, for $160<m_{D1}<180$ GeV, $1<\Delta m^+<280$ GeV, $\Delta m^0=1$ GeV.}
\label{tab:mfdm_exc_3}
\end{center}
\end{table}
\clearpage

\begin{table}[!ht]
	\begin{center}
		\resizebox{\columnwidth}{!}{%
			\begin{tabular}{|l|l|l||l|l|l|l|l|l||l|l|l|l|l|l|}
				$m_{D1}$ &  $\Delta m^+$ & $\Delta m^0$ & 2$\ell$ $\sigma^{95}_A$ (fb) &$\frac{100}{\sqrt{N_{MC}}}$ & 2$\ell$ $\sigma^{95}_B$ (fb)&$\frac{100}{\sqrt{N_{MC}}}$ & 2$\ell$ $\sigma^{95}_C$ (fb)&$\frac{100}{\sqrt{N_{MC}}}$ & 3$\ell$ $\sigma^{95}_A$ (fb)&$\frac{100}{\sqrt{N_{MC}}}$ & 3$\ell$ $\sigma^{95}_B$ (fb)&$\frac{100}{\sqrt{N_{MC}}}$ & 3$\ell$ $\sigma^{95}_C$ (fb)&$\frac{100}{\sqrt{N_{MC}}}$ \\
				\hline \hline
40	&	10	&	10	&	975	&	41	&	8.77$\times10^3$	&	71	&	2.39$\times10^3$	&	45	&	-	&	-	&	-	&	-	&	-	&	-\\
40	&	20	&	10	&	1.43$\times10^3$	&	32	&	1.43$\times10^4$	&	58	&	1.30$\times10^3$	&	21	&	-	&	-	&	5.13$\times10^4$	&	50	&	9.65$\times10^3$	&	27\\

40	&	40	&	10	&	1.92$\times10^3$	&	21	&	1.66$\times10^4$	&	35	&	3.70$\times10^3$	&	20	&	-	&	-	&	4.77$\times10^3$	&	15	&	1.15$\times10^3$	&	9\\

40	&	60	&	10	&	821	&	11	&	5.12$\times10^3$	&	16	&	1.86$\times10^3$	&	12	&	-	&	-	&	2.08$\times10^3$	&	17	&	461	&	10\\

40	&	80	&	10	&	438	&	8	&	2.73$\times10^3$	&	12	&	1.14$\times10^3$	&	9	&	-	&	-	&	1.67$\times10^3$	&	16	&	287	&	8\\

40	&	100	&	10	&	731	&	10	&	1.25$\times10^4$	&	25	&	5.06$\times10^3$	&	19	&	-	&	-	&	7.03$\times10^3$	&	41	&	304	&	10\\

40	&	120	&	10	&	366	&	7	&	4.87$\times10^3$	&	16	&	1.99$\times10^3$	&	12	&	-	&	-	&	-	&	100	&	90	&	6\\

40	&	140	&	10	&	211	&	6	&	2.63$\times10^3$	&	11	&	1.23$\times10^3$	&	10	&	-	&	-	&	5.93$\times10^3$	&	45	&	44	&	5\\

40	&	160	&	10	&	127	&	4	&	3.22$\times10^4$	&	38	&	2.32$\times10^3$	&	12	&	-	&	-	&	4.03$\times10^3$	&	30	&	24	&	3\\

40	&	180	&	10	&	103	&	4	&	2.25$\times10^4$	&	32	&	1.41$\times10^3$	&	10	&	-	&	-	&	3.16$\times10^3$	&	27	&	18	&	2\\

40	&	200	&	10	&	564	&	12	&	1.51$\times10^3$	&	11	&	217	&	5	&	-	&	-	&	2.11$\times10^3$	&	22	&	15	&	2\\

40	&	220	&	10	&	435	&	10	&	1.41$\times10^3$	&	11	&	133	&	4	&	-	&	-	&	1.23$\times10^3$	&	24	&	13	&	3\\

40	&	240	&	10	&	379	&	10	&	1.18$\times10^3$	&	10	&	86	&	3	&	-	&	-	&	870	&	20	&	11	&	3\\

40	&	260	&	10	&	413	&	10	&	1.03$\times10^3$	&	9	&	66	&	3	&	-	&	100	&	949	&	21	&	9	&	3\\

40	&	280	&	10	&	369	&	9	&	818	&	8	&	50	&	2	&	-	&	-	&	803	&	20	&	9	&	2\\

60	&	10	&	10	&	532	&	30	&	1.60$\times10^3$	&	30	&	415	&	19	&	-	&	-	&	-	&	-	&	-	&	-\\
60	&	20	&	10	&	749	&	26	&	2.59$\times10^3$	&	28	&	777	&	19	&	-	&	-	&	4.10$\times10^4$	&	45	&	8.16$\times10^3$	&	24\\

60	&	40	&	10	&	597	&	20	&	4.78$\times10^3$	&	33	&	1.08$\times10^3$	&	19	&	-	&	-	&	3.80$\times10^3$	&	14	&	1.06$\times10^3$	&	9\\

60	&	60	&	10	&	594	&	9	&	5.12$\times10^3$	&	16	&	1.90$\times10^3$	&	12	&	-	&	-	&	2.86$\times10^3$	&	20	&	375	&	9\\

60	&	80	&	10	&	370	&	7	&	3.02$\times10^3$	&	12	&	1.06$\times10^3$	&	9	&	-	&	100	&	1.85$\times10^3$	&	16	&	295	&	8\\

60	&	100	&	10	&	537	&	9	&	2.22$\times10^4$	&	33	&	2.82$\times10^3$	&	15	&	-	&	-	&	4.69$\times10^3$	&	33	&	288	&	10\\

60	&	120	&	10	&	288	&	6	&	2.51$\times10^4$	&	33	&	7.69$\times10^3$	&	23	&	-	&	-	&	4.69$\times10^3$	&	33	&	80	&	5\\

60	&	140	&	10	&	171	&	5	&	1.61$\times10^4$	&	27	&	2.15$\times10^3$	&	12	&	-	&	-	&	3.71$\times10^3$	&	35	&	39	&	4\\

60	&	160	&	10	&	122	&	4	&	1.61$\times10^4$	&	27	&	1.56$\times10^3$	&	10	&	-	&	100	&	1.23$\times10^3$	&	17	&	22	&	3\\

60	&	180	&	10	&	95	&	4	&	1.13$\times10^4$	&	22	&	1.43$\times10^3$	&	10	&	-	&	100	&	1.14$\times10^3$	&	16	&	15	&	2\\

60	&	200	&	10	&	74	&	3	&	9.81$\times10^3$	&	21	&	1.08$\times10^3$	&	8	&	-	&	-	&	746	&	19	&	13	&	3\\

60	&	220	&	10	&	409	&	10	&	2.05$\times10^3$	&	13	&	104	&	4	&	-	&	100	&	444	&	15	&	12	&	3\\

60	&	240	&	10	&	426	&	10	&	1.79$\times10^3$	&	12	&	70	&	3	&	-	&	-	&	354	&	13	&	10	&	3\\

60	&	260	&	10	&	348	&	9	&	1.51$\times10^3$	&	11	&	59	&	3	&	-	&	-	&	271	&	11	&	8	&	2\\

60	&	280	&	10	&	316	&	9	&	1.04$\times10^3$	&	9	&	48	&	2	&	-	&	100	&	444	&	18	&	8	&	3\\

80	&	10	&	10	&	450	&	28	&	1.25$\times10^3$	&	27	&	311	&	16	&	-	&	-	&	-	&	-	&	-	&	-\\
80	&	20	&	10	&	388	&	14	&	2.99$\times10^3$	&	22	&	758	&	13	&	-	&	-	&	3.42$\times10^4$	&	41	&	1.53$\times10^4$	&	33\\

80	&	40	&	10	&	960	&	15	&	7.80$\times10^3$	&	24	&	1.68$\times10^3$	&	14	&	-	&	-	&	2.97$\times10^3$	&	12	&	909	&	8\\

80	&	60	&	10	&	665	&	10	&	4.64$\times10^3$	&	15	&	1.85$\times10^3$	&	12	&	-	&	-	&	1.90$\times10^3$	&	17	&	361	&	9\\

80	&	80	&	10	&	368	&	7	&	2.89$\times10^3$	&	12	&	898	&	8	&	-	&	100	&	1.12$\times10^3$	&	13	&	259	&	8\\

80	&	100	&	10	&	447	&	8	&	3.33$\times10^4$	&	41	&	2.98$\times10^3$	&	15	&	-	&	-	&	3.84$\times10^3$	&	30	&	271	&	10\\

80	&	120	&	10	&	215	&	5	&	2.25$\times10^4$	&	32	&	4.42$\times10^3$	&	17	&	-	&	-	&	2.81$\times10^3$	&	26	&	85	&	6\\

80	&	140	&	10	&	139	&	4	&	1.50$\times10^4$	&	26	&	2.08$\times10^3$	&	12	&	-	&	-	&	1.70$\times10^3$	&	20	&	27	&	3\\

80	&	160	&	10	&	108	&	4	&	1.07$\times10^4$	&	22	&	1.48$\times10^3$	&	10	&	-	&	-	&	923	&	14	&	18	&	2\\

80	&	180	&	10	&	89	&	3	&	1.19$\times10^4$	&	23	&	1.32$\times10^3$	&	9	&	-	&	-	&	561	&	11	&	13	&	2\\

80	&	200	&	10	&	435	&	10	&	3.13$\times10^3$	&	16	&	116	&	4	&	-	&	100	&	522	&	16	&	13	&	3\\

80	&	220	&	10	&	360	&	9	&	2.50$\times10^3$	&	14	&	87	&	3	&	-	&	-	&	399	&	9	&	10	&	2\\

80	&	240	&	10	&	363	&	9	&	1.32$\times10^3$	&	10	&	60	&	3	&	-	&	-	&	326	&	12	&	9	&	2\\

80	&	260	&	10	&	342	&	9	&	1.13$\times10^3$	&	9	&	46	&	2	&	-	&	100	&	342	&	13	&	8	&	2\\

80	&	280	&	10	&	311	&	9	&	894	&	8	&	43	&	2	&	-	&	-	&	308	&	8	&	7	&	2
					\end{tabular}%
}
\caption{MFDM table of limits on the cross sections for A,B and C samples, for $40<m_{D1}<80$ GeV, $1<\Delta m^+<280$ GeV, $\Delta m^0=10$ GeV.}
\label{tab:mfdm_exc_4}
\end{center}
\end{table}
\clearpage

\begin{table}[!ht]
	\begin{center}
		\resizebox{\columnwidth}{!}{%
			\begin{tabular}{|l|l|l||l|l|l|l|l|l||l|l|l|l|l|l|}
				$m_{D1}$ &  $\Delta m^+$ & $\Delta m^0$ & 2$\ell$ $\sigma^{95}_A$ (fb) &$\frac{100}{\sqrt{N_{MC}}}$ & 2$\ell$ $\sigma^{95}_B$ (fb)&$\frac{100}{\sqrt{N_{MC}}}$ & 2$\ell$ $\sigma^{95}_C$ (fb)&$\frac{100}{\sqrt{N_{MC}}}$ & 3$\ell$ $\sigma^{95}_A$ (fb)&$\frac{100}{\sqrt{N_{MC}}}$ & 3$\ell$ $\sigma^{95}_B$ (fb)&$\frac{100}{\sqrt{N_{MC}}}$ & 3$\ell$ $\sigma^{95}_C$ (fb)&$\frac{100}{\sqrt{N_{MC}}}$ \\
				\hline \hline
100	&	10	&	10	&	532	&	30	&	1.35$\times10^3$	&	28	&	257	&	15	&	-	&	-	&	-	&	-	&	-	&	-\\
100	&	20	&	10	&	295	&	12	&	1.37$\times10^3$	&	15	&	598	&	12	&	-	&	-	&	4.10$\times10^4$	&	45	&	1.31$\times10^4$	&	31\\

100	&	40	&	10	&	789	&	13	&	3.58$\times10^3$	&	16	&	1.46$\times10^3$	&	13	&	-	&	100	&	3.66$\times10^3$	&	13	&	1.10$\times10^3$	&	9\\

100	&	60	&	10	&	821	&	12	&	5.40$\times10^3$	&	18	&	1.93$\times10^3$	&	13	&	-	&	-	&	1.49$\times10^3$	&	15	&	372	&	9\\

100	&	80	&	10	&	321	&	7	&	2.85$\times10^3$	&	12	&	947	&	8	&	-	&	100	&	1.40$\times10^3$	&	14	&	256	&	7\\

100	&	100	&	10	&	993	&	15	&	1.56$\times10^4$	&	35	&	2.69$\times10^3$	&	18	&	-	&	-	&	3.84$\times10^3$	&	30	&	236	&	9\\

100	&	120	&	10	&	250	&	6	&	1.41$\times10^4$	&	25	&	3.1$\times10^3$	&	14	&	-	&	-	&	3.01$\times10^3$	&	27	&	66	&	5\\

100	&	140	&	10	&	127	&	4	&	1.88$\times10^4$	&	29	&	1.99$\times10^3$	&	12	&	-	&	-	&	2.01$\times10^3$	&	21	&	24	&	3\\

100	&	160	&	10	&	96	&	4	&	1.25$\times10^4$	&	24	&	1.5$\times10^3$	&	10	&	-	&	-	&	886	&	14	&	16	&	2\\

100	&	180	&	10	&	401	&	10	&	4.47$\times10^3$	&	19	&	149	&	4	&	-	&	-	&	607	&	12	&	13	&	2\\

100	&	200	&	10	&	405	&	10	&	2.41$\times10^3$	&	14	&	100	&	3	&	-	&	-	&	435	&	14	&	12	&	3\\

100	&	220	&	10	&	430	&	10	&	1.65$\times10^3$	&	11	&	67	&	3	&	3.48$\times10^3$	&	71	&	674	&	18	&	10	&	3\\

100	&	240	&	10	&	345	&	9	&	1.11$\times10^3$	&	9	&	54	&	3	&	-	&	100	&	324	&	9	&	8	&	2\\

100	&	260	&	10	&	302	&	9	&	1.03$\times10^3$	&	9	&	43	&	2	&	-	&	100	&	278	&	12	&	8	&	2\\

100	&	280	&	10	&	314	&	9	&	802	&	8	&	38	&	2	&	-	&	-	&	275	&	8	&	7	&	2\\

120	&	10	&	10	&	292	&	22	&	1.03$\times10^3$	&	24	&	236	&	14	&	-	&	-	&	-	&	-	&	-	&	-\\
120	&	20	&	10	&	390	&	26	&	1.95$\times10^3$	&	33	&	359	&	18	&	-	&	100	&	-	&	-	&	2.16$\times10^3$	&	34\\

120	&	40	&	10	&	614	&	12	&	3.68$\times10^3$	&	17	&	1.26$\times10^3$	&	12	&	-	&	-	&	3.31$\times10^3$	&	13	&	1.14$\times10^3$	&	9\\

120	&	60	&	10	&	915	&	13	&	3.80$\times10^3$	&	15	&	1.61$\times10^3$	&	12	&	-	&	-	&	1.46$\times10^3$	&	15	&	312	&	8\\

120	&	80	&	10	&	285	&	7	&	3.12$\times10^3$	&	12	&	748	&	7	&	-	&	-	&	1.34$\times10^3$	&	14	&	223	&	7\\

120	&	100	&	10	&	888	&	15	&	1.14$\times10^4$	&	30	&	1.67$\times10^3$	&	14	&	-	&	-	&	2.34$\times10^3$	&	24	&	176	&	8\\

120	&	120	&	10	&	192	&	5	&	2.26$\times10^4$	&	32	&	2.33$\times10^3$	&	12	&	-	&	-	&	2.12$\times10^3$	&	27	&	39	&	4\\

120	&	140	&	10	&	122	&	4	&	1.61$\times10^4$	&	27	&	1.51$\times10^3$	&	10	&	-	&	-	&	1.08$\times10^3$	&	16	&	20	&	3\\

120	&	160	&	10	&	89	&	3	&	9.40$\times10^3$	&	20	&	1.49$\times10^3$	&	10	&	-	&	-	&	662	&	12	&	15	&	2\\

120	&	180	&	10	&	401	&	10	&	2.91$\times10^3$	&	15	&	134	&	4	&	-	&	100	&	674	&	18	&	13	&	3\\

120	&	200	&	10	&	366	&	9	&	2.24$\times10^3$	&	13	&	87	&	3	&	-	&	-	&	467	&	10	&	11	&	2\\

120	&	220	&	10	&	319	&	9	&	1.35$\times10^3$	&	10	&	62	&	3	&	-	&	100	&	380	&	13	&	10	&	3\\

120	&	240	&	10	&	296	&	8	&	963	&	9	&	50	&	2	&	2.37$\times10^3$	&	71	&	430	&	17	&	9	&	3\\

120	&	260	&	10	&	319	&	9	&	829	&	8	&	41	&	2	&	-	&	100	&	312	&	8	&	7	&	2\\

120	&	280	&	10	&	271	&	8	&	808	&	8	&	36	&	2	&	-	&	100	&	252	&	11	&	7	&	2\\

140	&	10	&	10	&	482	&	28	&	1.57$\times10^3$	&	29	&	295	&	15	&	-	&	-	&	-	&	-	&	-	&	-\\
140	&	20	&	10	&	223	&	10	&	1.18$\times10^3$	&	14	&	448	&	10	&	-	&	100	&	3.73$\times10^4$	&	58	&	1.26$\times10^4$	&	41\\

140	&	40	&	10	&	293	&	16	&	4.77$\times10^3$	&	38	&	773	&	19	&	-	&	-	&	3.81$\times10^3$	&	24	&	566	&	11\\

140	&	60	&	10	&	664	&	11	&	4.29$\times10^3$	&	16	&	1.13$\times10^3$	&	10	&	-	&	-	&	1.34$\times10^3$	&	14	&	270	&	8\\

140	&	80	&	10	&	298	&	7	&	2.17$\times10^3$	&	10	&	835	&	8	&	2.51$\times10^3$	&	71	&	884	&	24	&	202	&	14\\

140	&	100	&	10	&	787	&	14	&	9.63$\times10^3$	&	28	&	1.44$\times10^3$	&	13	&	-	&	100	&	2.81$\times10^3$	&	26	&	197	&	8\\

140	&	120	&	10	&	176	&	5	&	2.05$\times10^4$	&	30	&	2.97$\times10^3$	&	14	&	-	&	-	&	2.61$\times10^3$	&	24	&	32	&	3\\

140	&	140	&	10	&	110	&	4	&	1.07$\times10^4$	&	22	&	2.24$\times10^3$	&	12	&	-	&	100	&	1.64$\times10^3$	&	19	&	19	&	3\\

140	&	160	&	10	&	363	&	9	&	5.22$\times10^3$	&	20	&	173	&	5	&	-	&	-	&	726	&	13	&	14	&	2\\

140	&	180	&	10	&	345	&	9	&	2.32$\times10^3$	&	14	&	101	&	3	&	-	&	100	&	486	&	15	&	12	&	3\\

140	&	200	&	10	&	326	&	9	&	1.76$\times10^3$	&	12	&	76	&	3	&	-	&	-	&	403	&	10	&	9	&	2\\

140	&	220	&	10	&	309	&	9	&	1.28$\times10^3$	&	10	&	54	&	3	&	-	&	100	&	282	&	12	&	9	&	3\\

140	&	240	&	10	&	292	&	8	&	1.06$\times10^3$	&	9	&	46	&	2	&	-	&	100	&	364	&	16	&	9	&	3\\

140	&	260	&	10	&	307	&	9	&	818	&	8	&	38	&	2	&	7.39$\times10^3$	&	71	&	270	&	8	&	7	&	2\\

140	&	280	&	10	&	243	&	8	&	763	&	8	&	32	&	2	&	-	&	100	&	203	&	10	&	7	&	2
					\end{tabular}%
}
\caption{MFDM table of limits on the cross sections for A,B and C samples, for $100<m_{D1}<140$ GeV, $1<\Delta m^+<280$ GeV, $\Delta m^0=10$ GeV.}
\label{tab:mfdm_exc_5}
\end{center}
\end{table}
\clearpage

\begin{table}[!ht]
	\begin{center}
		\resizebox{\columnwidth}{!}{%
			\begin{tabular}{|l|l|l||l|l|l|l|l|l||l|l|l|l|l|l|}
				$m_{D1}$ &  $\Delta m^+$ & $\Delta m^0$ & 2$\ell$ $\sigma^{95}_A$ (fb) &$\frac{100}{\sqrt{N_{MC}}}$ & 2$\ell$ $\sigma^{95}_B$ (fb)&$\frac{100}{\sqrt{N_{MC}}}$ & 2$\ell$ $\sigma^{95}_C$ (fb)&$\frac{100}{\sqrt{N_{MC}}}$ & 3$\ell$ $\sigma^{95}_A$ (fb)&$\frac{100}{\sqrt{N_{MC}}}$ & 3$\ell$ $\sigma^{95}_B$ (fb)&$\frac{100}{\sqrt{N_{MC}}}$ & 3$\ell$ $\sigma^{95}_C$ (fb)&$\frac{100}{\sqrt{N_{MC}}}$ \\
				\hline \hline
160	&	10	&	10	&	418	&	26	&	783	&	20	&	295	&	15	&	-	&	-	&	-	&	-	&	-	&	-\\
160	&	20	&	10	&	241	&	11	&	951	&	12	&	325	&	9	&	-	&	100	&	6.84$\times10^4$	&	58	&	1.15$\times10^4$	&	29\\

160	&	40	&	10	&	469	&	9	&	3.28$\times10^3$	&	14	&	1.04$\times10^3$	&	10	&	-	&	-	&	3.66$\times10^3$	&	13	&	805	&	8\\

160	&	60	&	10	&	641	&	11	&	4.40$\times10^3$	&	16	&	1.32$\times10^3$	&	11	&	-	&	-	&	2.01$\times10^3$	&	17	&	344	&	9\\

160	&	80	&	10	&	596	&	12	&	3.48$\times10^3$	&	17	&	812	&	10	&	-	&	100	&	1.10$\times10^3$	&	27	&	184	&	13\\

160	&	100	&	10	&	787	&	14	&	9.63$\times10^3$	&	28	&	1.35$\times10^3$	&	13	&	7.03$\times10^3$	&	71	&	2.22$\times10^3$	&	23	&	162	&	8\\

160	&	120	&	10	&	160	&	5	&	1.88$\times10^4$	&	29	&	1.77$\times10^3$	&	11	&	-	&	-	&	2.11$\times10^3$	&	22	&	31	&	3\\

160	&	140	&	10	&	360	&	9	&	6.26$\times10^3$	&	22	&	260	&	6	&	-	&	-	&	1.06$\times10^3$	&	15	&	19	&	3\\

160	&	160	&	10	&	363	&	9	&	3.79$\times10^3$	&	17	&	146	&	4	&	-	&	100	&	653	&	18	&	14	&	3\\

160	&	180	&	10	&	360	&	9	&	2.85$\times10^3$	&	15	&	100	&	3	&	-	&	100	&	536	&	16	&	12	&	3\\

160	&	200	&	10	&	348	&	9	&	1.51$\times10^3$	&	11	&	61	&	3	&	-	&	-	&	382	&	9	&	9	&	2\\

160	&	220	&	10	&	294	&	8	&	1.10$\times10^3$	&	9	&	48	&	2	&	-	&	100	&	252	&	11	&	9	&	2\\

160	&	240	&	10	&	280	&	8	&	1.02$\times10^3$	&	9	&	43	&	2	&	-	&	100	&	338	&	9	&	7	&	2\\

160	&	260	&	10	&	248	&	8	&	703	&	7	&	36	&	2	&	3.48$\times10^3$	&	71	&	211	&	10	&	7	&	2\\

160	&	280	&	10	&	251	&	8	&	732	&	8	&	32	&	2	&	-	&	-	&	253	&	8	&	6	&	1\\

180	&	10	&	10	&	241	&	20	&	940	&	22	&	286	&	15	&	-	&	-	&	-	&	-	&	-	&	-\\
180	&	20	&	10	&	325	&	26	&	770	&	23	&	263	&	16	&	-	&	-	&	-	&	-	&	-	&	-\\

180	&	40	&	10	&	309	&	17	&	1.97$\times10^3$	&	24	&	646	&	17	&	-	&	100	&	4.19$\times10^3$	&	14	&	933	&	8
					\end{tabular}%
}
\caption{MFDM table of limits on the cross sections for A,B and C samples, for $160<m_{D1}<180$ GeV, $1<\Delta m^+<280$ GeV, $\Delta m^0=10$ GeV.}
\label{tab:mfdm_exc_6}
\end{center}
\end{table}
\clearpage

\begin{table}[!ht]
	\begin{center}
		\resizebox{\columnwidth}{!}{%
			\begin{tabular}{|l|l|l||l|l|l|l|l|l||l|l|l|l|l|l|}
				$m_{D1}$ &  $\Delta m^+$ & $\Delta m^0$ & 2$\ell$ $\sigma^{95}_A$ (fb) &$\frac{100}{\sqrt{N_{MC}}}$ & 2$\ell$ $\sigma^{95}_B$ (fb)&$\frac{100}{\sqrt{N_{MC}}}$ & 2$\ell$ $\sigma^{95}_C$ (fb)&$\frac{100}{\sqrt{N_{MC}}}$ & 3$\ell$ $\sigma^{95}_A$ (fb)&$\frac{100}{\sqrt{N_{MC}}}$ & 3$\ell$ $\sigma^{95}_B$ (fb)&$\frac{100}{\sqrt{N_{MC}}}$ & 3$\ell$ $\sigma^{95}_C$ (fb)&$\frac{100}{\sqrt{N_{MC}}}$ \\
				\hline \hline
40	&	10	&	100	&	73	&	35	&	1.25$\times10^3$	&	27	&	710	&	25	&	-	&	-	&	-	&	-	&	-	&	-\\
40	&	20	&	100	&	1.83$\times10^3$	&	33	&	9.86$\times10^3$	&	45	&	2.96$\times10^3$	&	30	&	-	&	-	&	9.77$\times10^3$	&	22	&	5.13$\times10^3$	&	19\\

40	&	40	&	100	&	1.53$\times10^3$	&	30	&	1.69$\times10^4$	&	58	&	3.60$\times10^3$	&	33	&	-	&	-	&	2.21$\times10^3$	&	18	&	513	&	11\\

40	&	60	&	100	&	377	&	8	&	2.66$\times10^3$	&	12	&	1.16$\times10^3$	&	9	&	-	&	100	&	1.35$\times10^3$	&	8	&	274	&	4\\

40	&	80	&	100	&	197	&	5	&	2.85$\times10^3$	&	12	&	540	&	6	&	1.14$\times10^4$	&	71	&	816	&	11	&	156	&	6\\

40	&	100	&	100	&	532	&	9	&	5.70$\times10^3$	&	17	&	2.47$\times10^3$	&	14	&	-	&	-	&	767	&	13	&	196	&	8\\

40	&	120	&	100	&	354	&	7	&	3.91$\times10^3$	&	14	&	2.14$\times10^3$	&	13	&	-	&	-	&	715	&	13	&	128	&	7\\

40	&	140	&	100	&	282	&	7	&	2.98$\times10^3$	&	12	&	1.82$\times10^3$	&	12	&	-	&	-	&	594	&	12	&	102	&	6\\

40	&	160	&	100	&	189	&	5	&	2.98$\times10^3$	&	12	&	1.38$\times10^3$	&	10	&	-	&	-	&	212	&	7	&	42	&	4\\

40	&	180	&	100	&	141	&	5	&	1.90$\times10^3$	&	10	&	903	&	8	&	-	&	-	&	198	&	7	&	30	&	3\\

40	&	200	&	100	&	318	&	9	&	1.44$\times10^3$	&	11	&	276	&	6	&	-	&	100	&	187	&	9	&	24	&	4\\

40	&	220	&	100	&	270	&	8	&	875	&	8	&	195	&	5	&	-	&	100	&	192	&	10	&	20	&	4\\

40	&	240	&	100	&	228	&	7	&	970	&	9	&	121	&	4	&	-	&	100	&	173	&	9	&	17	&	3\\

40	&	260	&	100	&	214	&	7	&	691	&	7	&	106	&	4	&	-	&	99	&	161	&	6	&	13	&	2\\

40	&	280	&	100	&	181	&	7	&	658	&	7	&	81	&	3	&	-	&	100	&	177	&	6	&	12	&	2\\

60	&	10	&	100	&	390	&	26	&	675	&	20	&	466	&	20	&	-	&	-	&	-	&	-	&	-	&	-\\
60	&	20	&	100	&	731	&	35	&	1.46$\times10^3$	&	29	&	573	&	22	&	-	&	-	&	8.54$\times10^3$	&	20	&	5.02$\times10^3$	&	19\\

60	&	40	&	100	&	920	&	14	&	8.83$\times10^3$	&	26	&	1.56$\times10^3$	&	13	&	-	&	-	&	3.31$\times10^3$	&	13	&	516	&	6\\

60	&	60	&	100	&	292	&	7	&	3.77$\times10^3$	&	14	&	820	&	8	&	-	&	-	&	1.05$\times10^3$	&	12	&	177	&	6\\

60	&	80	&	100	&	145	&	5	&	2.98$\times10^3$	&	12	&	485	&	6	&	-	&	-	&	845	&	11	&	157	&	6\\

60	&	100	&	100	&	416	&	8	&	3.70$\times10^3$	&	14	&	1.77$\times10^3$	&	12	&	-	&	100	&	1.06$\times10^3$	&	16	&	247	&	9\\

60	&	120	&	100	&	368	&	7	&	3.99$\times10^3$	&	14	&	2.35$\times10^3$	&	13	&	7.03$\times10^3$	&	71	&	603	&	12	&	135	&	7\\

60	&	140	&	100	&	232	&	6	&	2.89$\times10^3$	&	12	&	1.55$\times10^3$	&	11	&	-	&	100	&	326	&	10	&	60	&	6\\

60	&	160	&	100	&	161	&	5	&	2.77$\times10^3$	&	12	&	1.24$\times10^3$	&	10	&	-	&	100	&	193	&	7	&	35	&	3\\

60	&	180	&	100	&	366	&	9	&	1.29$\times10^3$	&	10	&	326	&	6	&	-	&	100	&	170	&	6	&	28	&	3\\

60	&	200	&	100	&	300	&	8	&	1.32$\times10^3$	&	10	&	222	&	5	&	-	&	-	&	166	&	9	&	21	&	4\\

60	&	220	&	100	&	216	&	7	&	1.16$\times10^3$	&	10	&	148	&	4	&	-	&	-	&	151	&	9	&	17	&	4\\

60	&	240	&	100	&	210	&	7	&	875	&	8	&	107	&	4	&	7.38$\times10^3$	&	71	&	174	&	6	&	14	&	2\\

60	&	260	&	100	&	196	&	7	&	652	&	7	&	79	&	3	&	2.37$\times10^3$	&	71	&	197	&	12	&	17	&	4\\

60	&	280	&	100	&	160	&	6	&	645	&	7	&	65	&	3	&	-	&	-	&	144	&	8	&	11	&	3\\

80	&	10	&	100	&	418	&	27	&	924	&	23	&	203	&	13	&	-	&	-	&	-	&	-	&	-	&	-\\
80	&	20	&	100	&	266	&	21	&	2.51$\times10^3$	&	38	&	404	&	19	&	-	&	-	&	1.14$\times10^4$	&	24	&	3.61$\times10^3$	&	16\\

80	&	40	&	100	&	736	&	13	&	5.76$\times10^3$	&	21	&	1.26$\times10^3$	&	12	&	-	&	-	&	2.86$\times10^3$	&	20	&	365	&	9\\

80	&	60	&	100	&	253	&	6	&	5.12$\times10^3$	&	16	&	814	&	8	&	-	&	100	&	1.07$\times10^3$	&	12	&	184	&	6\\

80	&	80	&	100	&	123	&	4	&	4.87$\times10^3$	&	16	&	447	&	6	&	-	&	-	&	1.08$\times10^3$	&	16	&	132	&	7\\

80	&	100	&	100	&	356	&	7	&	4.87$\times10^3$	&	16	&	2.32$\times10^3$	&	13	&	-	&	-	&	959	&	15	&	171	&	8\\

80	&	120	&	100	&	311	&	7	&	4.87$\times10^3$	&	16	&	1.8$\times10^3$	&	12	&	-	&	-	&	353	&	11	&	77	&	6\\

80	&	140	&	100	&	257	&	6	&	2.69$\times10^3$	&	11	&	1.83$\times10^3$	&	11	&	-	&	100	&	435	&	10	&	91	&	6\\

80	&	160	&	100	&	401	&	10	&	1.58$\times10^3$	&	11	&	480	&	8	&	-	&	-	&	154	&	6	&	31	&	3\\

80	&	180	&	100	&	296	&	8	&	1.72$\times10^3$	&	12	&	257	&	6	&	-	&	100	&	168	&	9	&	25	&	4\\

80	&	200	&	100	&	213	&	7	&	986	&	9	&	167	&	4	&	-	&	100	&	135	&	8	&	21	&	4\\

80	&	220	&	100	&	223	&	7	&	978	&	9	&	119	&	4	&	-	&	-	&	131	&	8	&	16	&	3\\

80	&	240	&	100	&	183	&	7	&	715	&	8	&	89	&	3	&	-	&	100	&	102	&	7	&	13	&	3\\

80	&	260	&	100	&	170	&	6	&	662	&	7	&	73	&	3	&	-	&	100	&	107	&	7	&	12	&	3\\

80	&	280	&	100	&	140	&	6	&	572	&	7	&	60	&	3	&	-	&	100	&	111	&	7	&	11	&	3
					\end{tabular}%
}
\caption{MFDM table of limits on the cross sections for A,B and C samples, for $40<m_{D1}<80$ GeV, $1<\Delta m^+<280$ GeV, $\Delta m^0=100$ GeV.}
\label{tab:mfdm_exc_7}
\end{center}
\end{table}
\clearpage

\begin{table}[!ht]
	\begin{center}
		\resizebox{\columnwidth}{!}{%
			\begin{tabular}{|l|l|l||l|l|l|l|l|l||l|l|l|l|l|l|}
				$m_{D1}$ &  $\Delta m^+$ & $\Delta m^0$ & 2$\ell$ $\sigma^{95}_A$ (fb) &$\frac{100}{\sqrt{N_{MC}}}$ & 2$\ell$ $\sigma^{95}_B$ (fb)&$\frac{100}{\sqrt{N_{MC}}}$ & 2$\ell$ $\sigma^{95}_C$ (fb)&$\frac{100}{\sqrt{N_{MC}}}$ & 3$\ell$ $\sigma^{95}_A$ (fb)&$\frac{100}{\sqrt{N_{MC}}}$ & 3$\ell$ $\sigma^{95}_B$ (fb)&$\frac{100}{\sqrt{N_{MC}}}$ & 3$\ell$ $\sigma^{95}_C$ (fb)&$\frac{100}{\sqrt{N_{MC}}}$ \\
				\hline \hline
100	&	10	&	100	&	522	&	29	&	1.88$\times10^3$	&	32	&	270	&	15	&	-	&	-	&	-	&	-	&	-	&	-\\
100	&	20	&	100	&	195	&	18	&	1.95$\times10^3$	&	33	&	247	&	15	&	-	&	-	&	1.03$\times10^4$	&	22	&	4.05$\times10^3$	&	17\\

100	&	40	&	100	&	764	&	12	&	3.49$\times10^3$	&	14	&	1.28$\times10^3$	&	11	&	-	&	100	&	2.47$\times10^3$	&	11	&	460	&	6\\

100	&	60	&	100	&	211	&	6	&	4.25$\times10^3$	&	15	&	632	&	7	&	1.14$\times10^4$	&	71	&	1.18$\times10^3$	&	13	&	162	&	6\\

100	&	80	&	100	&	119	&	4	&	3.70$\times10^3$	&	14	&	418	&	6	&	2.58$\times10^3$	&	71	&	418	&	16	&	103	&	10\\

100	&	100	&	100	&	261	&	6	&	3.84$\times10^3$	&	14	&	2.27$\times10^3$	&	13	&	-	&	-	&	827	&	14	&	182	&	8\\

100	&	120	&	100	&	348	&	7	&	2.43$\times10^3$	&	10	&	1.57$\times10^3$	&	10	&	-	&	-	&	349	&	11	&	74	&	6\\

100	&	140	&	100	&	162	&	5	&	2.85$\times10^3$	&	12	&	1.43$\times10^3$	&	10	&	-	&	-	&	172	&	6	&	41	&	4\\

100	&	160	&	100	&	302	&	9	&	1.87$\times10^3$	&	12	&	413	&	7	&	-	&	100	&	193	&	10	&	32	&	5\\

100	&	180	&	100	&	224	&	7	&	1.19$\times10^3$	&	10	&	193	&	5	&	-	&	100	&	132	&	5	&	21	&	3\\

100	&	200	&	100	&	221	&	7	&	941	&	9	&	136	&	4	&	-	&	-	&	117	&	7	&	17	&	3\\

100	&	220	&	100	&	186	&	7	&	715	&	8	&	104	&	4	&	-	&	100	&	115	&	7	&	15	&	3\\

100	&	240	&	100	&	163	&	6	&	673	&	7	&	85	&	3	&	-	&	-	&	98	&	7	&	12	&	3\\

100	&	260	&	100	&	154	&	6	&	666	&	7	&	65	&	3	&	-	&	100	&	123	&	5	&	11	&	2\\

100	&	280	&	100	&	134	&	6	&	526	&	6	&	53	&	3	&	-	&	100	&	104	&	7	&	10	&	3\\

120	&	10	&	100	&	202	&	19	&	548	&	18	&	132	&	11	&	-	&	-	&	-	&	-	&	-	&	-\\
120	&	20	&	100	&	266	&	21	&	1.60$\times10^3$	&	30	&	214	&	14	&	-	&	-	&	1.37$\times10^4$	&	26	&	2.60$\times10^3$	&	14\\

120	&	40	&	100	&	542	&	10	&	4.53$\times10^3$	&	16	&	1.01$\times10^3$	&	10	&	-	&	-	&	2.56$\times10^3$	&	11	&	433	&	6\\

120	&	60	&	100	&	208	&	6	&	3.77$\times10^3$	&	14	&	628	&	7	&	-	&	-	&	1.11$\times10^3$	&	13	&	173	&	6\\

120	&	80	&	100	&	110	&	4	&	3.33$\times10^3$	&	13	&	431	&	6	&	627	&	35	&	836	&	24	&	112	&	11\\

120	&	100	&	100	&	439	&	10	&	4.04$\times10^3$	&	18	&	1.38$\times10^3$	&	13	&	-	&	-	&	619	&	11	&	135	&	6\\

120	&	120	&	100	&	421	&	10	&	2.61$\times10^3$	&	14	&	914	&	10	&	-	&	-	&	258	&	9	&	66	&	6\\

120	&	140	&	100	&	306	&	9	&	1.84$\times10^3$	&	12	&	546	&	8	&	-	&	-	&	178	&	6	&	35	&	3\\

120	&	160	&	100	&	233	&	7	&	1.67$\times10^3$	&	12	&	331	&	6	&	-	&	-	&	138	&	6	&	26	&	3\\

120	&	180	&	100	&	218	&	7	&	1.08$\times10^3$	&	9	&	198	&	5	&	-	&	100	&	114	&	5	&	19	&	3\\

120	&	200	&	100	&	205	&	7	&	823	&	8	&	125	&	4	&	-	&	-	&	118	&	8	&	16	&	3\\

120	&	220	&	100	&	147	&	6	&	740	&	8	&	87	&	3	&	3.48$\times10^3$	&	71	&	105	&	7	&	14	&	3\\

120	&	240	&	100	&	136	&	6	&	593	&	7	&	69	&	3	&	-	&	-	&	117	&	5	&	12	&	2\\

120	&	260	&	100	&	144	&	6	&	564	&	7	&	58	&	3	&	-	&	-	&	121	&	5	&	11	&	2\\

120	&	280	&	100	&	137	&	6	&	499	&	6	&	50	&	2	&	2.36$\times10^3$	&	71	&	88	&	8	&	8	&	3\\

140	&	10	&	100	&	376	&	15	&	2.06$\times10^3$	&	21	&	259	&	9	&	-	&	-	&	-	&	-	&	-	&	-\\
140	&	20	&	100	&	271	&	24	&	2.09$\times10^3$	&	38	&	241	&	16	&	-	&	-	&	-	&	-	&	-	&	-\\

140	&	40	&	100	&	507	&	10	&	3.35$\times10^3$	&	14	&	731	&	8	&	-	&	-	&	3.11$\times10^3$	&	12	&	484	&	6\\

140	&	60	&	100	&	185	&	5	&	4.34$\times10^3$	&	15	&	739	&	7	&	-	&	100	&	939	&	12	&	148	&	6\\

140	&	80	&	100	&	103	&	4	&	3.12$\times10^3$	&	12	&	465	&	6	&	-	&	100	&	975	&	19	&	167	&	10\\

140	&	100	&	100	&	430	&	10	&	3.05$\times10^3$	&	16	&	1.34$\times10^3$	&	13	&	-	&	-	&	549	&	14	&	114	&	8\\

140	&	120	&	100	&	345	&	9	&	2.50$\times10^3$	&	14	&	677	&	9	&	7.38$\times10^3$	&	71	&	213	&	7	&	53	&	4\\

140	&	140	&	100	&	248	&	8	&	1.87$\times10^3$	&	12	&	382	&	7	&	-	&	100	&	151	&	6	&	32	&	3\\

140	&	160	&	100	&	223	&	7	&	1.35$\times10^3$	&	10	&	293	&	6	&	7.38$\times10^3$	&	71	&	137	&	6	&	26	&	3\\

140	&	180	&	100	&	184	&	7	&	1.06$\times10^3$	&	9	&	162	&	4	&	-	&	-	&	115	&	5	&	18	&	2\\

140	&	200	&	100	&	172	&	6	&	840	&	8	&	115	&	4	&	-	&	100	&	124	&	5	&	14	&	2\\

140	&	220	&	100	&	159	&	6	&	688	&	7	&	80	&	3	&	3.48$\times10^3$	&	71	&	110	&	7	&	13	&	3\\

140	&	240	&	100	&	142	&	6	&	590	&	7	&	62	&	3	&	-	&	-	&	111	&	5	&	11	&	2\\

140	&	260	&	100	&	120	&	5	&	614	&	7	&	54	&	3	&	-	&	-	&	129	&	5	&	10	&	2\\

140	&	280	&	100	&	124	&	5	&	474	&	6	&	47	&	2	&	-	&	100	&	127	&	8	&	10	&	3
					\end{tabular}%
}
\caption{MFDM table of limits on the cross sections for A,B and C samples, for $100<m_{D1}<140$ GeV, $1<\Delta m^+<280$ GeV, $\Delta m^0=100$ GeV.}
\label{tab:mfdm_exc_8}
\end{center}
\end{table}
\clearpage

\begin{table}[!ht]
	\begin{center}
		\resizebox{\columnwidth}{!}{%
			\begin{tabular}{|l|l|l||l|l|l|l|l|l||l|l|l|l|l|l|}
				$m_{D1}$ &  $\Delta m^+$ & $\Delta m^0$ & 2$\ell$ $\sigma^{95}_A$ (fb) &$\frac{100}{\sqrt{N_{MC}}}$ & 2$\ell$ $\sigma^{95}_B$ (fb)&$\frac{100}{\sqrt{N_{MC}}}$ & 2$\ell$ $\sigma^{95}_C$ (fb)&$\frac{100}{\sqrt{N_{MC}}}$ & 3$\ell$ $\sigma^{95}_A$ (fb)&$\frac{100}{\sqrt{N_{MC}}}$ & 3$\ell$ $\sigma^{95}_B$ (fb)&$\frac{100}{\sqrt{N_{MC}}}$ & 3$\ell$ $\sigma^{95}_C$ (fb)&$\frac{100}{\sqrt{N_{MC}}}$ \\
				\hline \hline
160	&	10	&	100	&	196	&	18	&	1.71$\times10^3$	&	30	&	196	&	13	&	-	&	-	&	-	&	-	&	-	&	-\\
160	&	20	&	100	&	217	&	19	&	1.17$\times10^3$	&	26	&	194	&	13	&	-	&	100	&	2.40$\times10^4$	&	71	&	7.19$\times10^3$	&	47\\

160	&	40	&	100	&	386	&	12	&	5.37$\times10^3$	&	27	&	725	&	12	&	-	&	-	&	3.36$\times10^3$	&	13	&	416	&	6\\

160	&	60	&	100	&	213	&	6	&	4.16$\times10^3$	&	14	&	659	&	7	&	-	&	-	&	1.05$\times10^3$	&	12	&	174	&	6\\

160	&	80	&	100	&	98	&	4	&	3.56$\times10^3$	&	13	&	409	&	6	&	386	&	28	&	716	&	22	&	129	&	11\\

160	&	100	&	100	&	394	&	10	&	3.05$\times10^3$	&	16	&	1.04$\times10^3$	&	11	&	-	&	-	&	396	&	9	&	101	&	6\\

160	&	120	&	100	&	267	&	8	&	2.66$\times10^3$	&	15	&	672	&	9	&	-	&	-	&	190	&	7	&	50	&	4\\

160	&	140	&	100	&	250	&	8	&	1.47$\times10^3$	&	11	&	397	&	7	&	3.48$\times10^3$	&	71	&	177	&	9	&	38	&	5\\

160	&	160	&	100	&	195	&	7	&	1.24$\times10^3$	&	10	&	243	&	5	&	-	&	-	&	132	&	5	&	22	&	3\\

160	&	180	&	100	&	170	&	6	&	1.00$\times10^3$	&	9	&	142	&	4	&	-	&	100	&	117	&	5	&	18	&	2\\

160	&	200	&	100	&	155	&	6	&	745	&	8	&	106	&	4	&	4.92$\times10^3$	&	58	&	121	&	5	&	14	&	2\\

160	&	220	&	100	&	133	&	6	&	649	&	7	&	72	&	3	&	-	&	-	&	122	&	5	&	13	&	2\\

160	&	240	&	100	&	125	&	5	&	569	&	7	&	60	&	3	&	2.37$\times10^3$	&	71	&	135	&	10	&	12	&	4\\

160	&	260	&	100	&	112	&	5	&	533	&	7	&	52	&	2	&	3.47$\times10^3$	&	71	&	113	&	7	&	10	&	3\\

160	&	280	&	100	&	111	&	5	&	483	&	6	&	44	&	2	&	-	&	100	&	81	&	8	&	8	&	3\\

180	&	10	&	100	&	310	&	14	&	1.58$\times10^3$	&	18	&	209	&	8	&	-	&	-	&	-	&	-	&	-	&	-\\
180	&	20	&	100	&	225	&	20	&	1.35$\times10^3$	&	28	&	142	&	11	&	1.25$\times10^5$	&	71	&	2.51$\times10^5$	&	58	&	2.26$\times10^5$	&	67\\

180	&	40	&	100	&	339	&	12	&	3.76$\times10^3$	&	22	&	566	&	11	&	-	&	-	&	2.44$\times10^3$	&	11	&	421	&	6
					\end{tabular}%
}
\caption{MFDM table of limits on the cross sections for A,B and C samples, for $160<m_{D1}<180$ GeV, $1<\Delta m^+<280$ GeV, $\Delta m^0=100$ GeV.}
\label{tab:mfdm_exc_9}
\end{center}
\end{table}
\clearpage

\section{Example Analyses and Cutflows
\label{app:cutflows}}
In appendix~\ref{app:cutflows} we present examples of the 13 TeV analyses used for the presented exclusion limits and their cutflows as implemented in CheckMATE.

\textbf{Analysis: cms\_sus\_16\_048,  Signal Region: SR1\_weakino\_1low\_mll\_1}

\begin{table}[!ht]
	\centering
	\begin{tabular}{l| c|}
		&Cuts\\
		\hline
		\hline
		$mT(\ell[0]/\ell[1],\ E_T^{miss})$& $< 70$ GeV\\
		$mT(p_{\ell[0]},\ E_T^{miss}$&$< 70$ GeV\\ 
		$mT(p_{\ell[1]},\ E_T^{miss})$&$< 70$ GeV\\ 
		$E_T^{miss}$   &$<200$ GeV\\
		$M_{\ell\ell}$&$<10$ GeV\\
		
	\end{tabular}
	\caption{Cutflow for cms\_sus\_16\_048, SR1\_weakino\_1low\_mll\_1}
	\label{tab:cms_sus_16_048_weakino_1low_mll_1}
\end{table}

\textbf{Analysis: {\tt cms\_sus\_16039}, Signal Region: {\tt SR\_A02}}

Two out of the three leptons ($e$ or $\mu$) will form an OSSF pair. This is signal region “A”
One of the $m_{\ell\ell}$ bins is defined to be below the Z mass ($M_ll < 75$ GeV).

\begin{table}[!ht]
	\centering
	\begin{tabular}{l| c|}
		&Cuts\\
		\hline
		\hline
		$MT$&$0 <MT< 100$ GeV\\
		$p_T^{miss}$&$200<p_T^{miss}< 250$ GeV\\
		$M_{\ell\ell}$&$<75$ GeV
		
	\end{tabular}
	\caption{Cutflow for {\tt cms\_sus\_16039}, {\tt SR\_A02}}
	\label{tab:cms_sus_16_039_A02}
\end{table}

\textbf{Analysis: cms\_sus\_16\_025, Signal Region: SR2\_stop\_1low\_pt\_1}

The second category corresponds to the two leptons stemming from two different particles, as
in the decays of two top squarks, or in two cascades like $\chi^\pm$ to $W,\ \chi_1$. In these cases, the leptons
are not required to have the same flavour.

\begin{table}[!ht]
	\centering
	\begin{tabular}{l| c|}
		&Cuts\\
		\hline
		\hline
		$E_T^{miss}$&$<200$GeV\\
		$p_T(\ell[0])$& $< 12$ GeV
	
\end{tabular}
\caption{Cutflow for {\tt cms\_sus\_16\_025}, {\tt SR2\_stop\_1low\_pt\_1}}
\label{tab:cms_sus_16_025_stop_1low_pt_1}
\end{table}

\textbf{Analysis: atlas\_conf\_2016\_096, Signal Region: 2LASF}

This analysis searches for $D^+,D^-$ with two lepton final states.

\begin{table}[!ht]
\centering
\begin{tabular}{l| c|}
	&Cuts\\
	\hline
	\hline
	$|mll-mZ|$&$> 10$ GeV\\
	$mT2$&$ > 90$ GeV
	
\end{tabular}
\caption{Cutflow for {\tt atlas\_conf\_2016\_096}, {\tt 2LASF}}
\label{tab:atlas_conf_2016_096_2LASF}
\end{table}

\textbf{Analysis: atlas\_conf\_2016\_096, Signal Region: 3LI}

This signal region consists of three leptons, targeting the intermediate mass splitting between D2, D1 of order ~$2\times m_Z$.
\begin{table}[!htb]
\centering
\begin{tabular}{l| c|}
	&Cuts\\
	\hline
	\hline
	$mT$&  $ > 110$ GeV\\
	
	$M(SFOS), E$&$ [81.2, 101.2]$ GeV\\
	
	$pT(3rd\ lep)$&$>30$ GeV\\
	
	$E_T^{miss} $&$>120$ GeV
\end{tabular}
\caption{Cutflow for {\tt atlas\_conf\_2016\_096}, {\tt 3LI}}
\label{tab:atlas_conf_2016_096_3LI}
\end{table}

\textbf{Analysis: cms\_sus\_16\_039, Signal Region: SR\_A03}

This analysis looks for electrowekinos in multilepton final states. This particular signal region requires three $e/\mu$ forming at least one opposite-sign same-flavour pair.
\begin{table}[!ht]
\centering
\begin{tabular}{l| c|}
	&Cuts\\
	\hline
	\hline
	$MT$&$0 <MT< 100$ GeV\\
	
	$p_T^{miss}$&$150 <p_T^{miss}< 200 $\\
	
	$M_{\ell\ell}$&$ < 75$ GeV
\end{tabular}
\caption{Cutflow for {\tt cms\_sus\_16\_039}, {\tt SR\_A03}}
\label{tab:cms_sus_16_039_A03}
\end{table}

\bibliographystyle{unsrt}
\bibliography{Bib}
\end{document}